\documentclass[12pt]{article}
\usepackage{graphicx}
\usepackage{dcolumn}
\usepackage{bm}
\usepackage{color}
\usepackage{amssymb}
\usepackage{amsmath}
\usepackage{natbib}
\usepackage{multicol,times,subfigure,multirow}

\begin{document}

\title{\textbf{Universal price impact functions of individual trades in an order-driven market}}
\author{WEI-XING ZHOU{\small$^{\mbox{\ref{SB}, \ref{SS},
\ref{RCE}, \ref{RCSE}, \footnote{Corresponding author. Email:
wxzhou@ecust.edu.cn}}}$}}

\date{}

\maketitle

\vspace{-9mm}

\begin{enumerate}
\item{School of Business, East China University of Science and Technology, Shanghai 200237, China}\vspace{-3mm}\label{SB}%
\item{School of Science, East China University of Science and Technology, Shanghai 200237, China}\vspace{-3mm}\label{SS}%
\item{Research Center for Econophysics, East China University of Science and Technology, Shanghai 200237, China}\vspace{-3mm}\label{RCE}%
\item{Research Center of Systems Engineering, East China University of Science and Technology, Shanghai 200237, China}\vspace{-3mm}\label{RCSE}%
\end{enumerate}

\vspace{1mm}
\begin{center}
({\em{Received 7 October 2007; in final form }})
\end{center}
\vspace{1mm}

\begin{quote}
The trade size $\omega$ has direct impact on the price formation of
the stock traded. Econophysical analyses of transaction data for the
US and Australian stock markets have uncovered market-specific
scaling laws, where a master curve of price impact can be obtained
in each market when stock capitalization $C$ is included as an
argument in the scaling relation. However, the rationale of
introducing stock capitalization in the scaling is unclear and the
anomalous negative correlation between price change $r$ and trade
size $\omega$ for small trades is unexplained. Here we show that
these issues can be addressed by taking into account the
aggressiveness of orders that result in trades together with a
proper normalization technique. Using order book data from the
Chinese market, we show that trades from filled and partially filled
limit orders have very different price impact. The price impact of
trades from partially filled orders is constant when the volume is
not too large, while that of filled orders shows power-law behavior
$r\sim \omega^\alpha$ with $\alpha\approx2/3$. When returns and
volumes are normalized by stock-dependent averages,
capitalization-independent scaling laws emerge for both types of
trades. However, no scaling relation in terms of stock
capitalization can be constructed. In addition, the relation
$\alpha=\alpha_\omega/\alpha_r$ is verified, where $\alpha_\omega$
and $\alpha_r$ are the tail exponents of trade sizes and returns.
These observations also enable us to explain the anomalous negative
correlation between $r$ and $\omega$ for small-size trades. We
anticipate that these regularities may hold in other order-driven
markets.

\end{quote}

\section{Introduction}
\label{s1:intro}


A well-known adage in the Wall Street states that it takes trading
volume to move stock prices. The price-volume relation has been
extensively studied, which composes of the positive correlations
between trading volume $\omega(t,\Delta{t})$ and the volatility (or
the absolute value of return) $|r(t,\Delta{t})|$ (the positive
volume-volatility relation), and between trading volume and the
return {\em{per se}} $r(t,\Delta{t})$ (the positive volume-return
relation) over fixed time intervals $(t-\Delta{t},t]$ in financial
markets \citep{Karpoff-1987-JFQA}. These relations are robust at
various time scales including minutely
\citep{Wood-McInish-Ord-1985-JF}, hourly \citep{Jain-Joh-1988-JFQA},
daily
\citep{Ying-1966-Em,Epps-1977-JFQA,Harris-1987-JFQA,Gallant-Rossi-Tauchen-1992-RFS},
weekly \citep{Richardson-Sefcik-Thompson-1986-JFE}, to monthly
horizon \citep{Rogalski-1978-RES,Saatcioglu-Starks-1998-IJF}. In
these studies, the main result is that the return $r(\omega)$ is
linearly correlated with the volume $\omega$ and the ratio of price
change to trading volume is greater for trades driving price down
than for those driving price up \citep{Karpoff-1987-JFQA}. The
price-volume relation is of significant relevance to the mixture of
distributions hypothesis of returns
\citep{Clark-1973-Em,Epps-Epps-1976-Em}.

In recent years, the price-volume relation has been studied at the
transaction level
\citep{Chan-Fong-2000-JFE,Lillo-Farmer-Mantegna-2003-Nature,Lim-Coggins-2005-QF,Naes-Skjeltorp-2006-JFinM}.
It is well-known that buyer-initiated trades drive the price up and
seller-initiated trades drive it down. The difference between the
volume-return relation and the volume-volatility relation vanishes
at the transaction level. Equivalently, investigating the
price-volume relation amounts to the study of the immediate price
impact of trades of size $\omega$. Contrary to the linear
volume-return assumption at aggregated timescales,
\citet{Lillo-Farmer-Mantegna-2003-Nature} found that the price
impact is nonlinear and concave for US stocks at the transaction
level.

Concerning the volume-return relation at the microscopic transaction
level, one is interested in the determination of the immediate
impact of trade size on the price. Using the Trade and Quote (TAQ)
database, \citet{Lillo-Farmer-Mantegna-2003-Nature} unveiled a
master curve for price impact function in the sense that the data
collapse onto a single curve (LFM scaling)
\begin{equation}\label{Eq:rC}
    r\left(\frac{\omega}{\langle\omega\rangle},C\right)
    =C^{-\gamma}f\left(\frac{\omega}{\langle\omega\rangle}\frac{1}{C^\delta}\right)~,
\end{equation}
where $r$ is the shift of logarithmic mid-quote prices right before
and after a trade of size $\omega$ occurs, $\langle\omega\rangle$ is
the average volume per trade for each stock, $C$ is the stock
capitalization, $\gamma$ and $\delta$ are two scaling parameters,
and $f(x)$ is found to be a concave function and has a power-law
form for large $\omega$. \citet{Lim-Coggins-2005-QF} performed a
similar analysis on the Australian Stock Exchange and a similar
scaling was obtained, where the return and trade size were defined
in a slightly different way. More interestingly, $r(\omega)$ is
found to decrease with increasing $\omega$ in the Australian market
when $\omega$ is smaller than certain value
\citep{Lim-Coggins-2005-QF}. However, the underlying mechanism
causing this counterintuitive behavior is unknown.

The power-law impact function for large trades plays an essential
role in an unified theory explaining power laws of financial market
fluctuations
\citep{Gabaix-Gopikrishnan-Plerou-Stanley-2003-Nature,Gabaix-Gopikrishnan-Plerou-Stanley-2003-PA,Farmer-Lillo-2004-QF,Plerou-Gopikrishnan-Gabaix-Stanley-2004-QF,Gabaix-Gopikrishnan-Plerou-Stanley-2006-QJE,Gabaix-Gopikrishnan-Plerou-Stanley-2007-JEEA}.
It is found that the distribution of equity returns obeys the
(inverse) cubic law:
\begin{equation}\label{Eq:Pr:r}
\Pr(|r|>x) \sim x^{-\alpha_r}~,
\end{equation}
where $\alpha_r\approx 3$
\citep{Gopikrishnan-Meyer-Amaral-Stanley-1998-EPJB,Gopikrishnan-Plerou-Amaral-Meyer-Stanley-1999-PRE,Plerou-Gopikrishnan-Amaral-Meyer-Stanley-1999-PRE}
and that of trading volumes fulfills the half-cubic law
\begin{equation}\label{Eq:Pr:w}
\Pr(\omega>x) \sim x^{-\alpha_\omega}
\end{equation}
with $\alpha_\omega \approx 3/2$
\citep{Gopikrishnan-Plerou-Gabaix-Stanley-2000-PRE}. If the price
impact function has a power-law form $|r| \sim \omega^\alpha$, we
have immediately that
\begin{equation}\label{Eq:alpha}
\alpha=\alpha_\omega/\alpha_r~,
\end{equation}
which bridges the two tail exponents for returns and volumes. The
unified theory of
\citep{Gabaix-Gopikrishnan-Plerou-Stanley-2003-Nature} can thus be
tested empirically.

In this work we report on the existence of two universal
price-impact functions of two types of trades in an order-driven
stock market, which do not depend on the stock capitalization. We
also show that the relation (\ref{Eq:alpha}) put forward originally
by \citet{Gabaix-Gopikrishnan-Plerou-Stanley-2003-Nature} can be
verified also at the transaction level. The rest of the paper is
organized as follows. Section \ref{S1:Data} describes the data used
in our study and briefly the trading system. Section \ref{S1:Zhou}
investigate the price impact functions and data collapse based on a
proper normalization approach. In section \ref{S1:LFM:LC}, we adopt
the methods utilized by \citet{Lillo-Farmer-Mantegna-2003-Nature}
and \citet{Lim-Coggins-2005-QF} for comparison. The relation between
different power-law exponents is tested in section \ref{S1:alpha}
using at the level of individual stocks and aggregate data of all
stocks as well. Section \ref{S1:Conc} summarizes and concludes.

\section{Description of data sets}
\label{S1:Data}

We use a database recording all orders of submission and cancelation
of 23 A-share stocks traded on the Shenzhen Stock Exchange in year
2003. The Shenzhen Stock Exchange (SZSE) was established on December
1, 1990 and started its operations on July 3, 1991. There are two
separate markets for A-shares and B-shares. A-shares are common
stocks issued by mainland Chinese companies, subscribed and traded
in Chinese currency {\em{Renminbi}} (RMB), listed on mainland
Chinese stock exchanges, bought and sold by Chinese nationals and
approved foreign investors. The A-share market was launched in 1990
and opened only to domestic investors in 2003. B-shares are issued
by mainland Chinese companies, traded in foreign currencies and
listed on mainland Chinese stock exchanges. B-shares carry a face
value denominated in RMB. The B Share Market was launched in 1992
and was restricted to foreign investors before February 19, 2001. It
has been opened to Chinese investors since.

The Chinese markets have grown rapidly together with China's
economy. At the end of 2003, there were 491 A-share stocks and 57
B-share stocks listed on the SZSE. Concerning the A-share market,
the total market capitalization was 173.28 billion shares and the
float market capitalization was 65.57 billion shares
\citep{SZSE-2004}. At the end of March 2008, the total market
capitalization for 686 A-share stocks increased to 290.26 billion
shares and the float market capitalization to 158.45 billion shares.
Our sample stocks were part of the 40 constituent stocks included in
the Shenshen Stock Exchange Component Index in 2003. The total
market capitalization $C_{\rm{tot}}$ and float capitalization $C$,
both in unit of million shares, of individual stocks are listed in
table \ref{Tb:BasicStat}. We note that $C_{\rm{tot}}$ varies from
275.9 to 1994.1 and $C$ ranges from 107.1 to 1406.5 spaning over one
order of magnitude.

\begin{table}[h!]
\centering \caption{\label{Tb:BasicStat}Basic statistics for the 23
SZSE. Shown in columns are the codes of stocks, the total market
capitalization ($C_{\rm{tot}}$, million shares), the float market
capitalization ($C$, million shares), the annual turnovers ($z\%$),
the average returns ($\langle{r_{\rm{PB}}}\rangle$,
$\langle{r_{\rm{PS}}}\rangle$, $\langle{r_{\rm{FB}}}\rangle$, and
$\langle{r_{\rm{FS}}}\rangle$, multiplied by 1000) caused by filled
and partially filled buys and sells, the numbers of trades ($N$,
thousand shares), and the industry sectors. }
\bigskip
\begin{tabular}{cr@{.}lr@{.}lr@{.}lccccr@{.}ll}
  \hline\hline
   Code & \multicolumn{2}{c}{$C_{\rm{tot}}$}&\multicolumn{2}{c}{$C$}&\multicolumn{2}{c}{$z$}
        & $\langle{r_{\rm{PB}}}\rangle$ & $\langle{r_{\rm{PS}}}\rangle$
        & $\langle{r_{\rm{FB}}}\rangle$ & $\langle{r_{\rm{FS}}}\rangle$ &\multicolumn{2}{c}{$N$}  & Industry\\  \hline
000001 &  1945&8 &  1406&5 & 149&9  &  1.19 & -1.21 &  0.03 & -0.05 &   889&7  &           Financials\\
000002 &  1152&3 &   929&5 & 166&8  &  1.55 & -1.54 &  0.04 & -0.05 &   509&4  &          Real estate\\
000009 &   958&8 &   579&1 & 210&5  &  2.23 & -2.25 &  0.05 & -0.06 &   448&0  &        Conglomerates\\
000012 &   377&9 &   107&1 & 529&5  &  1.63 & -1.61 &  0.08 & -0.10 &   290&4  &  Metals \& Nonmetals\\
000016 &   399&1 &   224&0 & 182&3  &  1.86 & -1.87 &  0.08 & -0.10 &   188&6  &          Electronics\\
000021 &   732&9 &   199&5 & 309&3  &  1.35 & -1.37 &  0.07 & -0.09 &   411&6  &          Electronics\\
000024 &   327&2 &   170&2 & 151&9  &  1.65 & -1.63 &  0.09 & -0.09 &   133&6  &          Real estate\\
000027 &  1202&5 &   486&0 & 206&0  &  1.63 & -1.63 &  0.06 & -0.05 &   313&9  &            Utilities\\
000063 &   667&3 &   250&8 & 227&6  &  1.27 & -1.24 &  0.07 & -0.06 &   265&5  &                   IT\\
000066 &   458&5 &   181&0 & 231&6  &  1.57 & -1.59 &  0.08 & -0.09 &   277&7  &          Electronics\\
000088 &   585&0 &   124&9 & 169&5  &  1.63 & -1.60 &  0.10 & -0.09 &    97&2  &       Transportation\\
000089 &   799&8 &   287&8 & 216&6  &  1.63 & -1.64 &  0.06 & -0.07 &   189&1  &       Transportation\\
000406 &   364&0 &   265&8 & 231&5  &  1.56 & -1.58 &  0.05 & -0.07 &   271&4  &       Petrochemicals\\
000429 &   953&4 &   274&1 & 118&2  &  2.35 & -2.37 &  0.08 & -0.09 &   117&4  &       Transportation\\
000488 &   526&1 &   235&6 & 136&8  &  1.72 & -1.62 &  0.14 & -0.17 &   120&1  &    Paper \& Printing\\
000539 &  1994&1 &   391&5 & 114&0  &  1.88 & -1.85 &  0.10 & -0.14 &   114&7  &            Utilities\\
000541 &   275&9 &   146&7 &  95&2  &  1.56 & -1.54 &  0.09 & -0.09 &    68&7  &          Electronics\\
000550 &   519&2 &   117&5 & 604&5  &  1.60 & -1.59 &  0.08 & -0.09 &   346&7  &        Manufacturing\\
000581 &   348&0 &   215&9 & 123&5  &  1.84 & -1.80 &  0.10 & -0.10 &    94&0  &        Manufacturing\\
000625 &   876&7 &   168&0 & 582&8  &  1.60 & -1.61 &  0.08 & -0.10 &   397&6  &        Manufacturing\\
000709 &  1955&0 &   565&8 & 125&7  &  2.02 & -2.04 &  0.04 & -0.05 &   207&8  &  Metals \& Nonmetals\\
000720 &   479&7 &   277&1 &  82&2  &  0.98 & -1.15 &  0.05 & -0.08 &   132&2  &            Utilities\\
000778 &   621&5 &   219&1 & 183&1  &  1.38 & -1.33 &  0.07 & -0.07 &   157&3  &        Manufacturing\\
  \hline\hline
\end{tabular}
\end{table}

The Exchange is open for trading from Monday to Friday except the
public holidays and other dates as announced by the China Securities
Regulatory Commission. On each trading day, the trading time period
is divided into three parts: opening call auction, cooling periods,
and continuous double auction. The market opens at 9:15 am and
entered the opening call auction till 9:25 am, during which the
trading system accepts order submission and cancelation, and all
matched transactions are executed at 9:25 am. It is followed by a
cooling period from 9:25 am to 9:30. During cooling periods, the
Exchange is open to orders routing from members, but does not
process orders or process cancelation of orders. The information
released to trading terminals also does not change during cooling
periods. All matched transactions are executed at the end of cooling
periods. The continuous double auction operates from 9:30 to 11:30
and 13:00 to 15:00 and transaction occurs based on automatic one to
one matching due to price-time priority. The time interval between
11:30 am to 13:00 pm is another cooling period. Outside these
opening hours, unexecuted orders will be removed by the system. Only
the trades during the continuous double auction are considered in
this work.

The 23 stocks investigated in this work are representative, which
belong to a variety of industry sectors as shown in table
\ref{Tb:BasicStat}. Although the Chinese stock market was in the
middle of a five-year bearish phase in 2003
\citep{Zhou-Sornette-2004a-PA}, the annual turnovers were still very
high. The annual turnovers of the 23 stocks are tabulated in table
\ref{Tb:BasicStat}. Hence, transactions are quite frequent,
resulting in a large number of trades in our analysis. The total
numbers $N$ of trades for individual stocks are also presented in
table \ref{Tb:BasicStat}.

Before July 1, 2007, only limit orders were allowed for submission
in the Chinese stock markets and the tick sizes of all stocks are
identical to 0.01 RMB. If a trade occurs at time $t$, we compute the
percentage return
\begin{equation}
 r(t+1)=[p(t+1)-p(t)]/p(t)~,
 \label{Eq:rt}
\end{equation}
where $p(t)$ and $p(t+1)$ are the mid-prices of the best bid and ask
right before and after the transaction. Following
\citet{Biais-Hillion-Spatt-1995-JF}, the trades are differentiated
into four types according to their directions (whether a trade is
seller-initiated or buyer-initiated) and aggressiveness:
buyer-initiated partially filled (PB) trades resulting from
partially filled buy orders, seller-initiated partially filled (PS)
trades resulting from partially filled sell orders, buyer-initiated
filled (FB) trades resulting from filled buy orders, and
seller-initiated filled (FS) trades resulting from filled sell
orders. Since the price rises following buyer-initiated trades and
falls following seller-initiated trades, the returns associated with
buyer-initiated trades are non-negative and that with
seller-initiated trades are non-positive.

The average returns corresponding to the four classes of trades for
individual stocks are shown in table \ref{Tb:BasicStat}. We observe
that partially filled trades have much larger impact on the price
than filled trades:
\begin{equation}
 \langle{r_{\rm{PB}}}\rangle \gg \langle{r_{\rm{FB}}}\rangle~~~
 {\rm{and}}~~~-\langle{r_{\rm{PS}}}\rangle\gg -\langle{r_{\rm{FS}}}\rangle~,
 \label{Eq:rr}
\end{equation}
which means that, on average, partially filled trades are much more
aggressive than filled trades. The magnitude of average return of a
partially filled trade is about $0.0016\pm0.0003$ and that of a
filled trade is about $0.00008\pm0.00003$. These very small values
also indicate that the percentage return in Eq.~(\ref{Eq:rt}) equals
to the logarithmic return. We notice that the average number of
shares per trade is
$\omega_{\rm{PB}}\approx\omega_{\rm{PS}}\approx4000$ for partially
filled trades and
$\omega_{\rm{FB}}\approx\omega_{\rm{FS}}\approx2100$ for filled
trades. The fact that partially filled trades have larger sizes
further confirms that they are more aggressive.

\section{Universal price impact functions}
\label{S1:Zhou}

\subsection{Price impact functions for individual stocks}

\begin{figure}[htb]
\centering
\includegraphics[width=6cm]{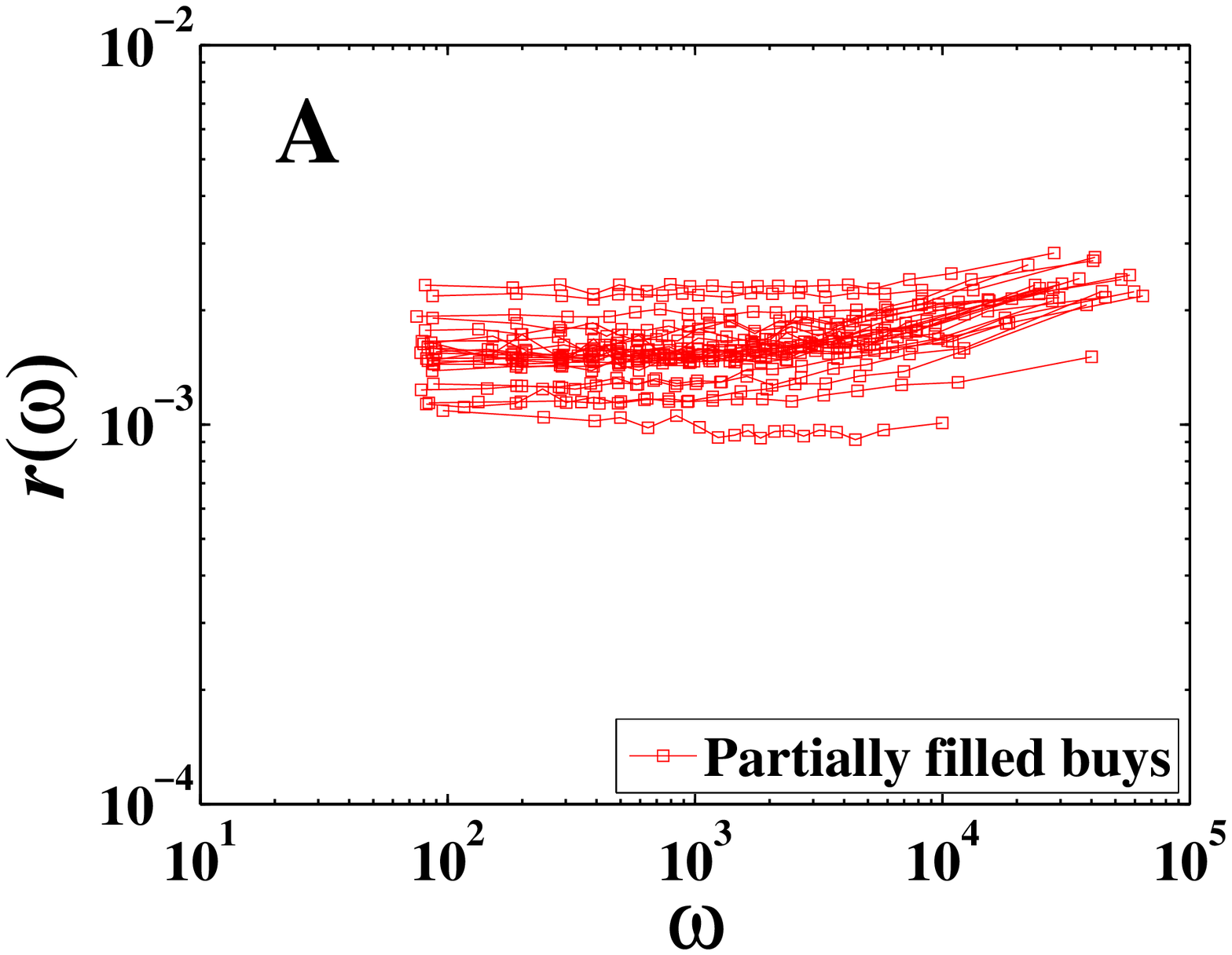}
\includegraphics[width=6cm]{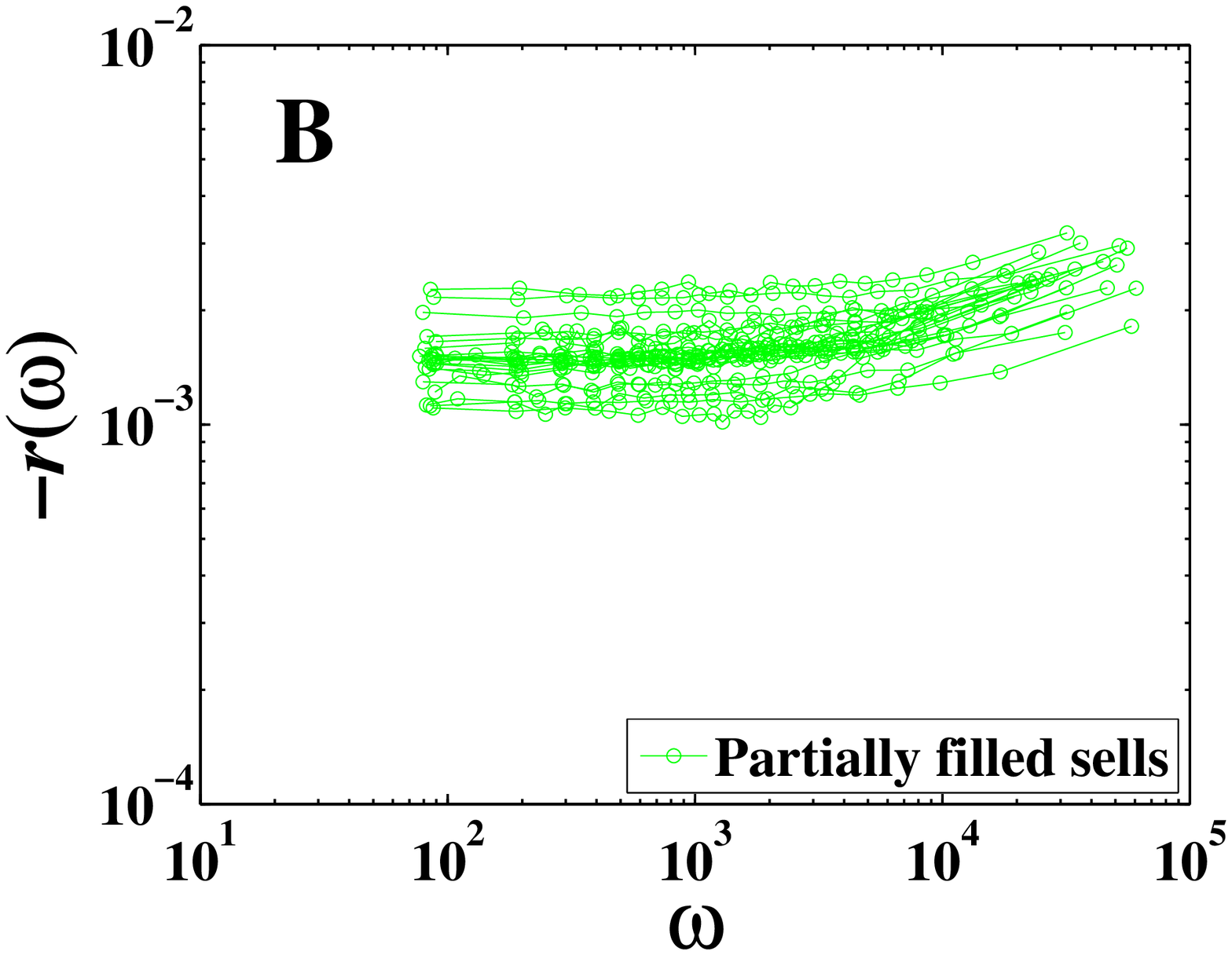}\\
\includegraphics[width=6cm]{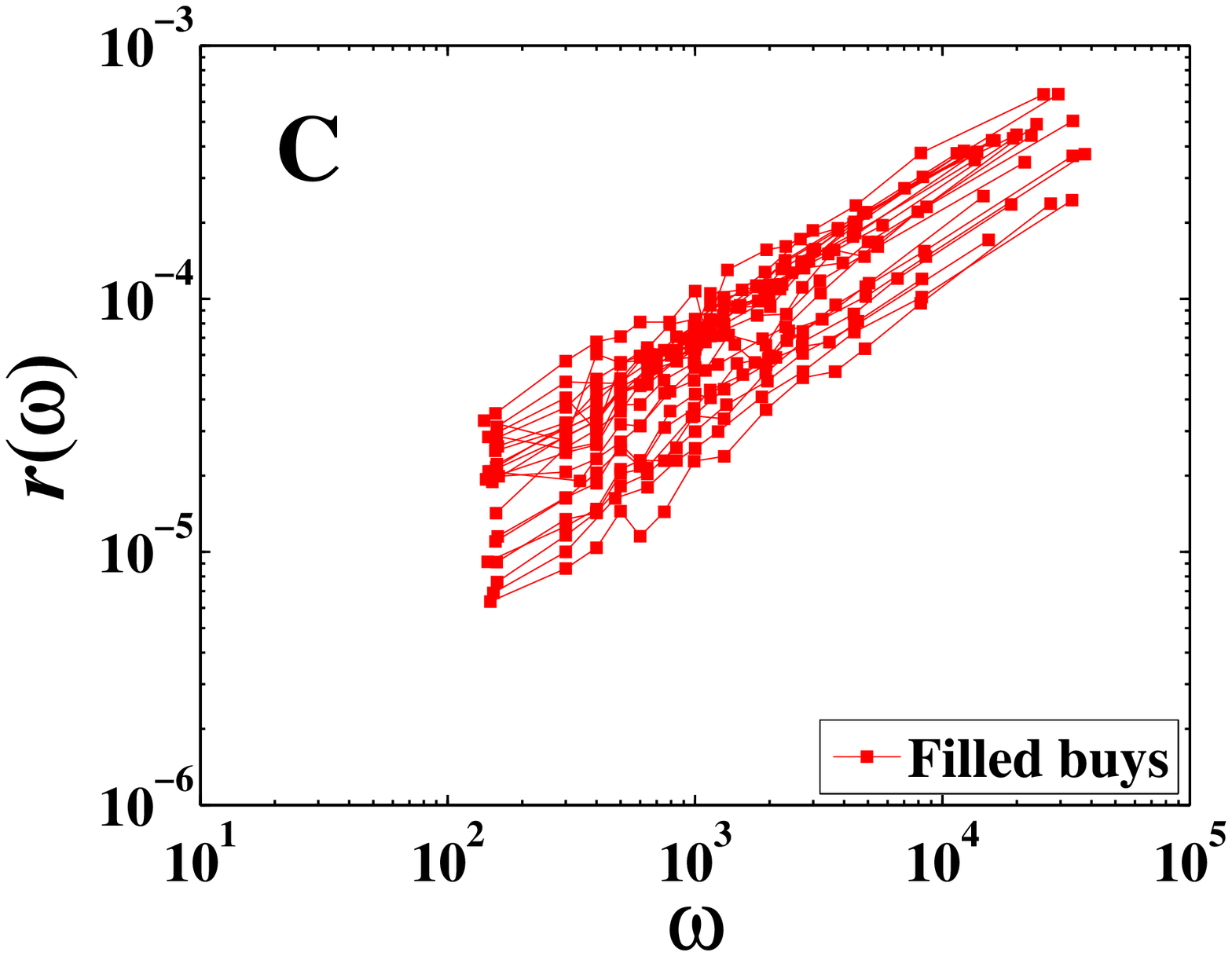}
\includegraphics[width=6cm]{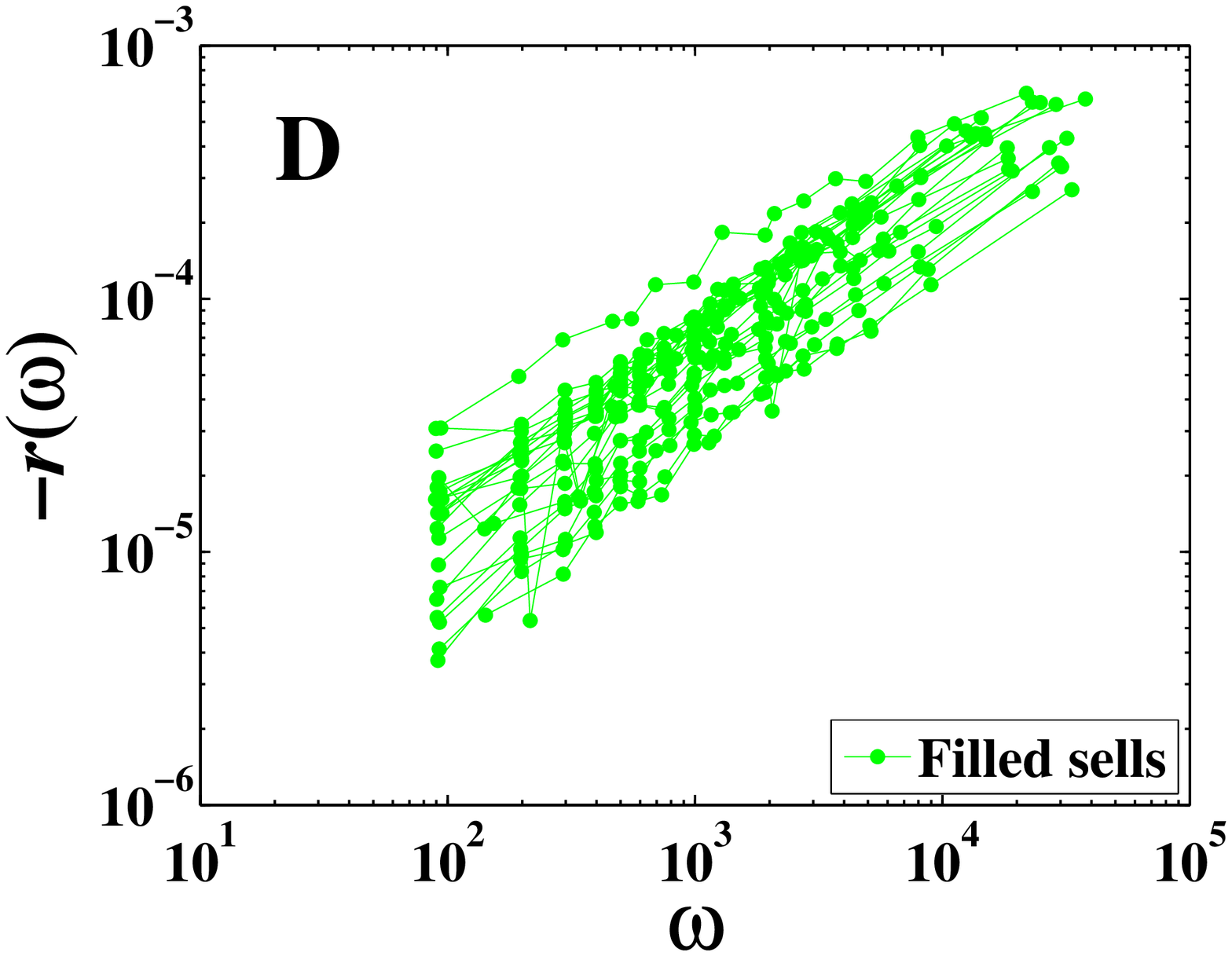}
\caption{The price impact functions of different types of trades for
23 individual stocks traded on the Shenzhen Stock Exchange. The
trades are classified into four types due to their directions and
aggressiveness: (A) buyer-initiated partially filled trades, (B)
seller-initiated partially filled trades, (C) buyer-initiated filled
trades, and (D) seller-initiated filled trades.}
\label{Fig:MPM:Trades:Raw}
\end{figure}

For each stock, we determine the four types of trades (PB, PS, FB
and FS). For each type of trades, we obtain a sequence of paired
points $(\omega, r)$ for the transaction sizes and returns. The data
points are divided into nonoverlapping groups by binning the
$\omega$-axis such that all groups have approximately same number of
points. The means of $r$ and $\omega$ in each group are calculated.
The resulting price impact functions of each type of trades for
individual stocks are illustrated in figure
\ref{Fig:MPM:Trades:Raw}. Several intriguing features arise. We find
that, for each type of trades, the price impact functions for
different stocks have similar shapes showing that these stocks share
similar underlying dynamic regularities. In addition, the shapes of
buy trades and sell trades are similar as well.

However, the shapes of filled trades and partially filled trades are
very different. For partially filled trades, $r(\omega)$ remains
constant when $\omega$ is less than about 10000 shares and then
increases. For filled trades, there are nice power-law relationships
\begin{equation}\label{Eq:r:w}
r(\omega)=A\omega^\alpha
\end{equation}
for the stocks under investigation, where $A$ is a stock-dependent
prefactor and $\alpha$ is the power-law exponent. We have fitted the
power-law relation for each stock, which gives the average values of
power-law exponents $\alpha_{\rm{FB}}=0.65\pm0.08$ and
$\alpha_{\rm{FS}}=0.69\pm0.06$. A closer scrutiny shows that the
stock 000720 does not exhibit nice power-law scaling especially for
FB trades \citep[see][Figure 8]{Zhou-2007-XXX}. A possible reason is
that this stock was argued to be controlled by large investors who
frequently manipulated the price. If we remove this stock, the
average exponents are $\alpha_{\rm{FB}}=0.66\pm0.05$ and
$\alpha_{\rm{FS}}=0.69\pm0.06$, where the standard deviation for FB
trades decreases remarkably.

We also find that partially filled trades have much larger price
impact than filled trades. In other words, partially filled trades
are more aggressive than filled trades. This observation has been
documented in (\ref{Eq:rr}) according to table \ref{Tb:BasicStat}.
There is a trivial ``mechanical'' explanation for this.  Consider a
buy order submitted at time $t$ with price $\pi$ no less than the
best ask price $a_1(t^-)$, before which the ask prices at each price
level of the sell order book were
$a_1(t^-)<a_2(t^-)<a_3(t^-)<\cdots$ and the best bid price was
$b_1(t^-)$. The buy order eats orders waiting on the sell order book
and the lowest price of the remaining orders submitted before $t$ is
$a_n(t^-)$. If the buy order is filled, then the new best ask and
bid prices are $a_1(t^+)=a_n(t^-)$ and $b_1(t^+)=b_1(t^-)$ and the
mid-price change is
\begin{equation}
 \Delta{p}_{\rm{FB}} 
 = [a_n(t^-)-a_1(t^-)]/2~.
\end{equation}
If the buy order is partially filled, then the new best ask and bid
prices are $a_1(t^+)=a_n(t^-)$ and $b_1(t^+)=\pi\geqslant a_1(t^-)$
since the unfilled part is left on the new best bid price. The
mid-price change reads
\begin{equation}
 \Delta{p}_{\rm{PB}} = \Delta{p}_{\rm{FB}} + [\pi-b_1(t^-)]/2~.
\end{equation}
It follows immediately that $\Delta{p}_{\rm{PB}}-
\Delta{p}_{\rm{FB}} = [\pi-b_1(t^-)]/2$, which is larger than half
of the bid-ask spread. On the other hand, $91.05\%$ of the FB trades
have $\Delta{p}_{\rm{FB}}=0$ and all PB trades have
$\Delta{p}_{\rm{PB}}\geqslant 0.005$ RMB, half of the tick size.
These considerations explain why PB trades have much larger price
impact than FB trades. This analysis applies for seller-initiated
trades as well. We note that $89.42\%$ of the FS trades have
$\Delta{p}_{\rm{FS}}=0$.

\subsection{Data collapse of normalized returns and trade sizes}

\begin{figure}[tb]
\centering
\includegraphics[width=8cm]{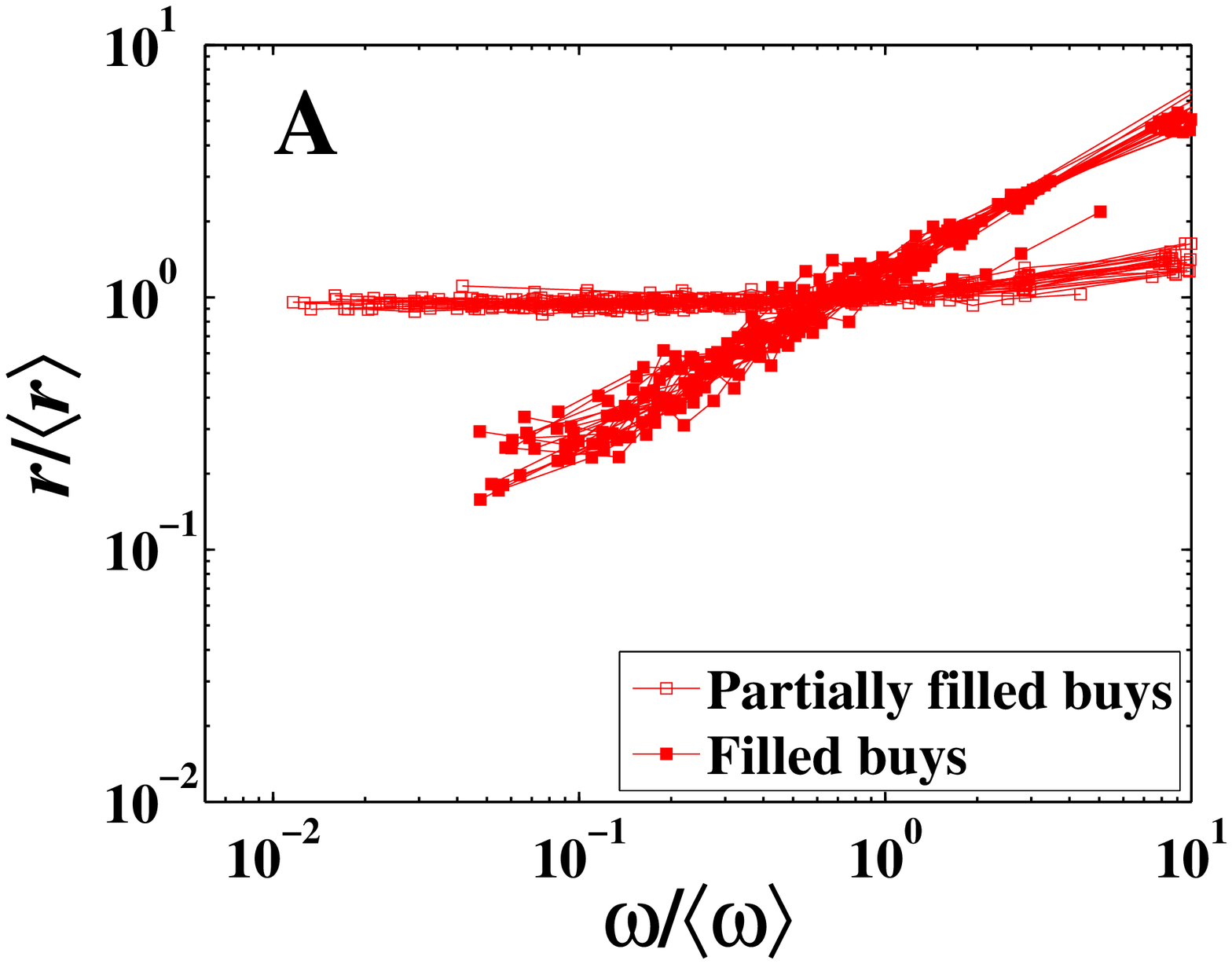}
\includegraphics[width=8cm]{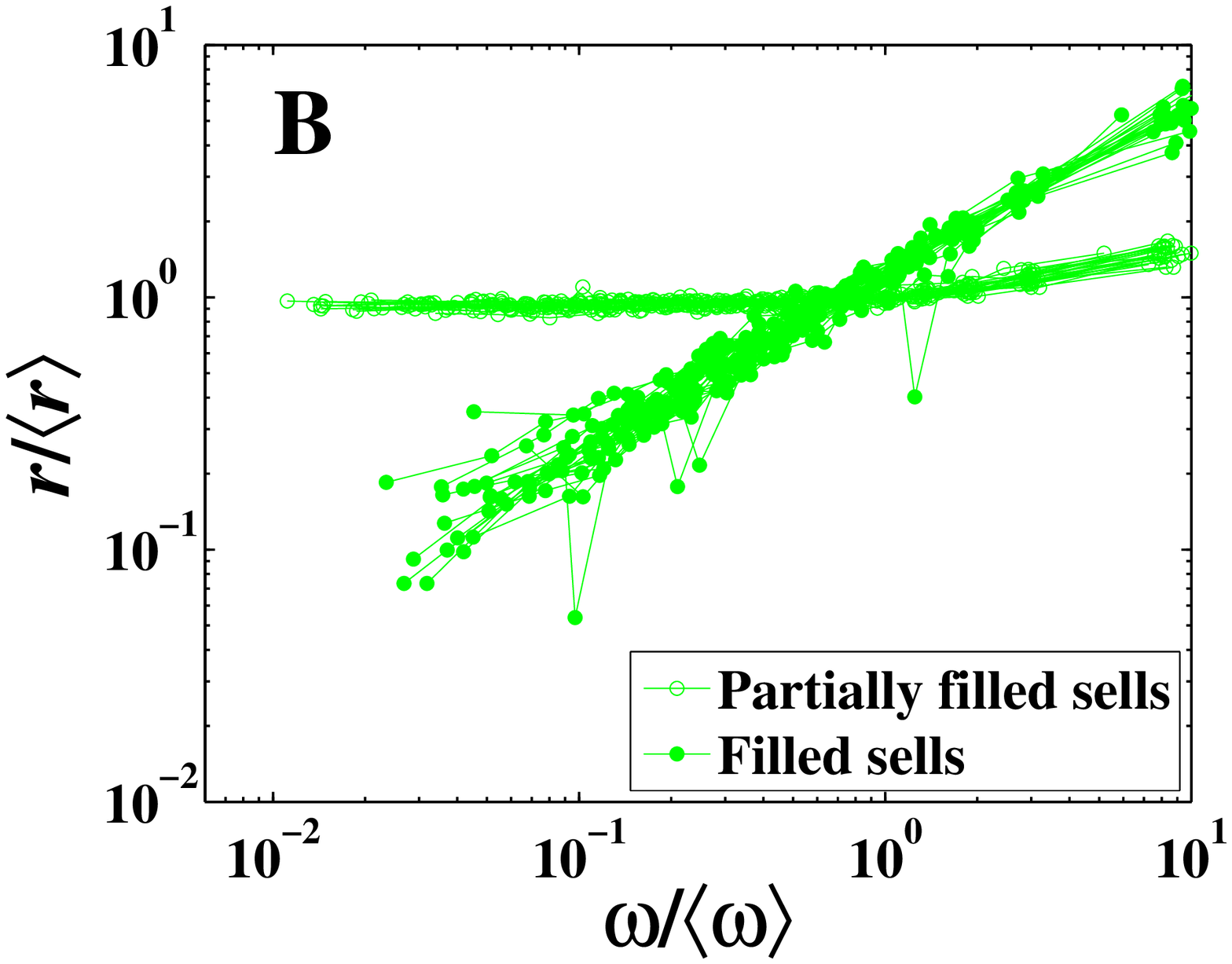}
\caption{Scaling of price impact functions for different types of
trades. For each trade, the price shift $r(\omega)$ and transaction
size $\omega$ are normalized by the averages $\langle{r}\rangle$ and
$\langle\omega\rangle$ of all the returns and sizes of trades
belonging to the same type of trade for each stock.}
\label{Fig:MPM:Trades:Normalized}
\end{figure}

We then adopt a simple method to normalize $r$ and $\omega$ for each
type of trades by their averages $\langle{r}\rangle$ and
$\langle\omega\rangle$ for each stock. We stress that the main
difference between our method and that of
\citet{Lillo-Farmer-Mantegna-2003-Nature} and
\citet{Lim-Coggins-2005-QF} is that they did not normalize the
returns. The results are depicted in figure
\ref{Fig:MPM:Trades:Normalized}. The data points of different stocks
collapse onto a single curve for each type of trades. The scaling
laws can be expressed in the following form
\begin{equation}\label{Eq:dpw}
    r/\langle{r}\rangle=f_{\rm{type}}(\omega/\langle\omega\rangle)
\end{equation}
for different types of trades. For partially filled trades, the
scaling function $f$ is almost a constant when the independent
argument $\omega/\langle\omega\rangle$ is not too large. For filled
trade, the scaling function $f$ has a power-law form.

\subsection{Averaged price impact functions}

\begin{figure}[htp]
\centering
\includegraphics[width=8cm]{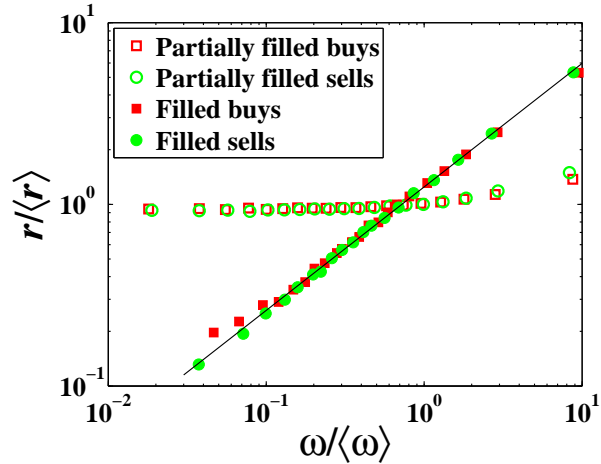}
\caption{Market averaged price impact curves for the four types of
trades. The solid line is a power-law function with the exponent
$\alpha=0.68$ to guide the eye.} \label{Fig:MPM:Trades}
\end{figure}

Based on the remarkable scaling of price impact for the four types,
we are able to determine the market-averaged price impact curves.
For each type of trades, we put together all normalized returns and
trade sizes of the 23 stocks. Then the paired points are divided
into groups with roughly the same size by binning the
$\omega/\langle\omega\rangle$-axis. Figure \ref{Fig:MPM:Trades}
plots the normalized return as a function of the normalized
transaction size for each type of trades. It is found that there is
no significant asymmetry between the buyer- and seller-initiated
trades at the transaction level, which is in line with the cases of
the USA and Australian stock markets
\citep{Lillo-Farmer-Mantegna-2003-Nature,Lim-Coggins-2005-QF} and
does not support the well-known conclusion that seller-initiated
trades have stronger impact on price than buyer-initiated trades
\citep{Epps-1975-AER,Epps-1977-JFQA,Karpoff-1987-JFQA}. Speaking
alternatively, we find that
$f_{\rm{FB}}(\omega/\langle\omega\rangle)=f_{\rm{FS}}(\omega/\langle\omega\rangle)$
and
$f_{\rm{PB}}(\omega/\langle\omega\rangle)=f_{\rm{PS}}(\omega/\langle\omega\rangle)$.
The power-law exponents are $\alpha_{\rm{FB}}=0.66\pm0.03$ for FB
trades and $\alpha_{\rm{FS}}=0.69\pm0.03$ for FS trades.

\begin{figure}[htp]
\centering
\includegraphics[width=8cm]{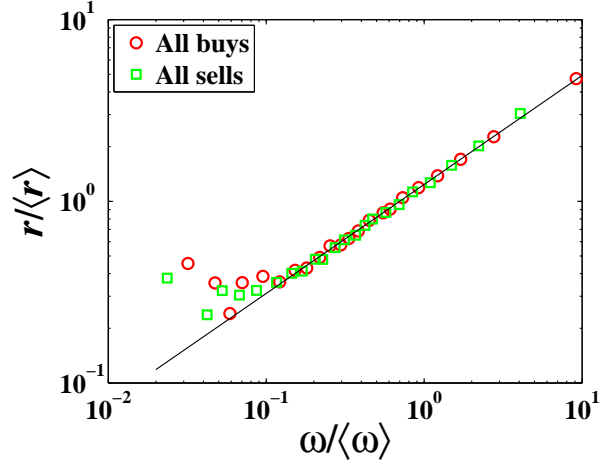}
\caption{Anomalous price impact curves for all buyer- and
seller-initiated trades. Smaller trades move the price more when
$\omega$ is smaller than about 100 shares. For large trades, the
price impact behaves as a power law with an exponent
$\alpha\approx0.60\pm0.03$.} \label{Fig:MPM:Trades:Mixed}
\end{figure}

The averaged price impact functions of all buyer- and
seller-initiated trades are computed separately and presented in
figure \ref{Fig:MPM:Trades:Mixed}. The most interesting feature is
that there is a negative correlation between $r$ and $\omega$ for
small-size trades, which exists for individual stocks as well. To
the best of our knowledge, this anomalous phenomenon was first
reported by \citet{Lim-Coggins-2005-QF} for Australian stocks. We
can derive the condition causing this anomalous negative
correlation. Consider two adjacent groups 1 and 2 with mean returns
$r_1$ and $r_2$ and mean trade sizes $\omega_1<\omega_2$. Assume
that the proportion of filled trade in group $i$ is $x_i$ and the
average returns of filled and partially filled trades are
$r_{{\rm{F}},i}$ and $r_{{\rm{P}},i}$, where $i=1,2$. We have
\begin{equation}
 r_i = r_{{\rm{P}},i}(1-x_i) + r_{{\rm{F}},i}x_i~.
 \label{Eq:rix}
\end{equation}
Since $r_{{\rm{P}},1} \approx r_{{\rm{P}},2}$ and $r_{{\rm{F}},1} <
r_{{\rm{F}},2}$ for small trades, the necessary condition of
$r_1>r_2$ is
\begin{equation}
 x_2 > \frac{r_{{\rm{P}},2} - r_{{\rm{F}},2}}{r_{{\rm{P}},1}-r_{{\rm{F}},1}} x_1~,%
 \label{Eq:x1:x2}
\end{equation}
where
$({r_{{\rm{P}},2}-r_{{\rm{F}},2}})/({r_{{\rm{P}},1}-r_{{\rm{F}},1}})$
is less than but also close to 1. For individual stocks, we have
$r_{{\rm{P}},i}\gg r_{{\rm{F}},i}$, as is shown in figure
\ref{Fig:MPM:Trades:Raw}, and condition (\ref{Eq:x1:x2}) can be
simplified to $x_2>x_1$. For normalized returns and trade sizes
illustrated in figure \ref{Fig:MPM:Trades}, condition
(\ref{Eq:x1:x2}) becomes roughly that $x_2>0.9 x_1$.

In the Chinese stock market, the size of a buy order is limited to
100 shares or an integer multiple of 100, while a seller can place
an sell order with any size. It follows that the transaction size of
a filled buy is 100 shares or its multiples and it is impossible to
be less than 100. For a partially filled buy, its transaction size
could be less than 100, if the volume of sell orders waiting on the
opposite book it can eat is less than 100. Therefore, condition
(\ref{Eq:x1:x2}) is satisfied for buy-initiated trades, since the
relation $x_2\approx1\gg x_1$ holds.

For seller-initiated trades, the situation is a little more
complicated. Here we provide a qualitative explanation. Since the
market is not frictionless, it is costly to sell a very small amount
of shares. We note that the majority of individual investors hold
only several hundred shares \citep{Mu-Chen-Zhou-2008-XXX}, which is
not irrational since the wealth distribution of people follows the
Pareto law. All FS trades with the sizes less than 100 are caused by
sellers who placed very small orders, which is very rare.
Empirically, we find that there are 7029 partially filled trades
with the sizes less than 100 out of the total 334516 PS trades, and
22651 filled trades with the sizes less than 100 out of the total
2527070 FS trades. In other words, $x_1=0.9790 < x_2 = 0.9910$.
Again, condition (\ref{Eq:x1:x2}) holds.

\section{Testing the LFM scaling}
\label{S1:LFM:LC}

\subsection{Scaling exponents determination approaches}

Since the LFM scaling was verified in the US stock market by
\citet{Lillo-Farmer-Mantegna-2003-Nature} and in the Australian
stock market by \citet{Lim-Coggins-2005-QF}, it is natural to test
weather it holds or not in the Chinese stock market. We notice that
there is a difference in the normalization of trade sizes in the two
studies. Hence, we perform this test following
\citet{Lillo-Farmer-Mantegna-2003-Nature} and
\citet{Lim-Coggins-2005-QF}, respectively. To avoid possible
confusion, we rewrite the notations in the LFM scaling as follows
\begin{equation}\label{Eq:xy:scaling}
    y(x,C)=C^{-\gamma}f(x/C^\delta)~,
\end{equation}
where $y$ stands for the unnormalized return, $x$ represents the
normalized trade size and $C$ is the capitalization.

To determine the scaling exponents $\delta$ and $\gamma$,
\citet{Lillo-Farmer-Mantegna-2003-Nature} sorted and divided the
$x/C^{\delta}$ data into bins and determined the values of $\delta$
and $\gamma$ that minimize the mean $\langle\epsilon\rangle$ of the
two-dimensional variance
\begin{equation}
 \epsilon(\delta,\gamma) = \left[\frac{\sigma(yC^{\gamma})}{\mu(yC^{\gamma})}\right]^2
         +\left[\frac{\sigma(x/C^{\delta})}{\mu(x/C^{\delta})}\right]^2
\end{equation}
where $\sigma$ denotes the standard deviation and $\mu$ denotes the
mean. This approach has also been adopted by \citet{Niwa-2005-JTB}
in the scaling analysis of probability distribution of dimensions
and biomass of fish schools. Alternatively,
\citet{Lim-Coggins-2005-QF} adopted the method proposed by
\citet{Bhattacharjee-Seno-2001-JPA}. Both approaches worked well in
the scaling analysis of US and Australian stocks. Here we utilize
the one proposed by \citet{Lillo-Farmer-Mantegna-2003-Nature}.

In addition, we propose a quantity $R$ to quantify the goodness of
data collapse, which is motivated by a similar measure used by
\citet{Lim-Coggins-2005-QF}. The quantity is calculated as follows
\begin{equation}
 R = 1- \frac{\epsilon(\hat\delta,\hat\gamma)}{\epsilon(0,0)}~,
 \label{Eq:R}
\end{equation}
where $\hat\delta$ and $\hat\gamma$ are the estimates of the scaling
exponents.

\subsection{The Lillo-Farmer-Mantegna approach}

\begin{figure}[htb]
\centering
\includegraphics[width=6cm]{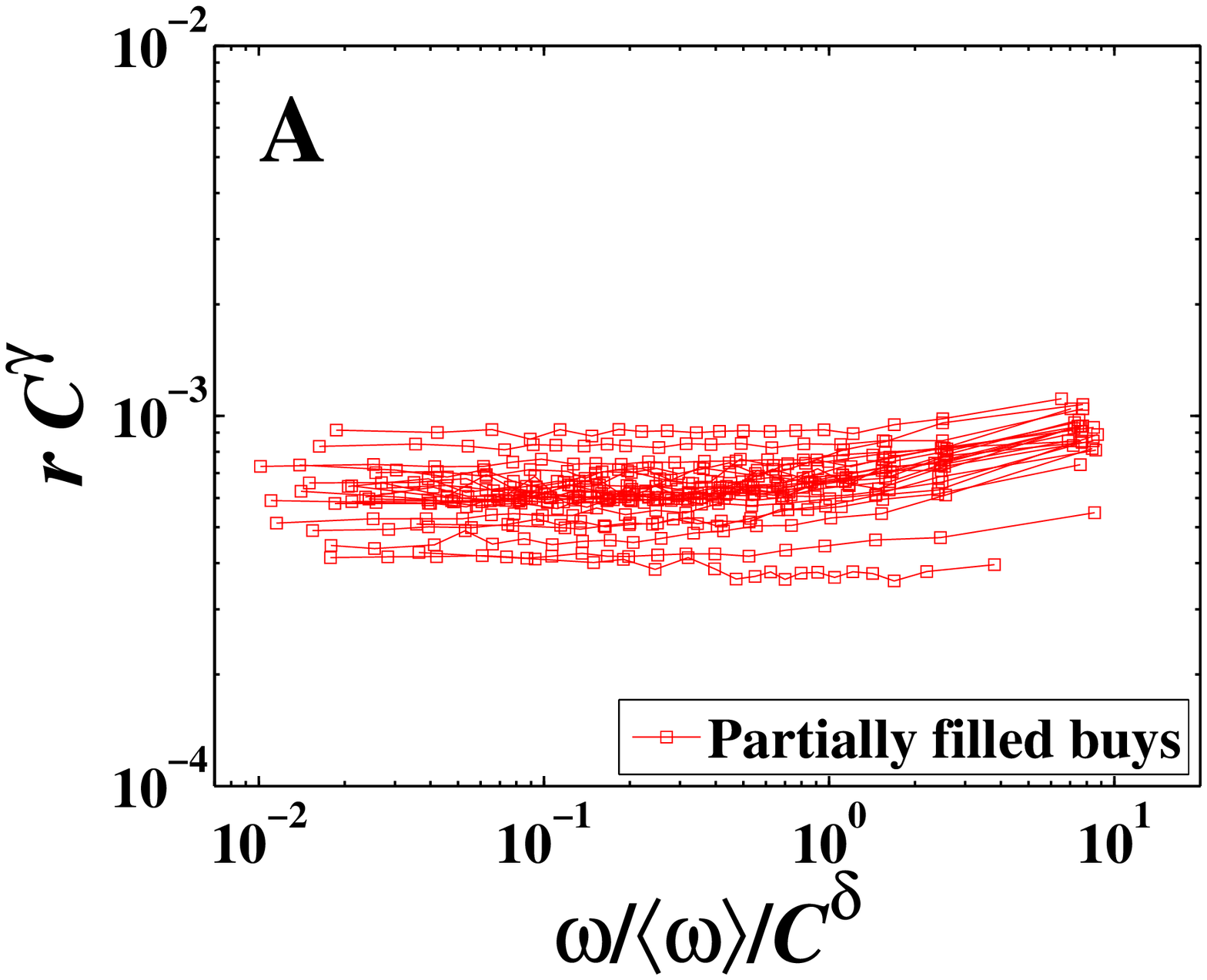}
\includegraphics[width=6cm]{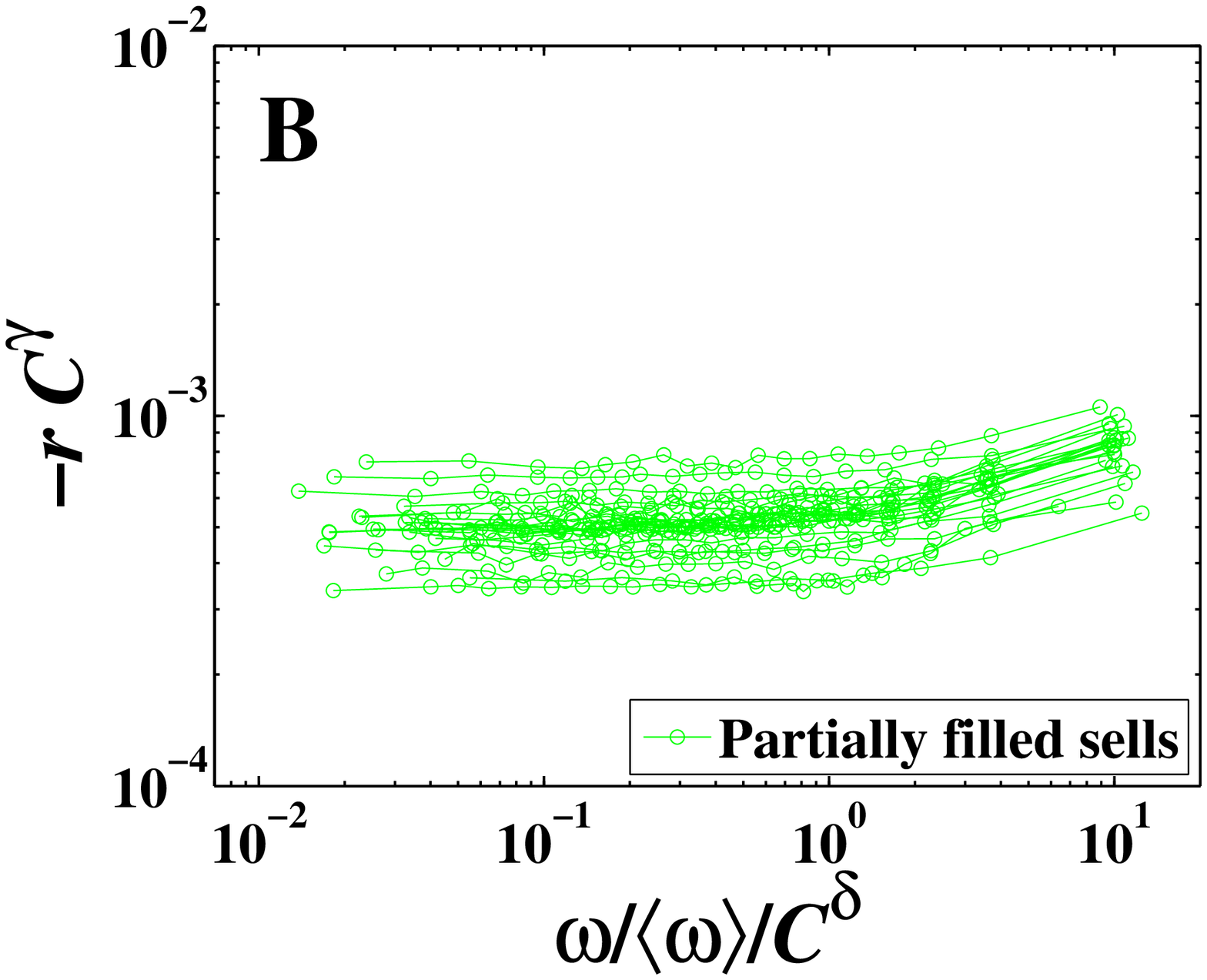}\\
\includegraphics[width=6cm]{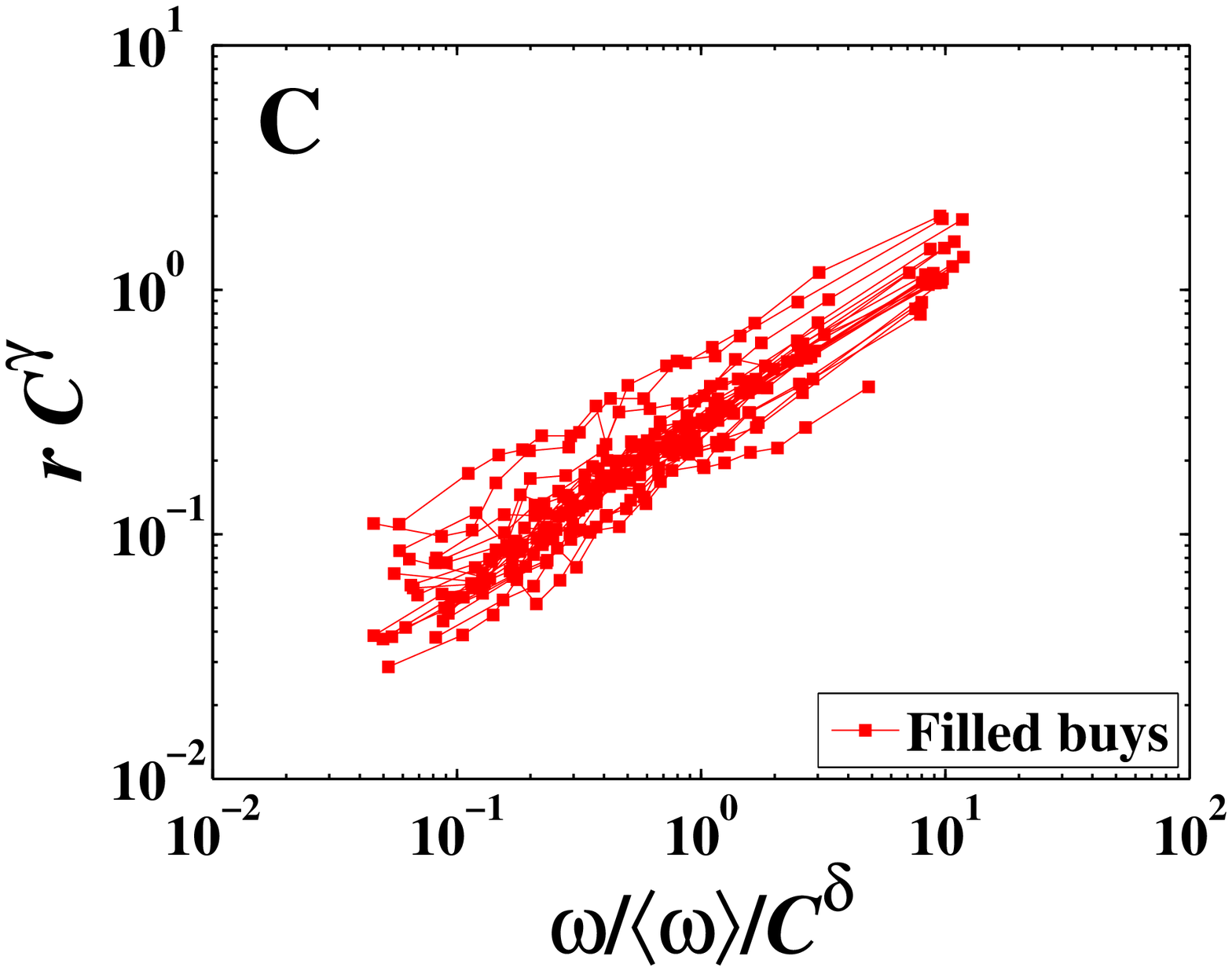}
\includegraphics[width=6cm]{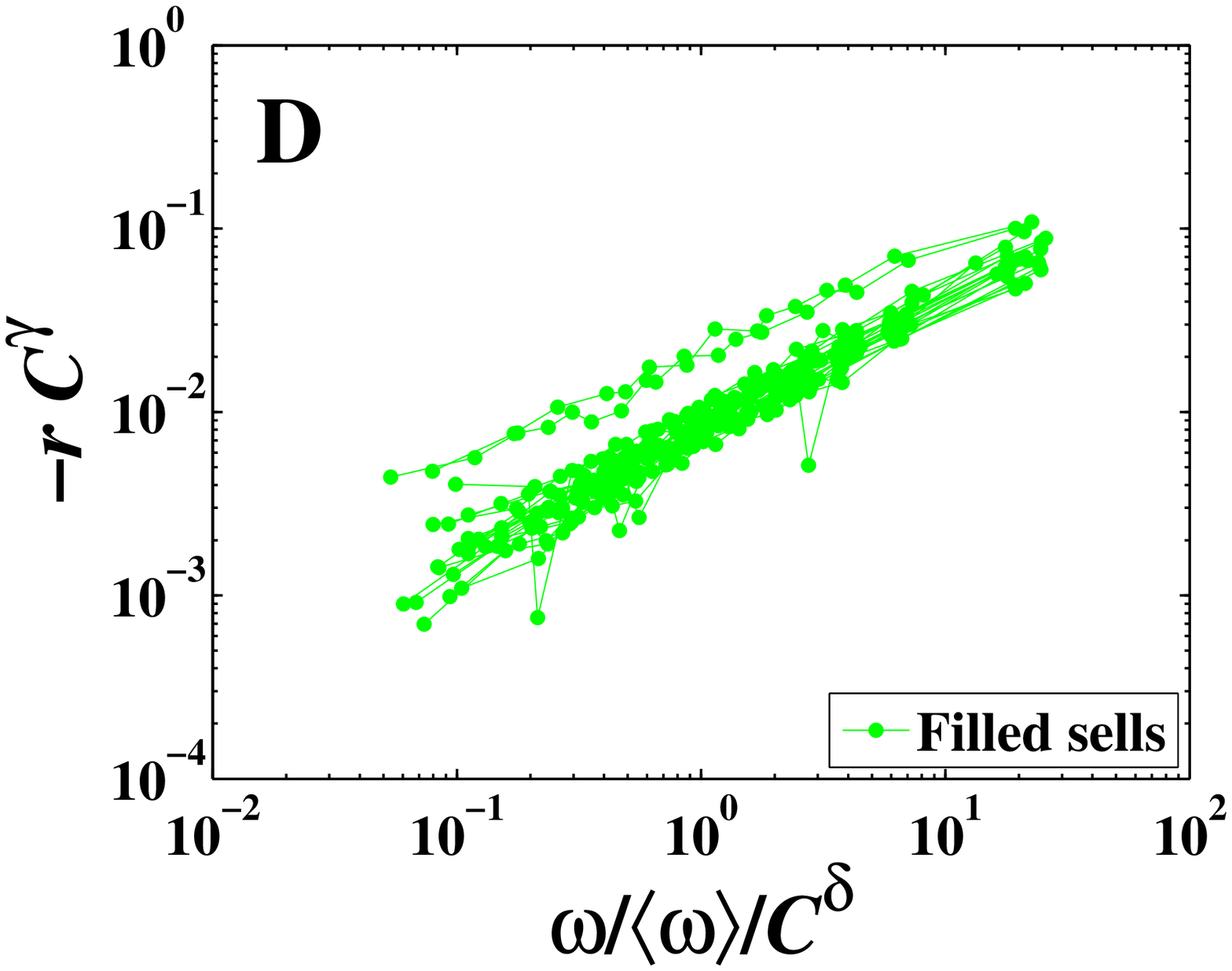}
\caption{Scaling analysis of the price impact functions of different
types of trades for the 23 individual stocks, following
\citet{Lillo-Farmer-Mantegna-2003-Nature}. The trades are classified
into four types due to their directions and aggressiveness. The
values of $\delta$ and $\gamma$ for each type of trades are listed
in table \ref{Tb:FLM}.} \label{Fig:MPM:LFM:Trades:20-23}
\end{figure}

We follow exactly the same approach of
\citet{Lillo-Farmer-Mantegna-2003-Nature} to perform the scaling
analysis in which stock capitalization is included as an
independent, except that we do not divide the stocks into groups. In
this case, the variables $y$ and $x$ in Eq.~(\ref{Eq:xy:scaling})
are $r$ and $\omega/\langle\omega\rangle$, respectively. The
analysis is conducted for the four types of trades. The results are
illustrated in figure \ref{Fig:MPM:LFM:Trades:20-23}.

The estimated values of scaling exponents $\delta$ and $\gamma$ for
each type of trades are listed in table \ref{Tb:FLM}. The results
show that it is hard to collapse the data onto a single curve for
each type of trades. This is also supported by the rather small
values of $R$. When we analyze all buyer- or seller initiated
trades, the LFM scaling does not work either.

\begin{table}[htb]
 \centering
 \caption{\label{Tb:FLM} Estimates of the scaling exponents following \citet{Lillo-Farmer-Mantegna-2003-Nature}.}
 \bigskip
 \begin{tabular}{cccccc}
  \hline\hline
            &    PB    &    PS    &    FB    &    FS \\  \hline
 $\delta$   &    0.0069 &  -0.0106  &  0.0020  & -0.0417\\
 $\gamma$   &   -0.0481 &  -0.0570  &  0.4174  &  0.2615\\
 $R$        &    0.0157 &   0.0239  &  0.1969  &  0.0739\\
 \hline\hline
 \end{tabular}
\end{table}

\subsection{The Lim-Coggins approach}

We also follow exactly the same approach of
\citet{Lim-Coggins-2005-QF} to perform the scaling analysis in which
stock capitalization is included as an independent, except that we
do not divide the stocks into groups. In this case, the variables
$y$ in Eq.~(\ref{Eq:xy:scaling}) is $r$ and $x$ is the
{\em{normalized daily-normalized}} volume
\begin{equation}
 x_{ij} = \frac{\omega_{ij}}{\sum_{m=1}^{T_i}}
 \left(\frac{N}{\sum_{d=1}^N T_d} \right)^{-1}~,
 \label{Eq:LC:yij}
\end{equation}
where $\omega_{ij}$ is the size of trade $j$ on day $i$, $T_i$ is
the total number of trades on day $i$, and $N$ is the total number
of days of a stock. The analysis is carried out for the four types
of trades. The estimated values of scaling exponents $\delta$ and
$\gamma$ for each type of trades are listed in table \ref{Tb:FLM}.
The rather small values of $R$ imply that it is hard to collapse the
data onto a single curve for each type of trades. When we analyze
all buyer- or seller initiated trades, no scaling is obtained
\citep[see][Figure 11]{Zhou-2007-XXX}.

\begin{table}[h!]
 \centering
 \caption{\label{Tb:LC} Estimates of the scaling exponents following \citet{Lim-Coggins-2005-QF}.}
 \bigskip
 \begin{tabular}{cccccc}
  \hline\hline
            &    PB    &    PS    &    FB    &    FS \\  \hline
 $\delta$   &  0.100  &  0.074  &  0.291  &  0.432 \\
 $\gamma$   & -0.036  & -0.029  &  0.196  & -0.026 \\
 $R$        &  0.017  &  0.018  &  0.096  &  0.113 \\
 \hline\hline
 \end{tabular}
\end{table}

It is noteworthy that, for each stock, the price impact curve for
partially filled trades remains constant for not too large trades,
while that of filled orders exhibits a nice power law
\citep[see][Figures 9-10]{Zhou-2007-XXX}. We have fitted the
power-law price impact functions and the average power-law exponent
is $\alpha_{\rm{FB}}=0.52\pm0.04$ for buyer-initiated filled trades
and $\alpha_{\rm{FS}}=0.53\pm0.05$ for seller-initiated filled
trades. These exponents are significantly smaller than those in
section \ref{S1:Zhou}. In addition, when we plot
$r/\langle{r}\rangle$ against $x$, nice scaling appears again
\citep[see][Figure 12]{Zhou-2007-XXX}.

\section{Price impacts and the distributions of returns and trade sizes}
\label{S1:alpha}

\begin{figure}[htp]
\centering
\includegraphics[width=8cm]{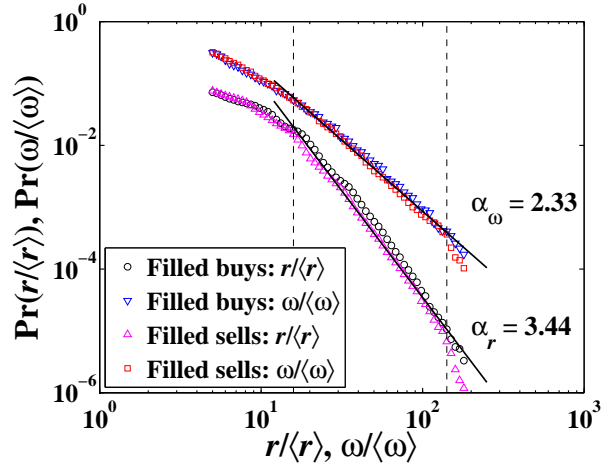}
\caption{Empirical complementary cumulative distributions of the
normalized returns and trade sizes for buyer- and seller-initiated
filled trades. The solid lines are power laws. The two curves for
trade sizes have been shifted upwards by a factor of 10 for
clarity.} \label{Fig:MPM:PDF}
\end{figure}

We now turn to investigate the relation between the power-law
exponents of price impacts, returns and trade sizes for filled
trades. For the normalized returns and trade sizes, we estimate the
empirical complementary cumulative distribution functions for either
buyer-initiated or seller-initiated trades. The results are
illustrated in figure \ref{Fig:MPM:PDF}. We find that each of the
four probability distributions has a power-law tail. The tail
exponents are estimated within the scaling range $[15.9,141]$
delimited by the two vertical dashed lines. We find that,
$\alpha_r=3.45 \pm 0.17$ and $\alpha_\omega = 2.30\pm 0.10$ for
buyer-initiated trades and $\alpha_r=3.44 \pm 0.13$ and
$\alpha_\omega = 2.36\pm 0.06$ for seller-initiated trades. Again,
the difference between the buyer- and seller-initiated trades are
negligible. Although the two distributions deviate respectively from
the inverse cubic law and the half-cubic law, the relation between
the tail exponents $\alpha_\omega/\alpha_r=\alpha$ is nevertheless
validated. Indeed, the two ratios are 0.666 for buy trades and 0.686
for sell trades, which are in excellent agreement with the
corresponding power-law exponents $\alpha_{\rm{FB}}=0.66\pm0.03$ and
$\alpha_{\rm{FS}}=0.69\pm0.03$ of the price impact function.

The above analysis is conducted based on the data aggregated across
different stocks. It is natural to ask if the relation
$\alpha_\omega/\alpha_r=\alpha$ holds at the level of individual
stocks. We perform such a test separately for buyer- and
seller-initiated filled trades. For each type of trades of each
stock, the tail exponents $\alpha_\omega$ and $\alpha_r$ of the
distributions of returns and trade sizes and the power-law exponent
$\alpha$ of the price impact function are determined. The value of
$\alpha_\omega/\alpha_r$ is then calculated, which should be
compared with $\alpha$. To quality the extend to which
$\alpha_\omega/\alpha_r$ is equal to $\alpha$, we calculate the
relative deviation as follows
\begin{equation}
 D = \left(\frac{\alpha_\omega}{\alpha_r}-\alpha\right)\frac{1}{\alpha}~.
\end{equation}
All these values ($\alpha_{\omega}$, $\alpha_{r}$,
$\alpha_{\omega}/\alpha_{r}$, $\alpha$, $D$) are depicted in table
\ref{Tb:alpha}. We also show the means and standard deviations of
$\alpha_{\omega}$, $\alpha_{r}$, $\alpha_{\omega}/\alpha_{r}$, and
$\alpha$ and the corresponding values extracted from figure
\ref{Fig:MPM:Trades}. According to table \ref{Tb:alpha}, we see that
most of the $D$ values are reasonably small. The worst case is
recognized for the buyer-initiated filled trades of stock 000720,
which is not surprising since its price impact function does not
exhibit power-law form. We thus figure that the relation between the
exponents of price impacts, returns and trade sizes also holds for
individual stocks.

\begin{table}[h!]
\centering \caption{\label{Tb:alpha}Testing the relation
$\alpha_\omega/\alpha_r=\alpha$ for buyer- and seller-initiated
filled trades. All values of $D$ are calculated as the relative
deviations of $\alpha_\omega/\alpha_r$ from $\alpha$. Each value
(except $D$) in the row ``{\em{MEAN}}'' is the mean of the values in
the corresponding column over the 23 individual stocks and the row
``{\em{STD}}'' shows the standard deviations. The values in the row
``{\em{ALL}}'' are determined from the normalized returns and trade
sizes aggregated across different stocks.}
\bigskip
\begin{tabular}{c c ccccr@{.}l c ccccr@{.}l}
  \hline\hline
  \multirow{3}*[2mm]{Code}&&\multicolumn{6}{c}{Buyer-initiated filled trades}&&\multicolumn{6}{c}{Seller-initiated filled trades}\\
    \cline{3-8}  \cline{10-15}
   &&$\alpha_{\omega}$&$\alpha_{r}$&$\alpha_{\omega}/\alpha_{r}$&$\alpha$&\multicolumn{2}{c}{$D$}&
    &$\alpha_{\omega}$&$\alpha_{r}$&$\alpha_{\omega}/\alpha_{r}$&$\alpha$&\multicolumn{2}{c}{$D$} \\\hline
000001 && 2.12 & 2.97 & 0.71 & 0.69 &  0&04  && 2.21 & 3.38 & 0.65 & 0.71 & -0&07 \\
000002 && 2.09 & 3.17 & 0.66 & 0.67 & -0&01  && 2.04 & 3.12 & 0.65 & 0.70 & -0&07 \\
000009 && 2.28 & 3.46 & 0.66 & 0.71 & -0&06  && 2.48 & 3.22 & 0.77 & 0.74 &  0&04 \\
000012 && 2.72 & 3.09 & 0.88 & 0.66 &  0&33  && 2.41 & 3.37 & 0.71 & 0.69 &  0&03 \\
000016 && 2.38 & 3.32 & 0.72 & 0.69 &  0&04  && 2.35 & 3.33 & 0.71 & 0.69 &  0&01 \\
000021 && 2.34 & 3.48 & 0.67 & 0.65 &  0&03  && 2.47 & 3.73 & 0.66 & 0.72 & -0&07 \\
000024 && 1.95 & 2.96 & 0.66 & 0.61 &  0&08  && 2.51 & 3.32 & 0.76 & 0.69 &  0&10 \\
000027 && 1.90 & 3.41 & 0.56 & 0.71 & -0&21  && 2.58 & 3.44 & 0.75 & 0.75 &  0&00 \\
000063 && 2.17 & 3.07 & 0.70 & 0.62 &  0&12  && 2.47 & 3.66 & 0.68 & 0.63 &  0&07 \\
000066 && 2.53 & 3.24 & 0.78 & 0.67 &  0&15  && 2.75 & 3.11 & 0.88 & 0.72 &  0&23 \\
000088 && 2.22 & 3.37 & 0.66 & 0.60 &  0&09  && 2.35 & 3.44 & 0.68 & 0.53 &  0&29 \\
000089 && 1.84 & 2.93 & 0.63 & 0.67 & -0&06  && 1.71 & 2.94 & 0.58 & 0.74 & -0&21 \\
000406 && 2.04 & 3.03 & 0.67 & 0.70 & -0&03  && 2.51 & 3.26 & 0.77 & 0.74 &  0&03 \\
000429 && 2.15 & 3.15 & 0.68 & 0.73 & -0&07  && 2.15 & 2.69 & 0.80 & 0.75 &  0&06 \\
000488 && 2.18 & 3.17 & 0.69 & 0.56 &  0&22  && 2.24 & 3.90 & 0.58 & 0.57 &  0&01 \\
000539 && 1.78 & 2.69 & 0.66 & 0.63 &  0&05  && 1.60 & 2.77 & 0.58 & 0.58 & -0&00 \\
000541 && 1.81 & 2.69 & 0.67 & 0.59 &  0&14  && 2.11 & 3.28 & 0.64 & 0.69 & -0&07 \\
000550 && 2.28 & 3.18 & 0.72 & 0.68 &  0&04  && 2.39 & 3.49 & 0.69 & 0.68 &  0&00 \\
000581 && 1.91 & 2.87 & 0.66 & 0.64 &  0&03  && 2.00 & 2.90 & 0.69 & 0.66 &  0&03 \\
000625 && 2.10 & 2.78 & 0.76 & 0.65 &  0&16  && 2.36 & 3.25 & 0.73 & 0.73 & -0&00 \\
000709 && 2.27 & 3.67 & 0.62 & 0.74 & -0&16  && 2.47 & 3.26 & 0.76 & 0.77 & -0&01 \\
000720 && 3.45 & 3.34 & 1.03 & 0.35 &  1&98  && 2.38 & 2.84 & 0.84 & 0.73 &  0&14 \\
000778 && 1.65 & 2.52 & 0.65 & 0.72 & -0&09  && 2.53 & 3.27 & 0.78 & 0.71 &  0&08
\\\hline
 MEAN  && 2.18 & 3.11 & 0.70 & 0.65 &  0&07  && 2.31 & 3.26 & 0.71 & 0.69 &  0&03 \\
  STD  && 0.37 & 0.29 & 0.10 & 0.08 &\multicolumn{2}{c}{/}&& 0.27 & 0.30 & 0.08 & 0.06 &\multicolumn{2}{c}{/}        \\
  ALL  && 2.30 & 3.45 & 0.67 & 0.66 &  0&01  && 2.36 & 3.44 & 0.69 & 0.69 &  0&00 \\
  \hline\hline
\end{tabular}
\end{table}

Several relevant remarks are in order. According to table
\ref{Tb:alpha}, the returns of Chinese stocks at the transaction
level follow the well-known cubic law. The cubic law of returns was
well-established for US stocks by
\citet{Gopikrishnan-Meyer-Amaral-Stanley-1998-EPJB} and
\citet{Plerou-Gopikrishnan-Amaral-Meyer-Stanley-1999-PRE} and for US
indexes by \citet{Gopikrishnan-Plerou-Amaral-Meyer-Stanley-1999-PRE}
at different timescales from 1 minute to a few days.
\citet{Plerou-Stanley-2008-PRE} further showed that the cubic law is
universal across three distinct markets, NYSE, London Stock Exchange
(LSE) and Paris Bourse, at the timescale of 5 minutes. For the
Chinese stocks, \citet{Gu-Chen-Zhou-2008a-PA} reported that the
cubic law holds at the transaction level and the distribution of
returns has power-law tails, in which the tail exponent increases
from 3 to 4 when the timescale increases from 1 minute to 32
minutes.

The issue of the distribution of trade sizes is more controversial.
\citet{Gopikrishnan-Plerou-Gabaix-Stanley-2000-PRE} analyzed the
transaction data for the largest 1000 stocks traded on the three
major US markets and found that the distribution of trade sizes
follow a power-law tail with the exponent being $1.53\pm0.07$, known
as the half-cubic law, and that of the aggregated share volumes at
timescales from a few minutes to several hundred minutes has a
power-law tail exponent $1.7\pm0.1$.
\citet{Plerou-Gopikrishnan-Gabaix-Amaral-Stanley-2001-QF,Plerou-Gopikrishnan-Gabaix-Stanley-2004-QF}
confirmed this analysis. \citet{Maslov-Mills-2001-PA} found that the
trade sizes of several NASDAQ stocks have power-law tails with the
exponent close to 1.4. \citet{Plerou-Stanley-2007-PRE} extended this
analysis to other two markets (LSE and Paris Bourse) and found
quantitatively similar results across the three distinct markets.
All these tail exponents are well below 2, within the L\'evy regime.
Alternatively,
\citet{Eisler-Kertesz-2006-EPJB,Eisler-Kertesz-2007-PA} reported
that the tail exponents of the traded volumes at a timescale of 15
minutes for six NYSE stocks are 2.2 and 2.8.
\cite{Racz-Eisler-Kertesz-2008-XXX} argued that the tail exponents
of the traded volume were underestimated. They investigated the 1000
most liquid stocks traded on the NYSE for the same period as studied
by \citet{Plerou-Stanley-2007-PRE} and found that the average tail
exponent is $2.02\pm0.45$. There is additional evidence supporting
the point that the tail exponent is outside the L\'evy regime for
stocks in other markets, such as the Korean stocks
\citep{Lee-Lee-2007-PA} and the Chinese stocks
\citep{Mu-Chen-Zhou-2008-XXX}. The tail exponents reported in table
\ref{Tb:alpha} are in line with these results.

From table \ref{Tb:alpha}, the power-law exponent of price impact is
roughly constant across distinct stocks. Similar phenomenon was
observed in the Australia market for large trades by
\citet{Lim-Coggins-2005-QF}. However, the exponents were found to
vary between 0.21 and 0.41, much smaller than those in table
\ref{Tb:alpha}. For the US stocks,
\citet{Lillo-Farmer-Mantegna-2003-Nature} found that the exponent is
around 0.2 for different groups of stocks.
\citet{Farmer-Lillo-2004-QF} investigated several LSE stocks and the
exponent is found to be about 0.26. Alternatively,
\citet{Bouchaud-Potters-2001-PA} and
\citet{Bouchaud-Gefen-Potters-Wyart-2004-QF} proposed a logarithmic
price impact relation $r\propto\ln\omega$ for several stocks listed
on the Paris Bourse. These two formalisms are not inconsistent since
the power-law exponent is small.
\citet{Gabaix-Gopikrishnan-Plerou-Stanley-2003-Nature,Gabaix-Gopikrishnan-Plerou-Stanley-2006-QJE,Gabaix-Gopikrishnan-Plerou-Stanley-2007-JEEA,Gabaix-Gopikrishnan-Plerou-Stanley-2008-JEDC}
studies the price impact function of aggregate trading volumes
theoretically and empirically, which leads to a self-consistent
exponent of about 0.5.

Combining all these results, several remarks follow. Although the
values of $\alpha_\omega$ and  $\alpha$ in our paper are remarkably
different from those of
\citet{Gabaix-Gopikrishnan-Plerou-Stanley-2003-Nature}, the relation
(\ref{Eq:alpha}) still holds. We would like to stress that the
unified theory of
\citet{Gabaix-Gopikrishnan-Plerou-Stanley-2003-Nature} was tested
with variables at fixed time intervals, while our study is carried
out at the transaction level. Therefore, our work has empirically
confirmed and extended the theory of
\citet{Gabaix-Gopikrishnan-Plerou-Stanley-2003-Nature}. In addition,
different values of $\alpha_\omega$ and $\alpha$ does not falsify
the theory. Indeed the relation (\ref{Eq:alpha}) can be derived, if
the three power laws in the distributions of returns and trade sizes
and the price impact hold, no matter what is the underlying
mechanism causing these power laws. Our work also suggests that it
might be better to test the relation (\ref{Eq:alpha}) in
order-driven markets by properly classified trades.

\section{Conclusion}
\label{S1:Conc}

In previous studies concerning the price-volume relationship using
transaction data, the trades are usually differentiated into two
types being buyer- or seller- initiated. We have found that there is
no difference in the price impact functions between buyer- or
seller-initiated trades. In contrast, the price impact function  is
more sensitive to the aggressiveness of the trades. In addition, two
universal price impact curves for filled trades and partially filled
trades appear when both returns and transaction size are normalized
by their stock-dependent averages. The scaling analysis is
independent of the capitalizations of stocks. To be more
conservative, our study does not deny the possibility that the
scaling relation in terms of stock capitalization when we group
stocks. Unfortunately, we do not have the whole database of all
stocks traded in the A-share markets. It is also interesting to see
if the B-share stocks have different behavior in the immediate price
impact. These researches can be carried out when the data are
available.

We figure that these universal price impact functions are
unambiguous targets that any empirical model of order-driven market
must hit. This result calls for further extension of the existing
models
\citep{Challet-Stinchcombe-2001-PA,Daniels-Farmer-Gillemot-Iori-Smith-2003-PRL,Farmer-Patelli-Zovko-2005-PNAS,Mike-Farmer-2008-JEDC}.
The observed phenomena may be present in other order-driven markets
as well. This conjecture can be tested when the order book data are
available for other markets, which may extend the universality of
the two impact functions.

There are also open problems unsolved in this work. The most
interesting one is to understand why the price impact of partially
filled trades is roughly the same for different trade sizes and why
the price impact of filled trades follows a power law. We hope to
address these questions in future research.

\bigskip
\noindent{\bf{Acknowledgments}}
\medskip

\noindent{We thank Didier Sornette and Liang Guo for valuable
discussions, Wei Chen for kindly providing the data, and Gao-Feng Gu
for preprocessing the data. This work was partially supported by the
National Natural Science Foundation of China (70501011), the Fok
Ying Tong Education Foundation (101086), and the Program for New
Century Excellent Talents in University (NCET-07-0288).}

\bibliographystyle{WXZHOU_QuantFin}
\bibliography{E:/papers/Auxiliary/Bibliography}


\newpage

\pagestyle{plain}

\pagenumbering{Roman}

\setcounter{page}{0}

\begin{center}
RCE Working Paper Series
\end{center}

\vspace{10cm}

\begin{center}
{\color{red}{\Huge{SUPPLEMENTARY INFORMATION}}}
\end{center}

\newpage
\begin{figure}
\includegraphics[width=5cm]{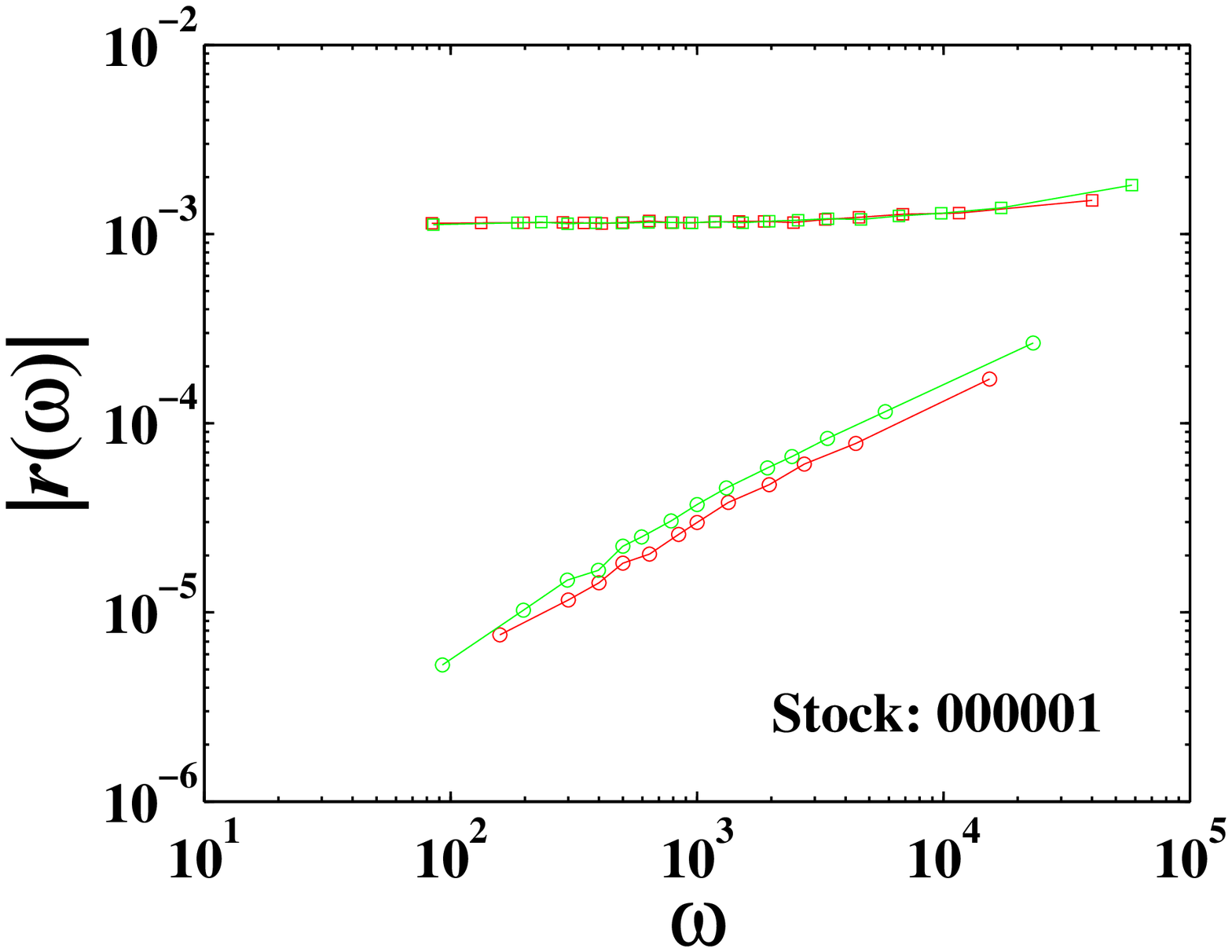}
\includegraphics[width=5cm]{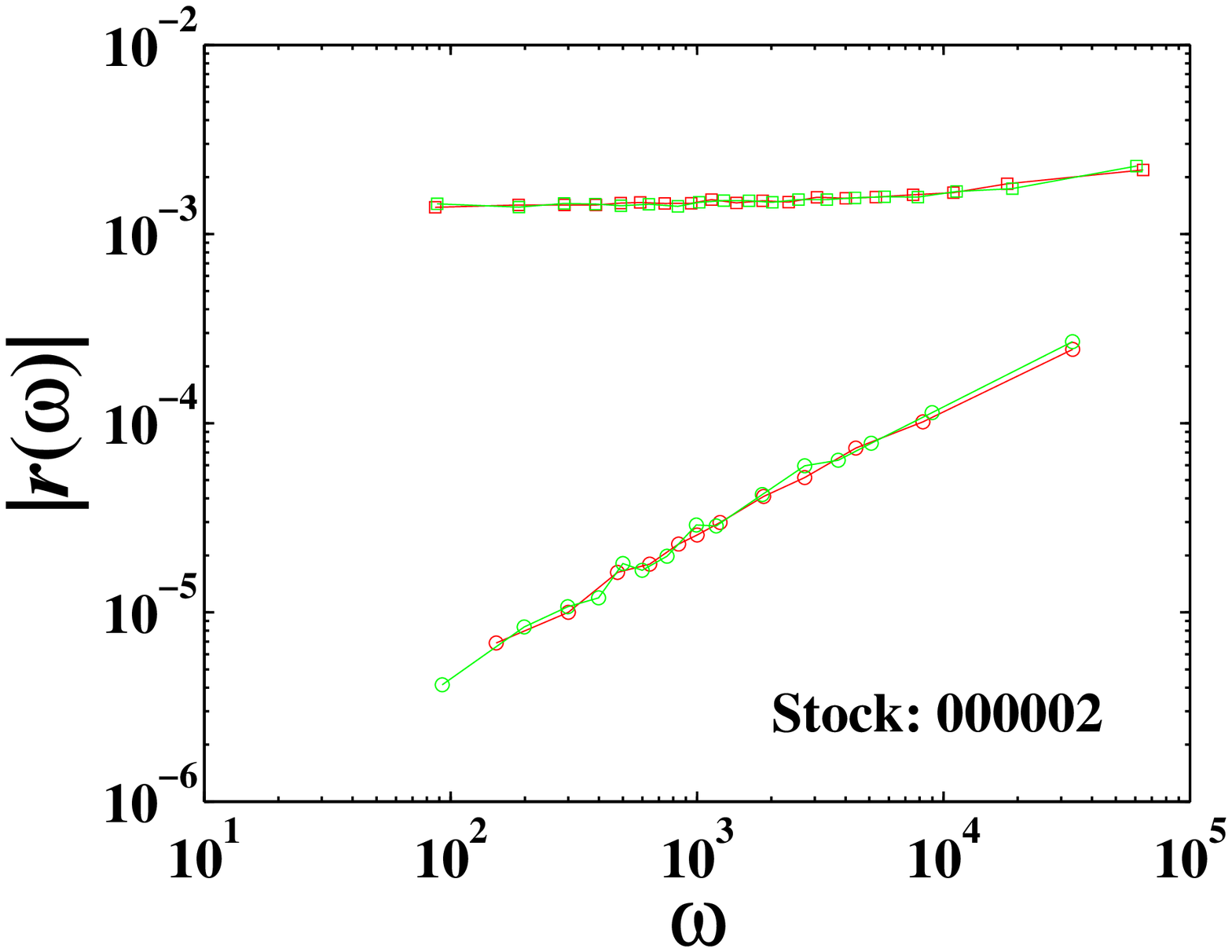}
\includegraphics[width=5cm]{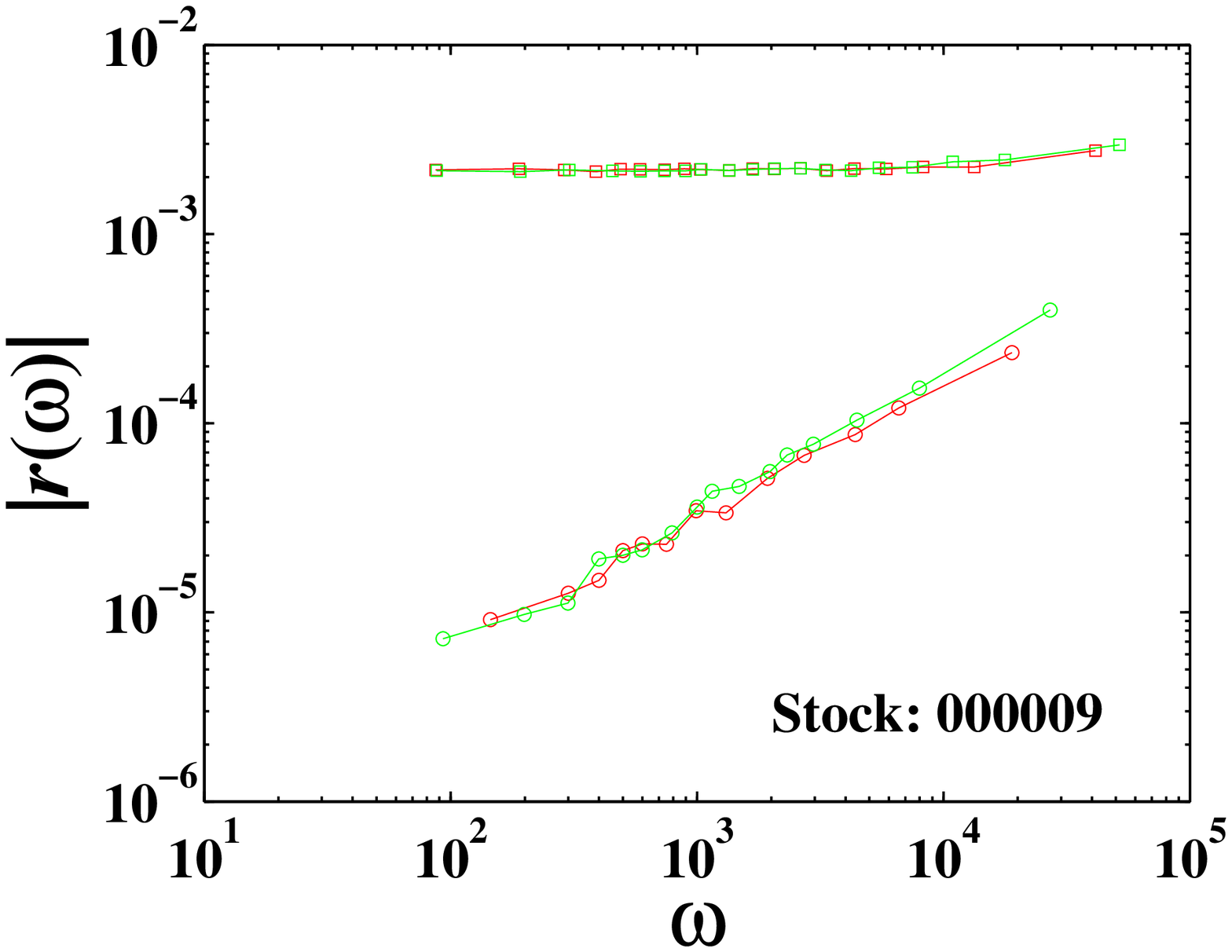}
\includegraphics[width=5cm]{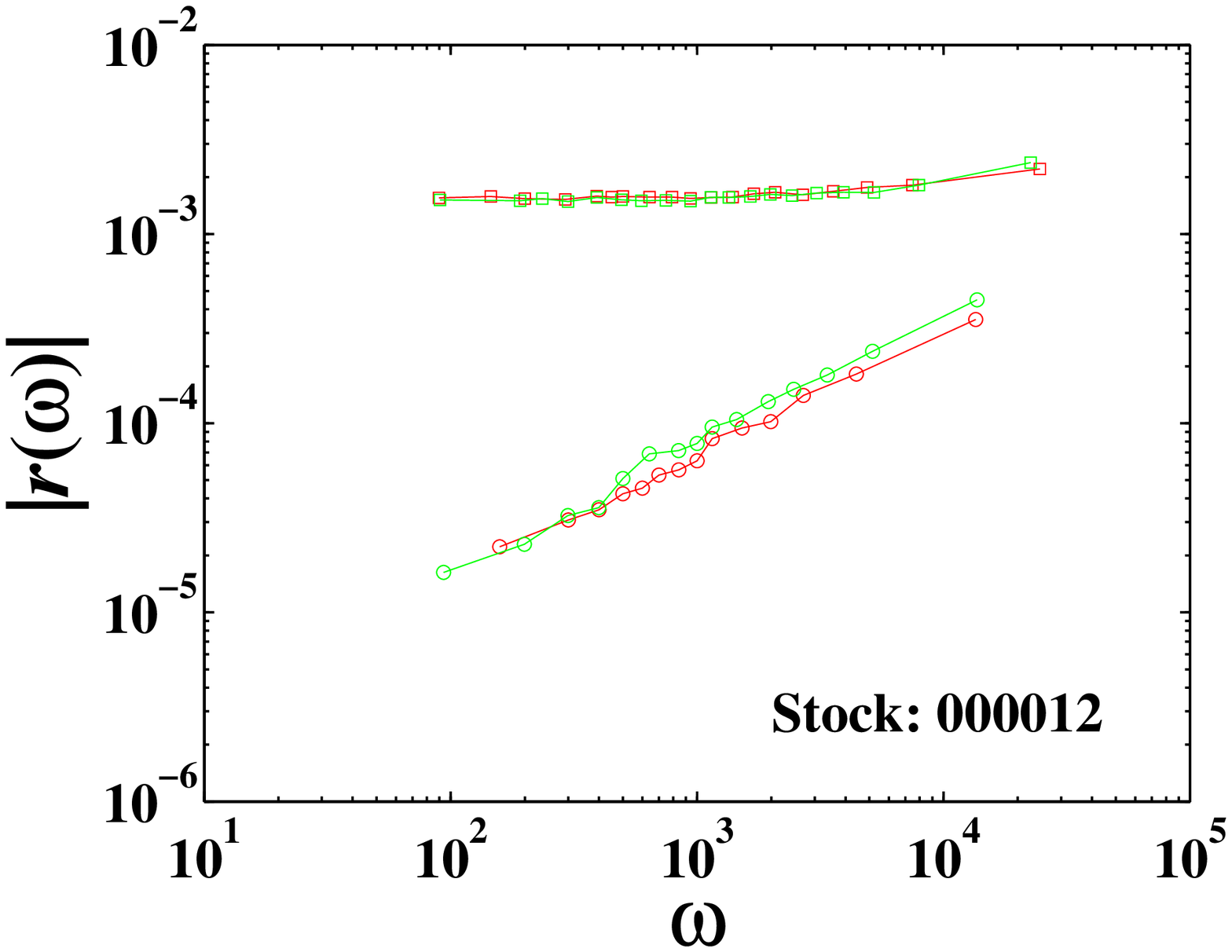}
\includegraphics[width=5cm]{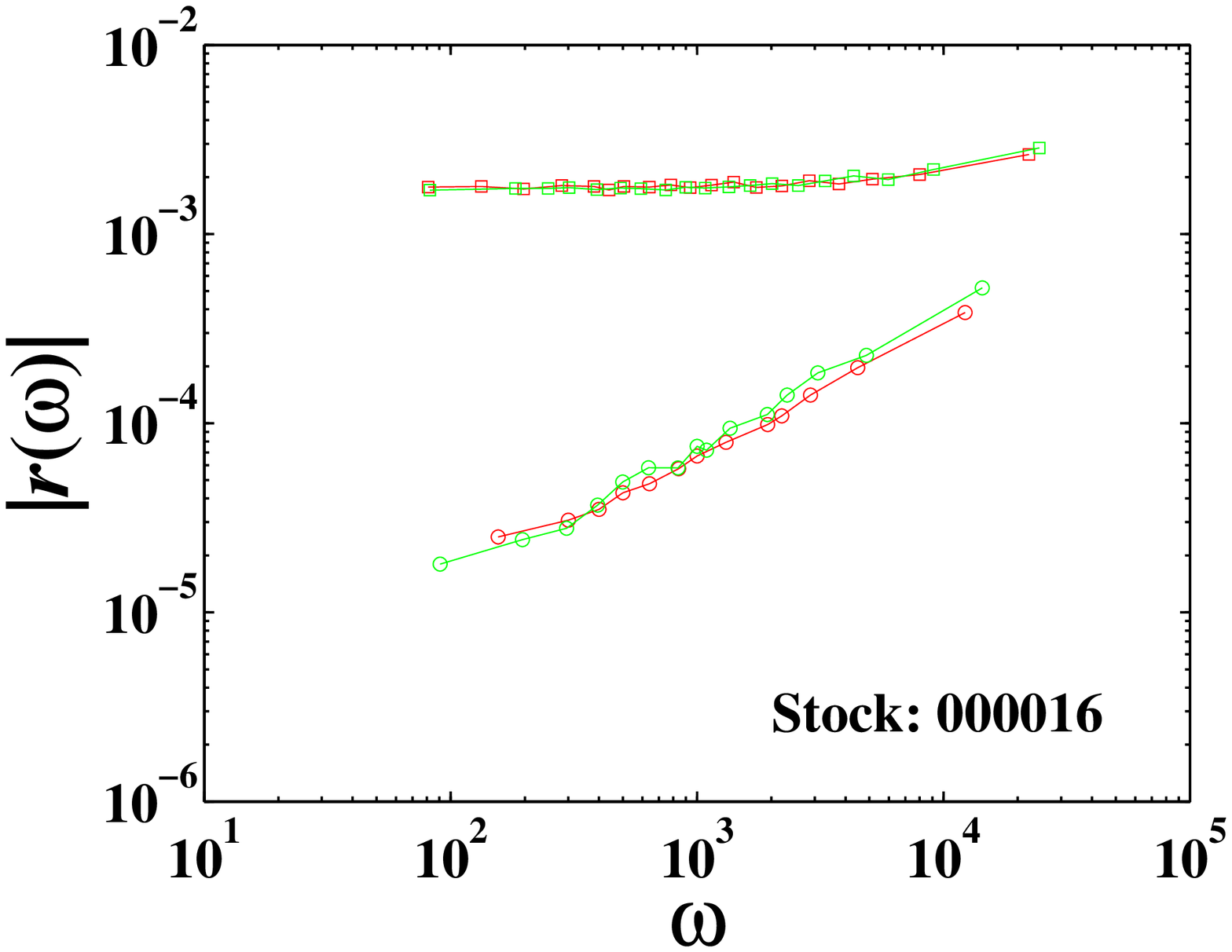}
\includegraphics[width=5cm]{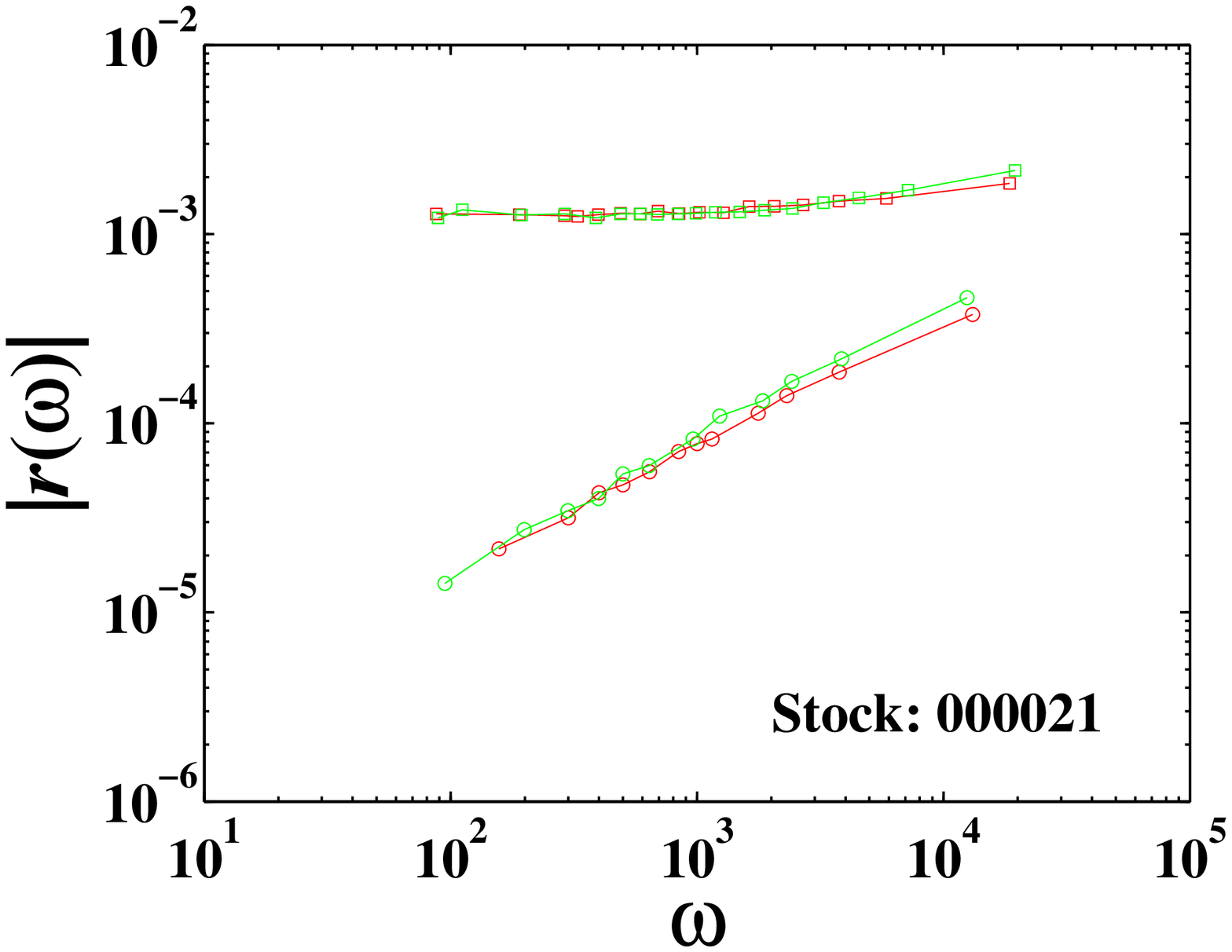}
\includegraphics[width=5cm]{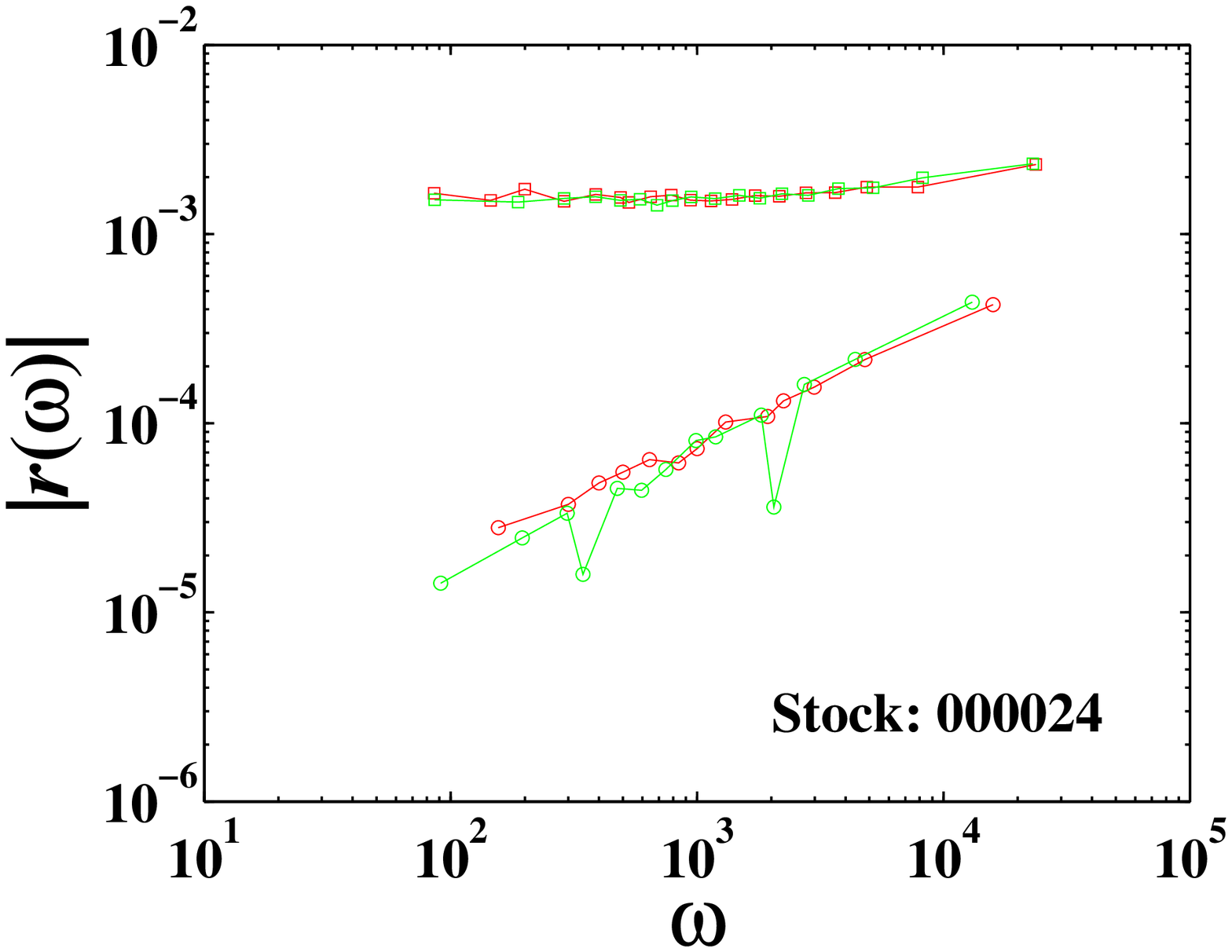}
\includegraphics[width=5cm]{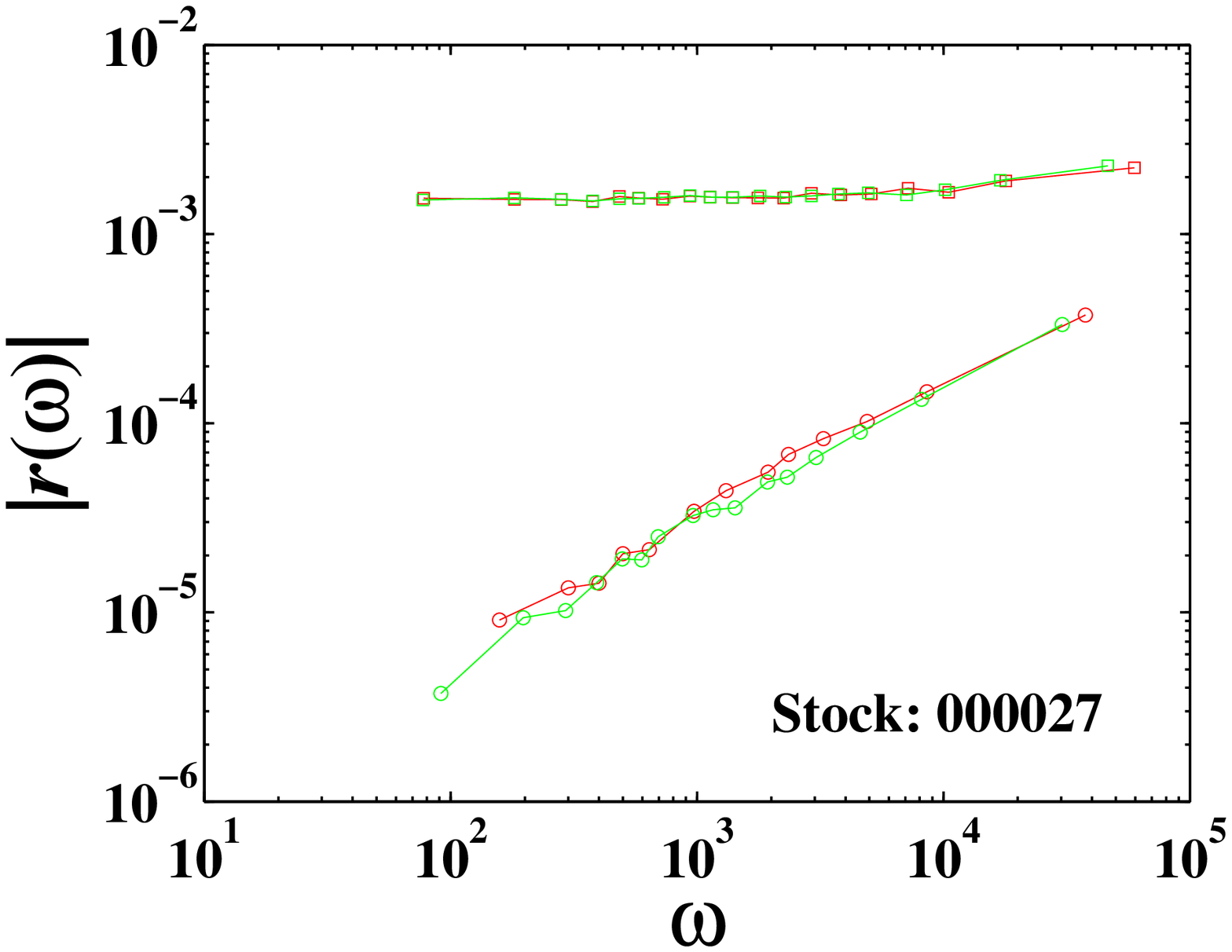}
\includegraphics[width=5cm]{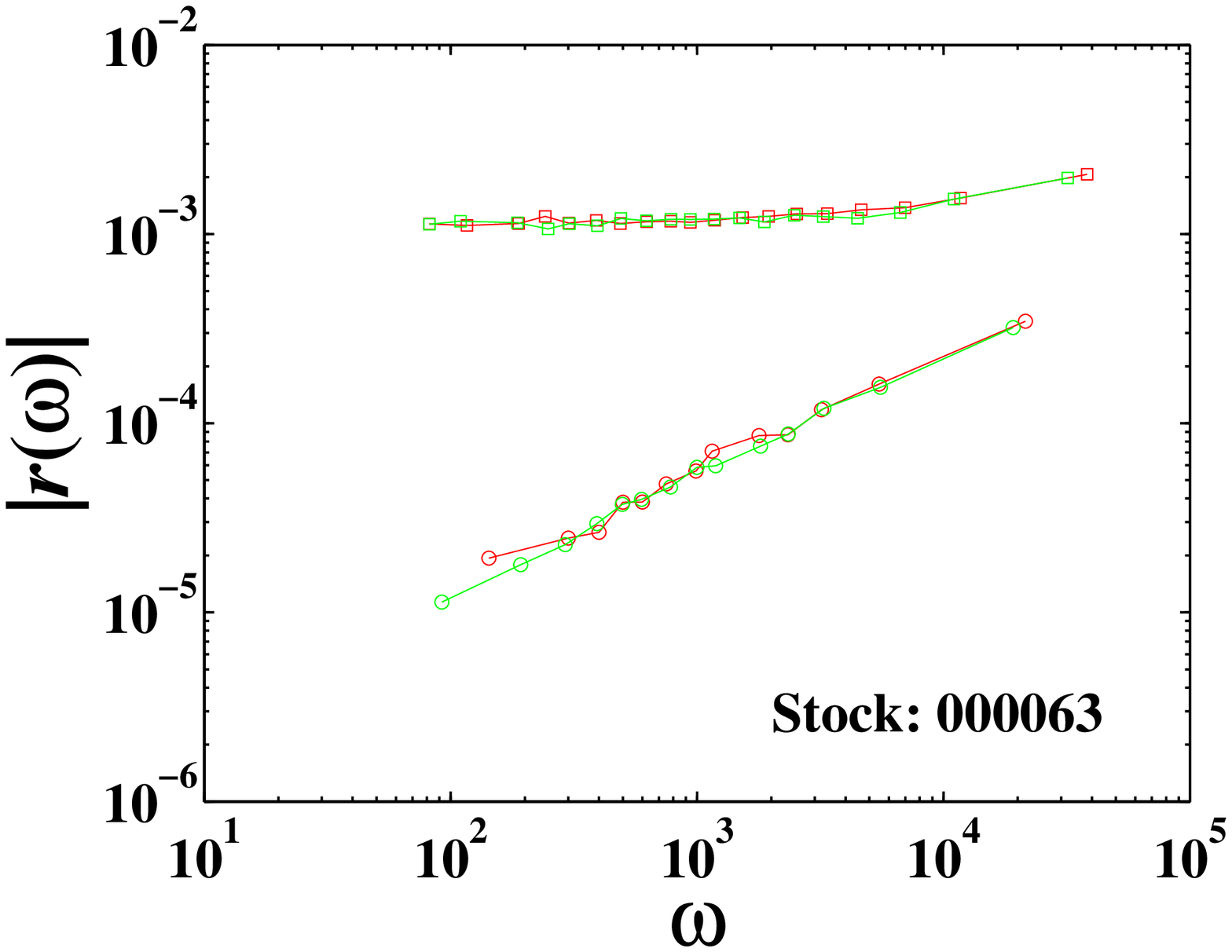}
\includegraphics[width=5cm]{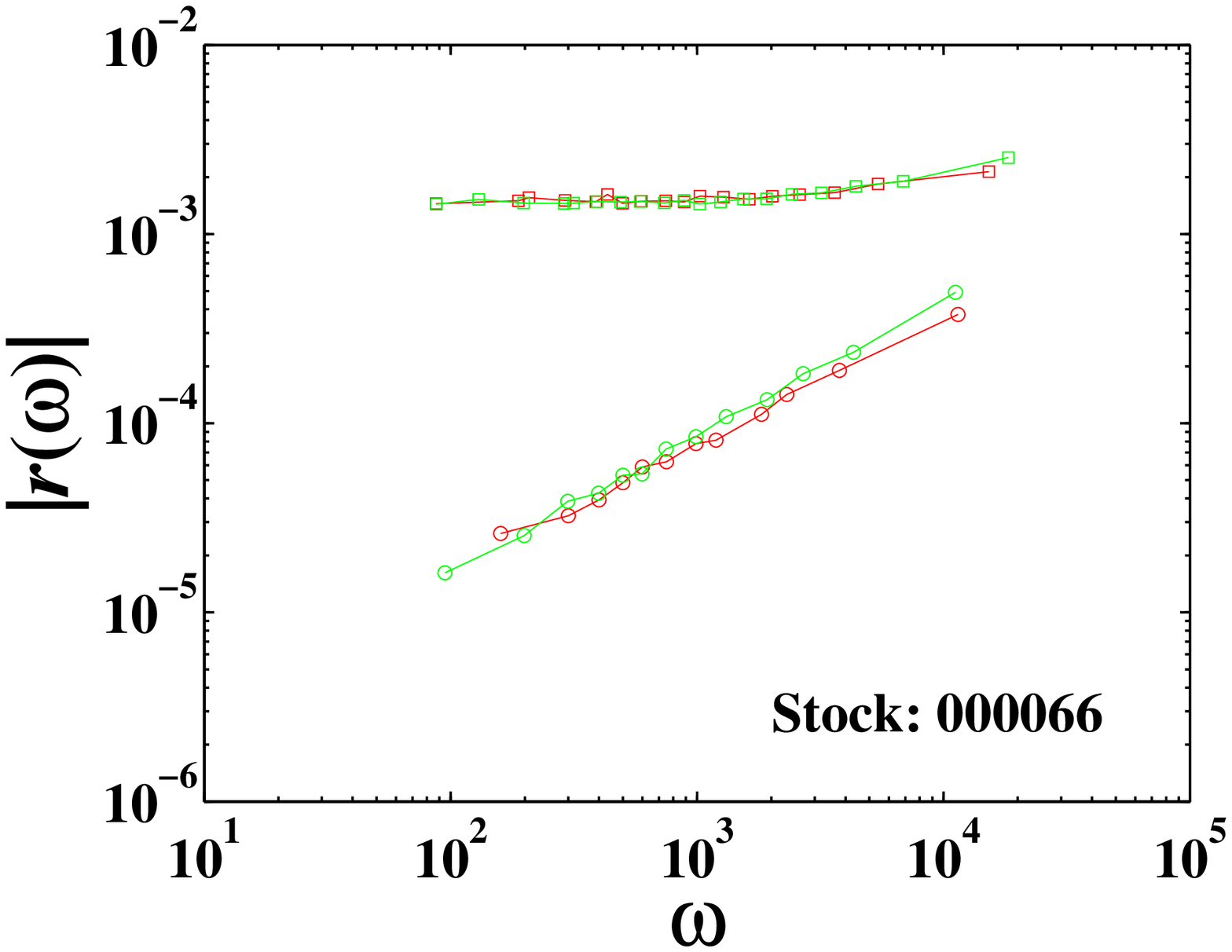}
\includegraphics[width=5cm]{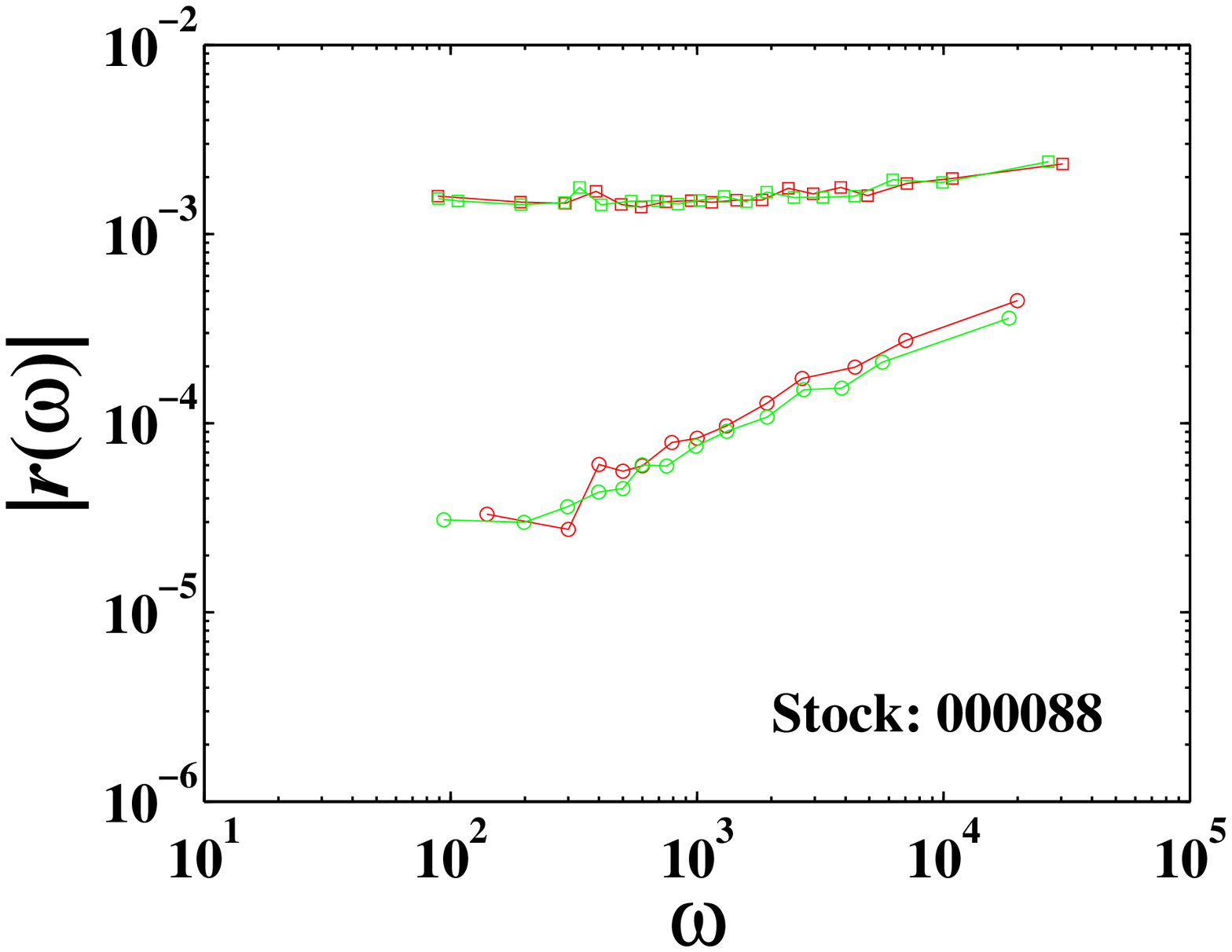}
\includegraphics[width=5cm]{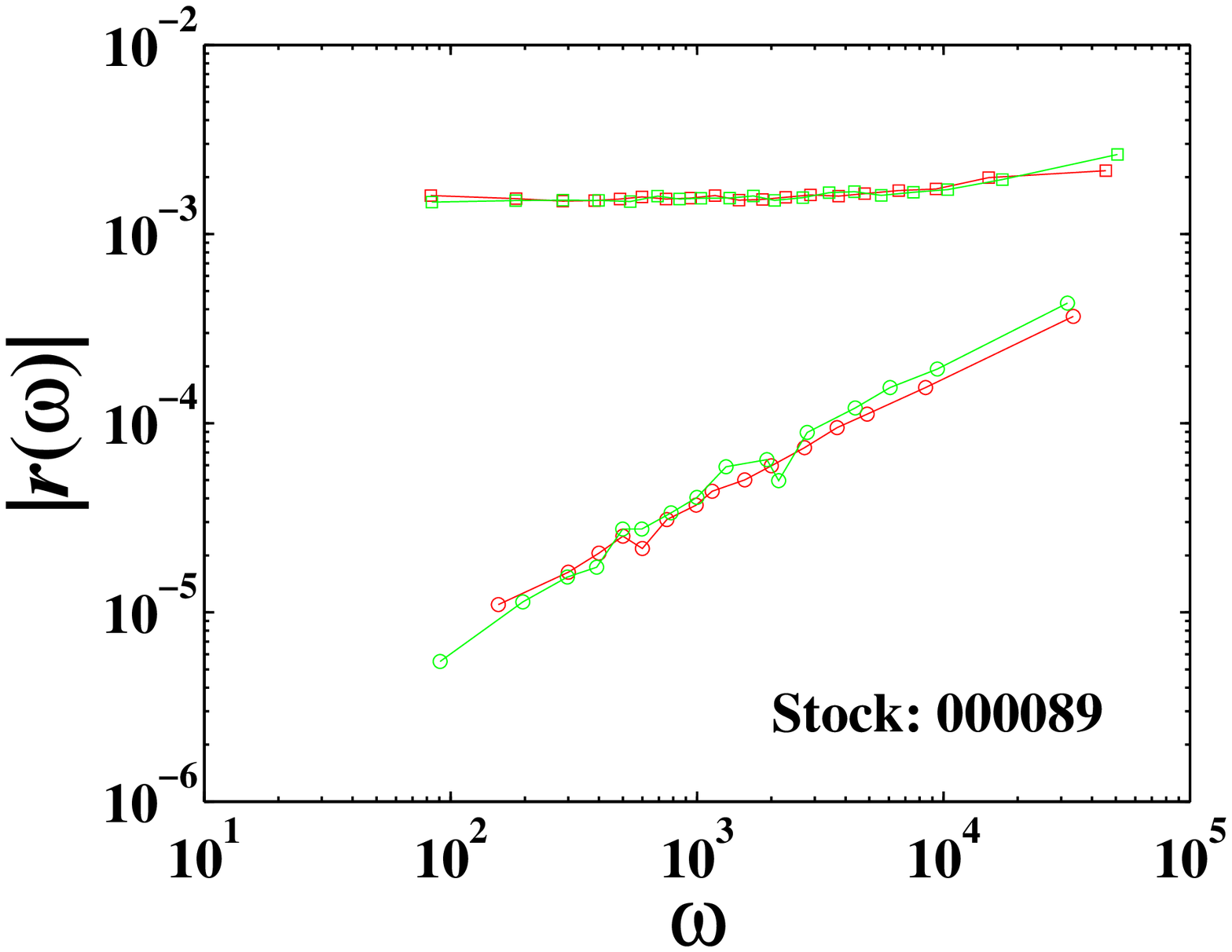}
\centering \caption{Dependence of $|r|$ with respect to $\omega$ for
the four types of trades for individual stocks. Note that the price
impact of buyer- and seller-initiated trades is symmetric and
unfilled trades have greater price impact than filled trades.}
\label{Fig:MPM:Raw1}
\end{figure}

\clearpage
\begin{figure}
\centering
\includegraphics[width=5cm]{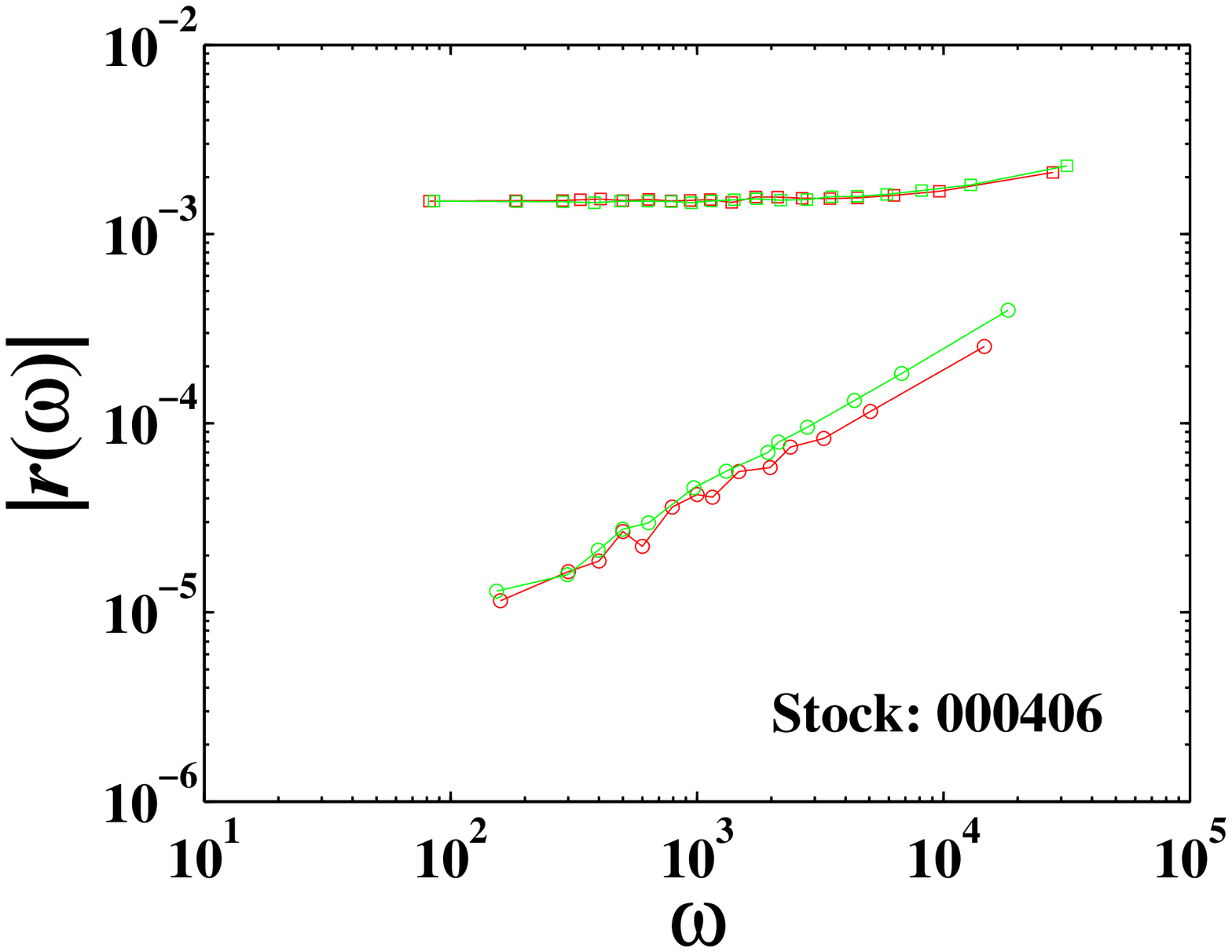}
\includegraphics[width=5cm]{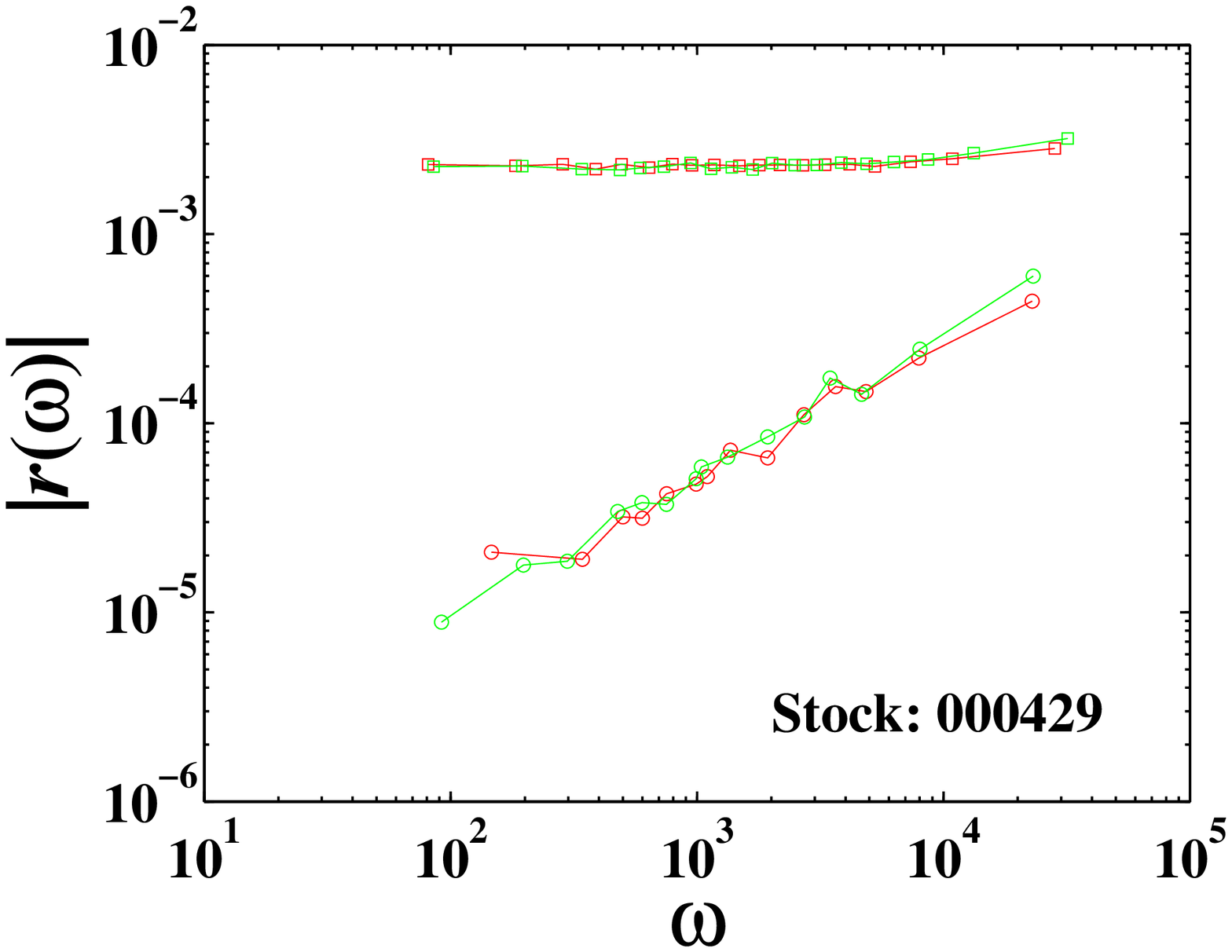}
\includegraphics[width=5cm]{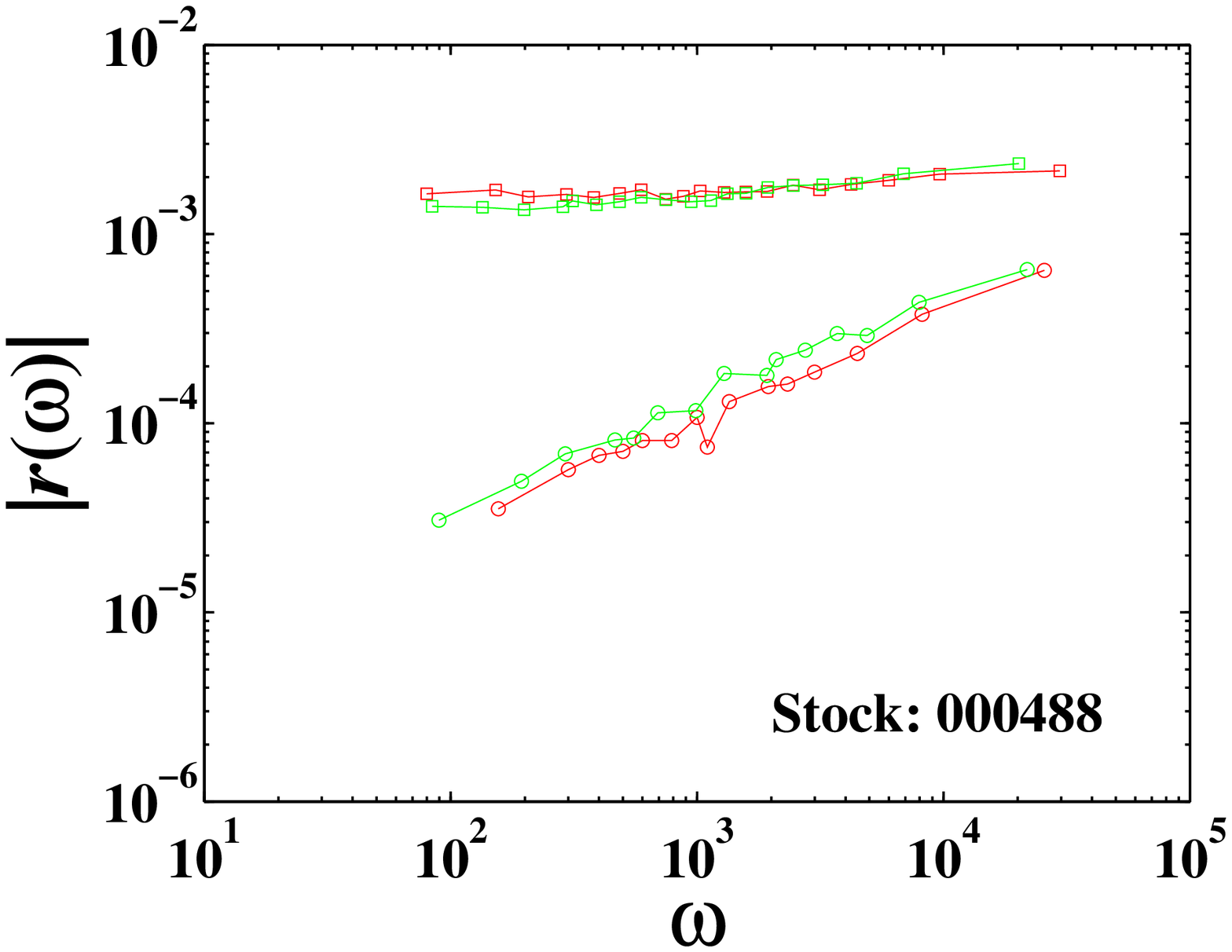}
\includegraphics[width=5cm]{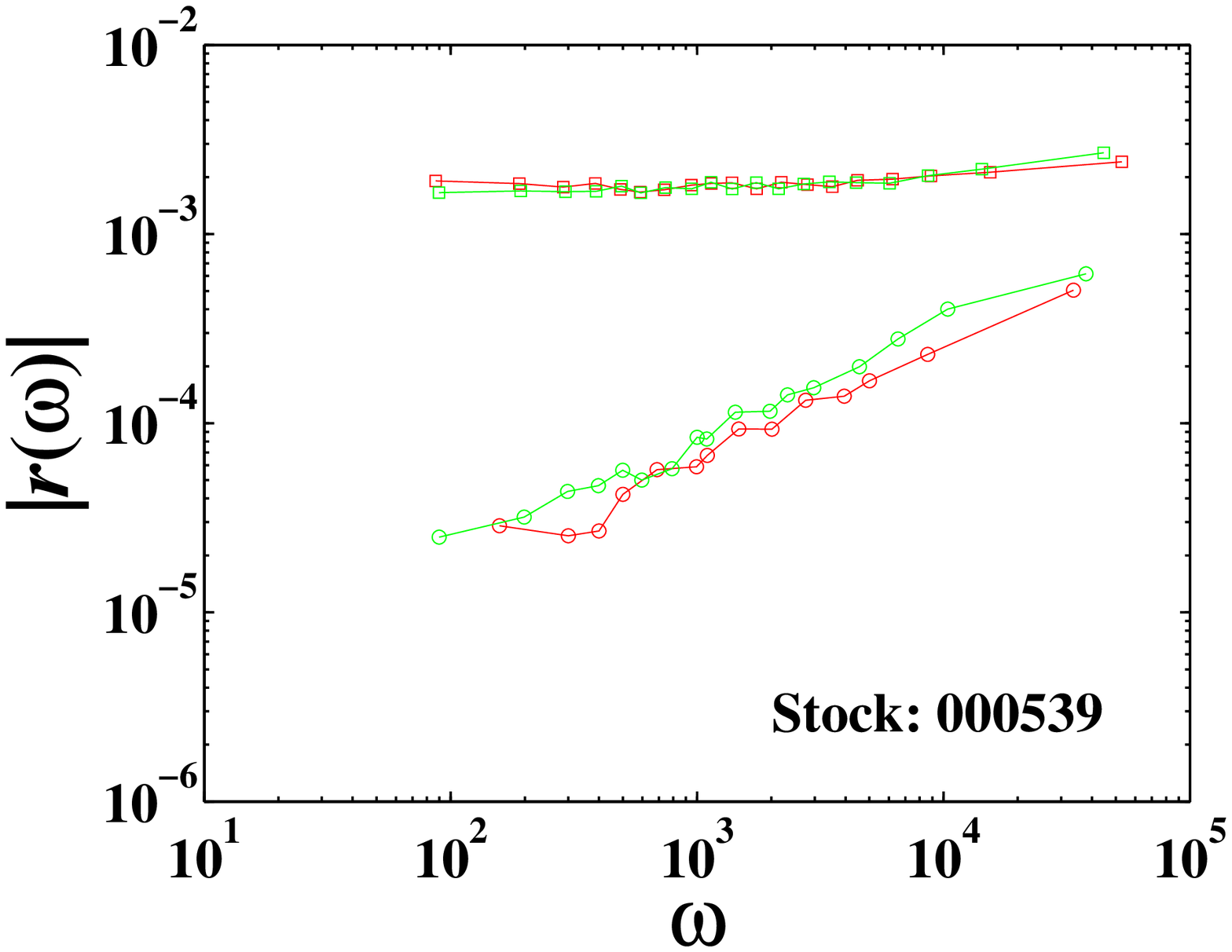}
\includegraphics[width=5cm]{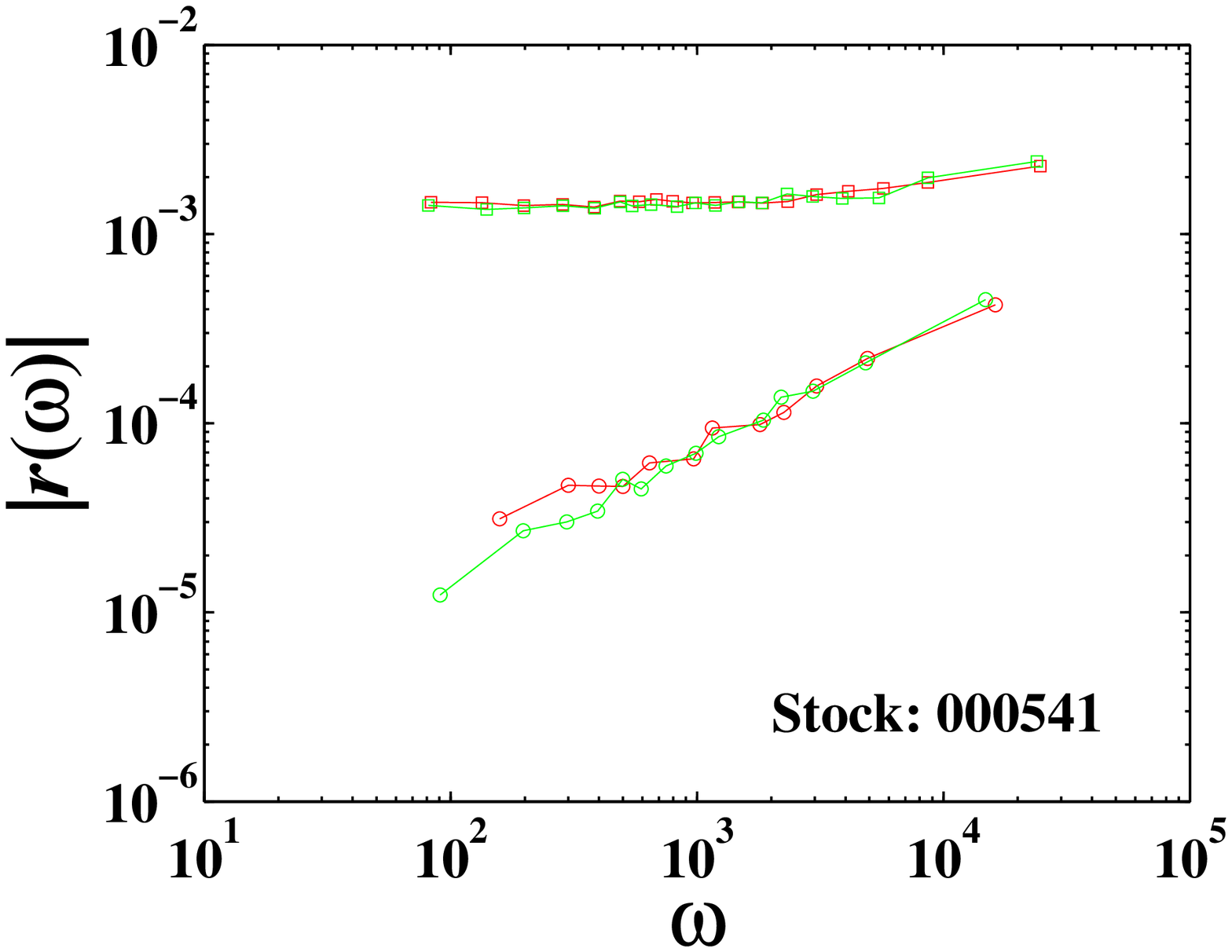}
\includegraphics[width=5cm]{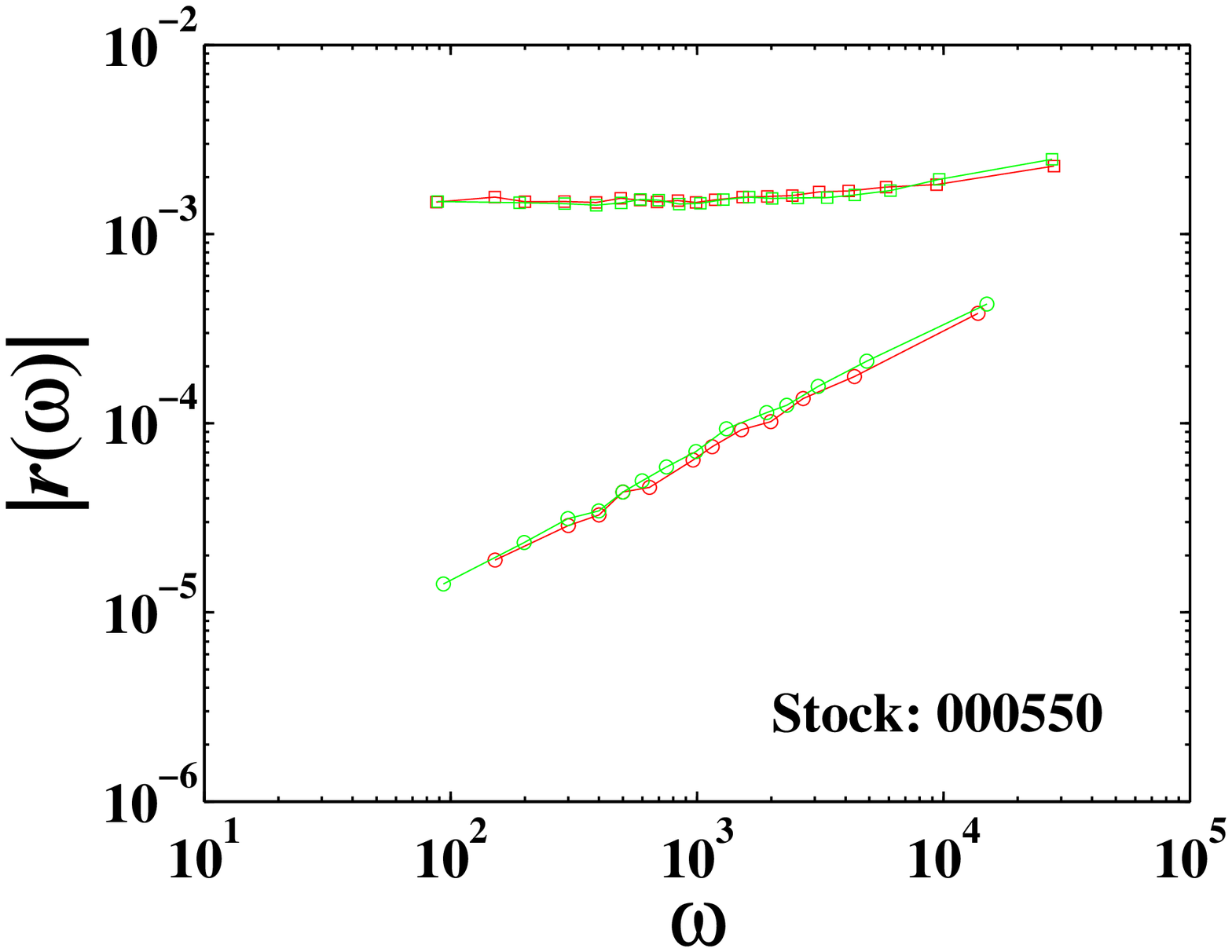}
\includegraphics[width=5cm]{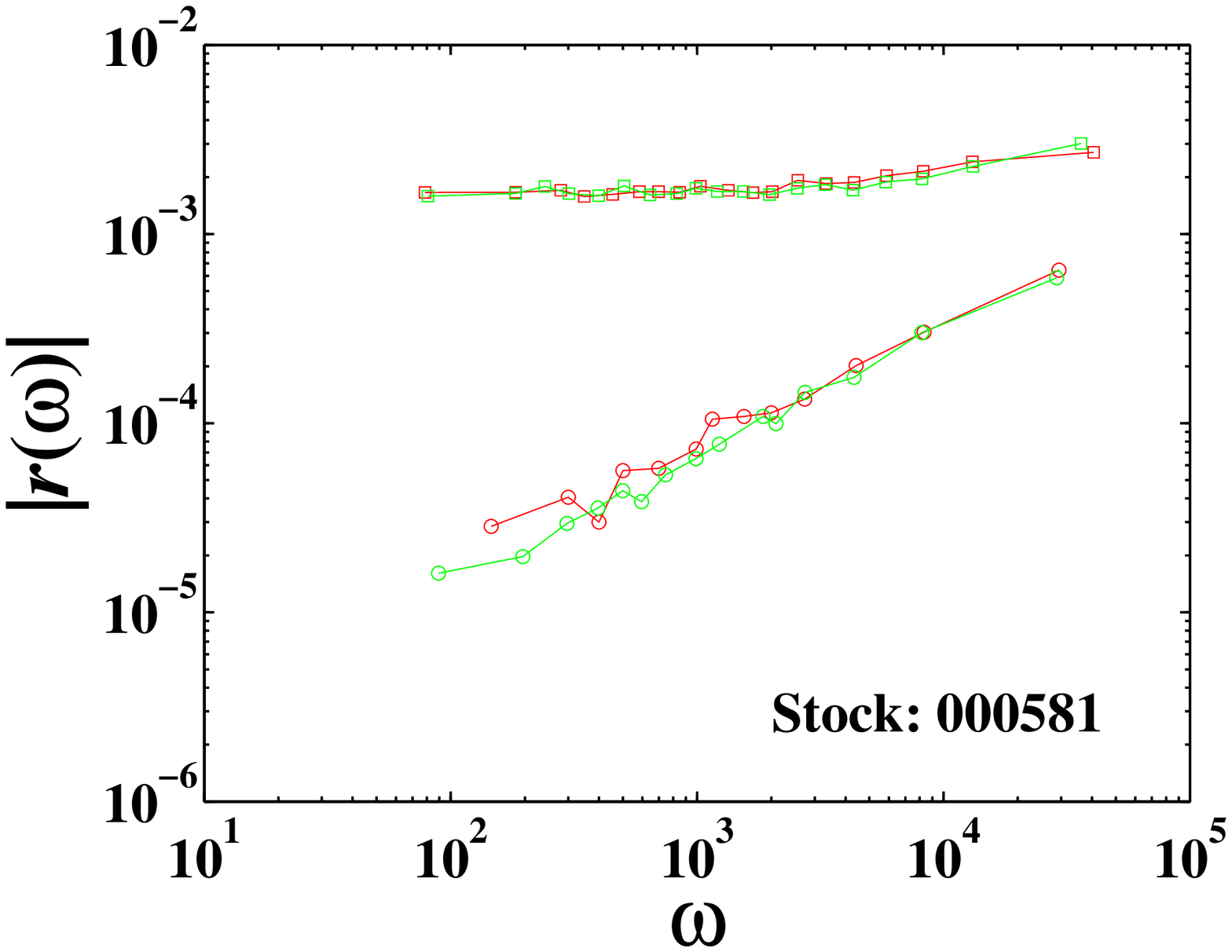}
\includegraphics[width=5cm]{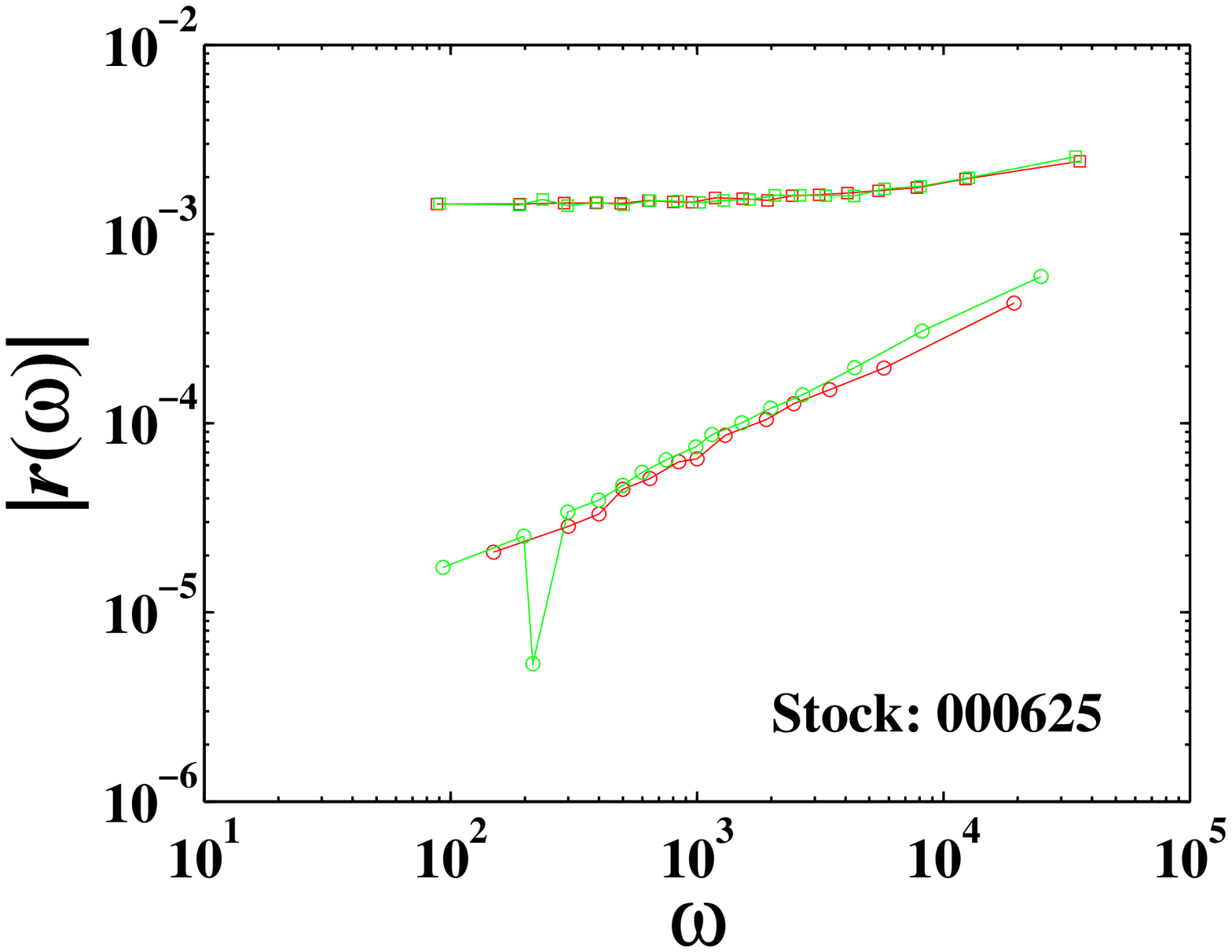}
\includegraphics[width=5cm]{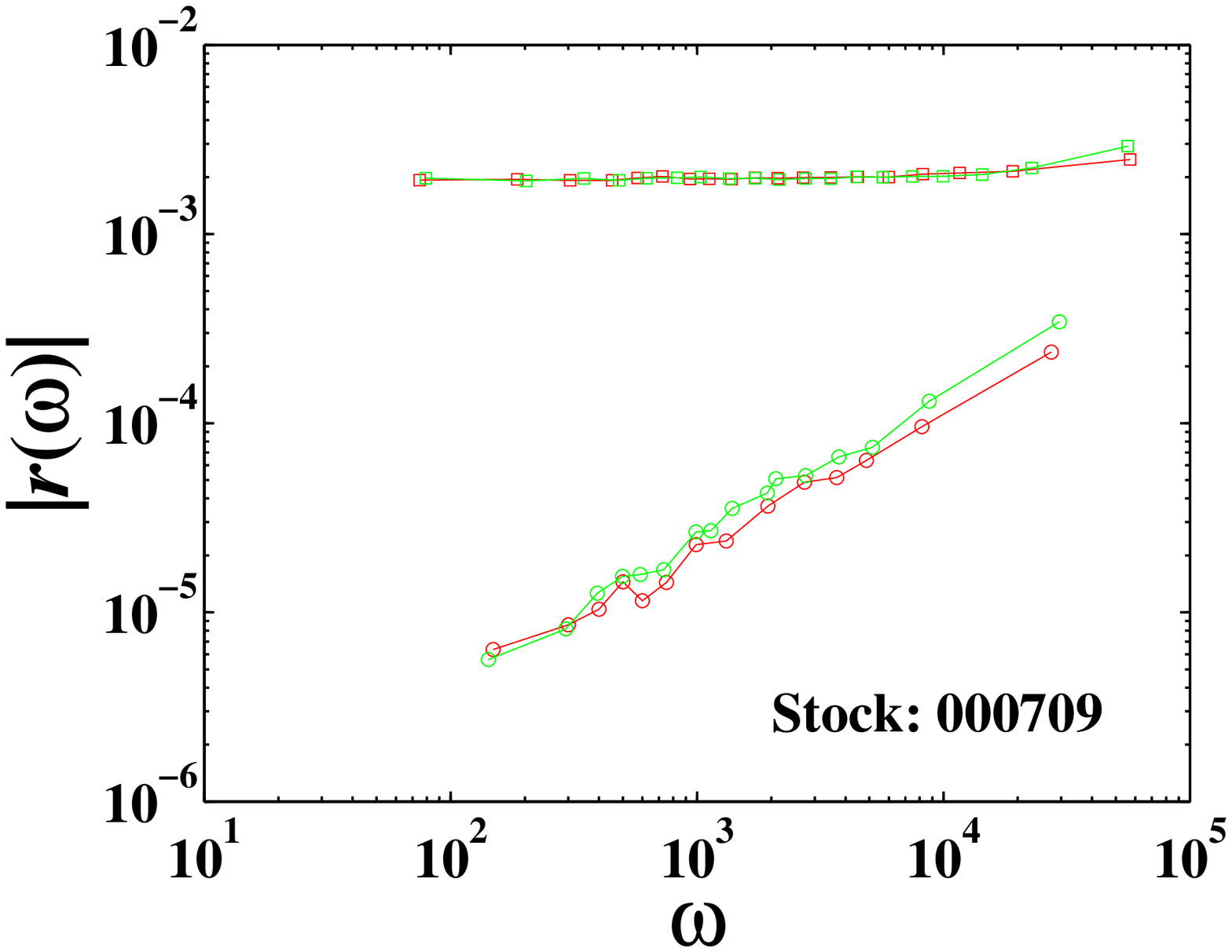}
\includegraphics[width=5cm]{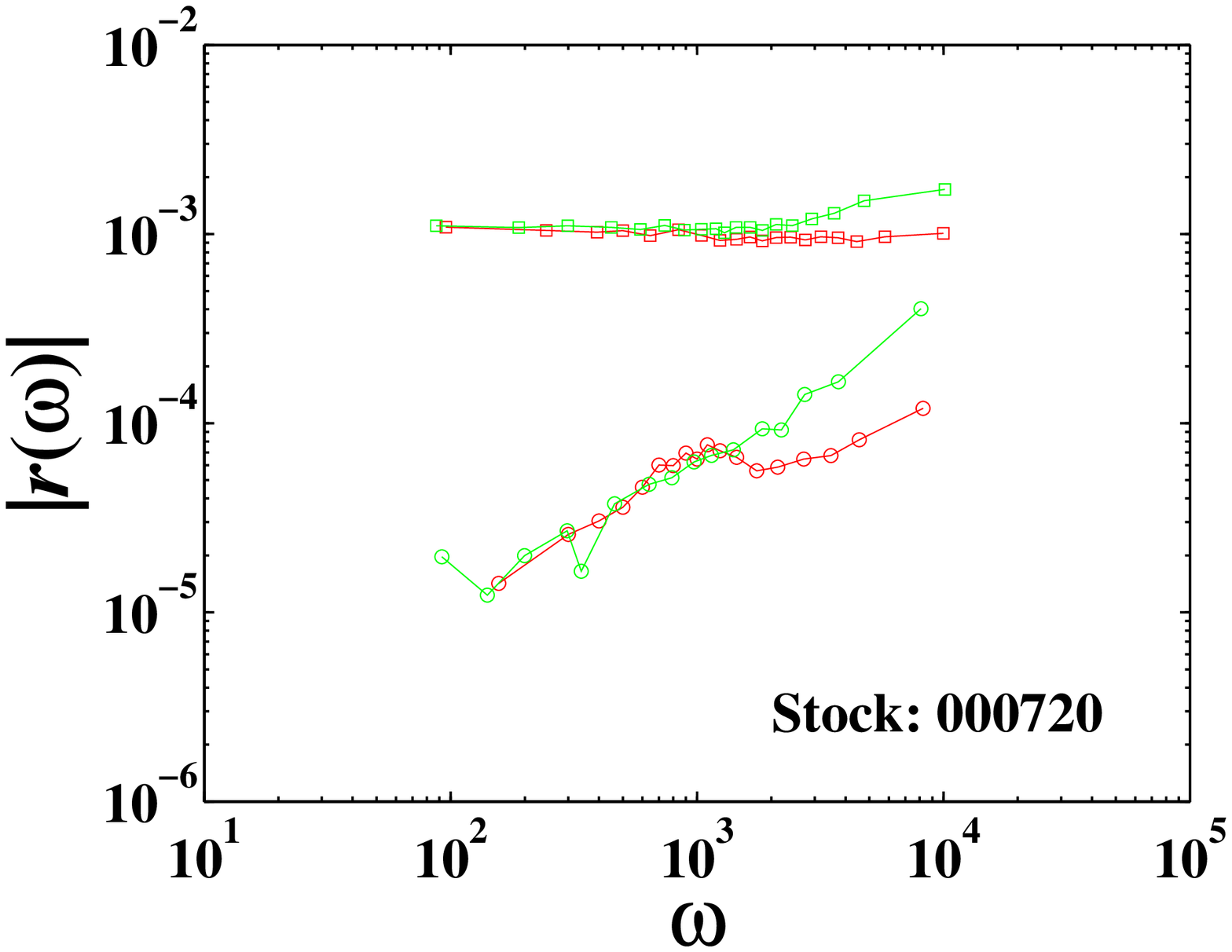}
\includegraphics[width=5cm]{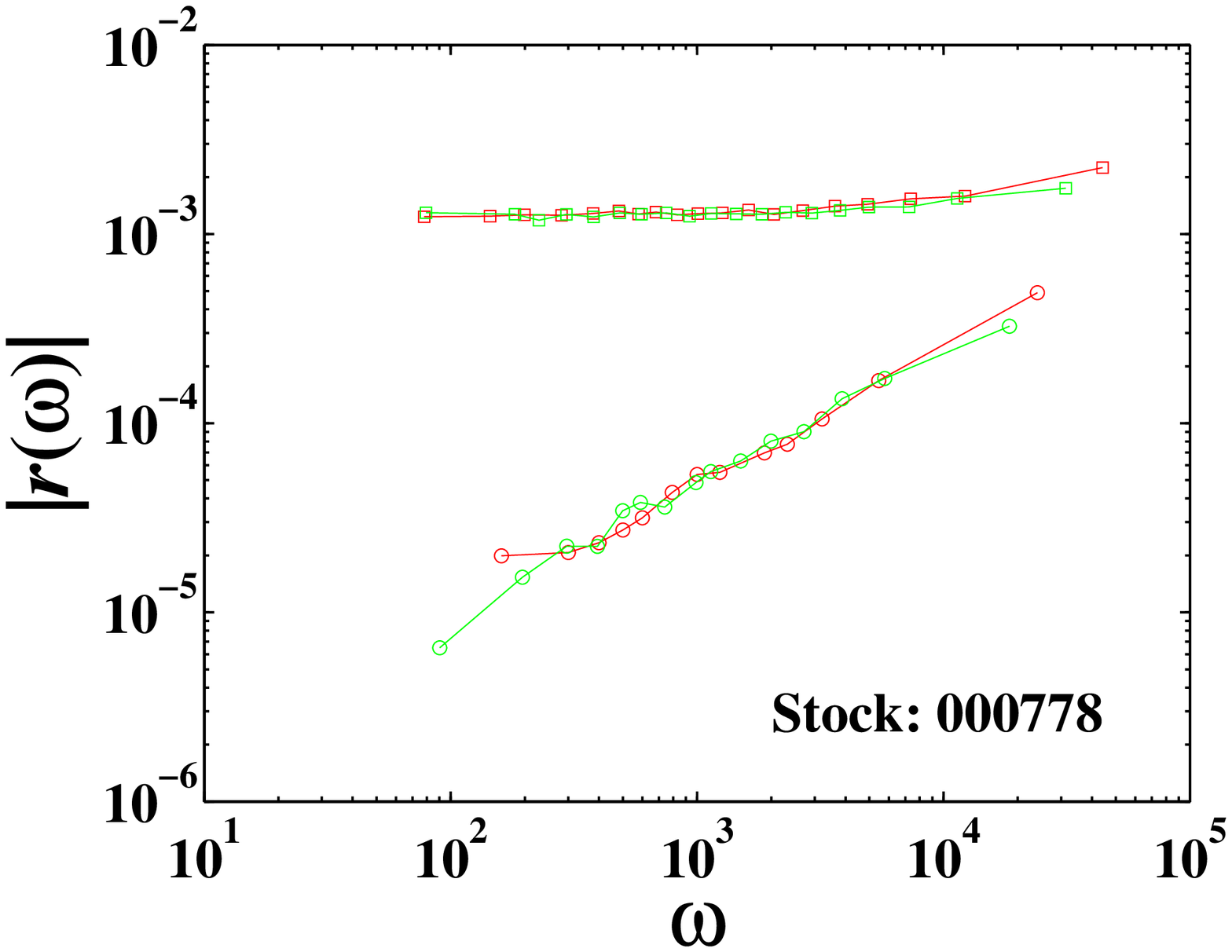}
\caption{Dependence of $|r|$ with respect to $x$ for the four types
of trades for individual stocks. Note that the price impact of
buyer- and seller-initiated trades is symmetric and unfilled trades
have greater price impact than filled trades ({\em{continued}}).}
\label{Fig:MPM:Raw2}
\end{figure}

\newpage
\begin{figure}
\includegraphics[width=5cm]{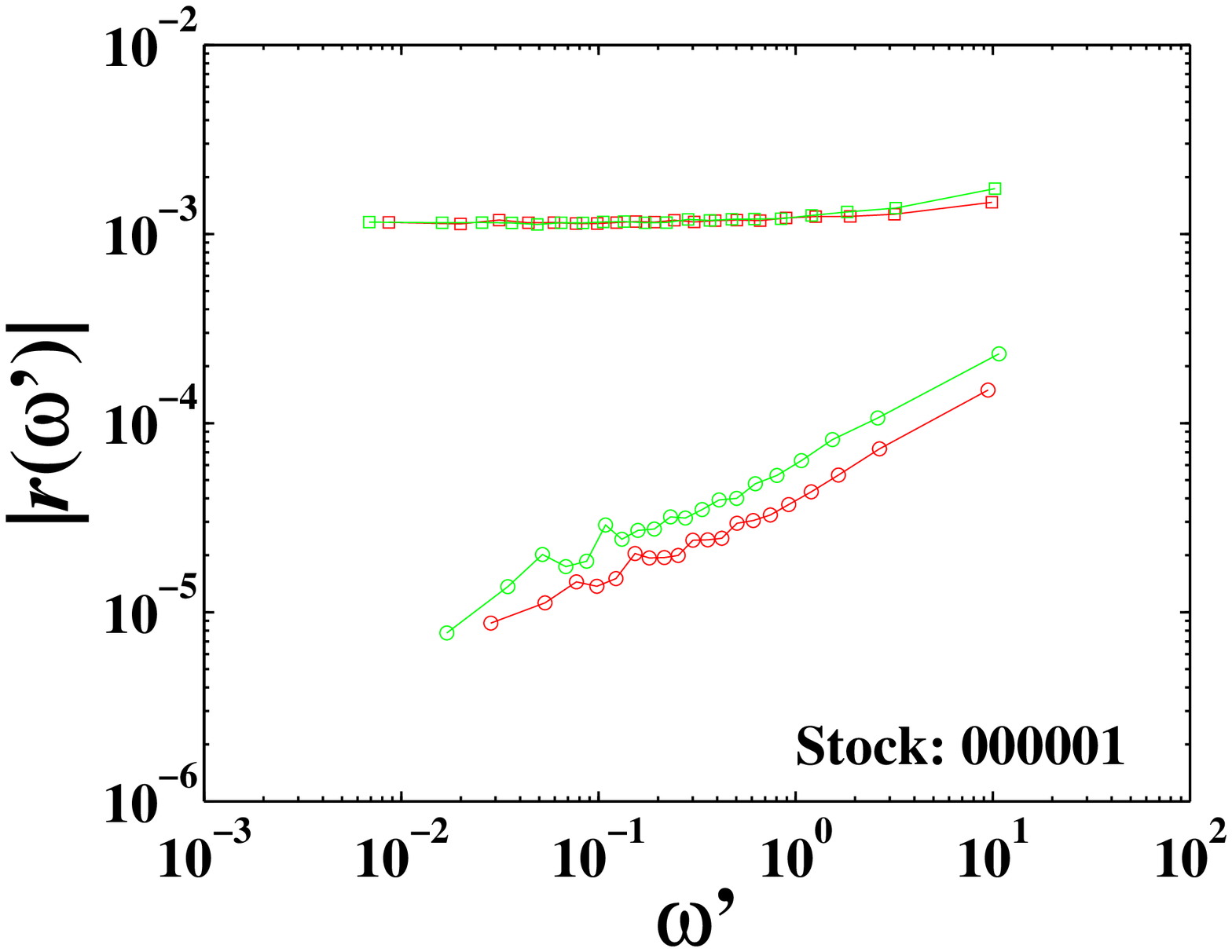}
\includegraphics[width=5cm]{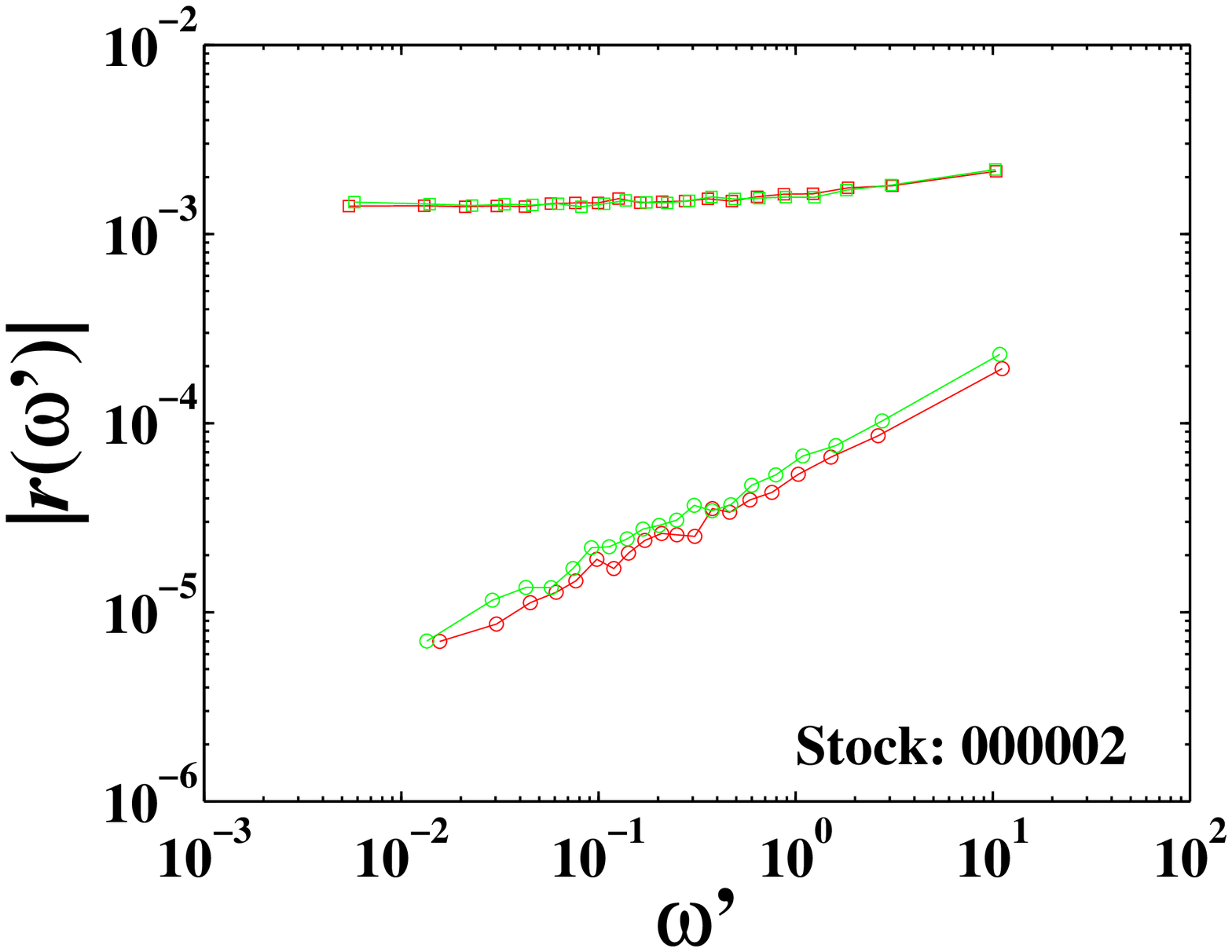}
\includegraphics[width=5cm]{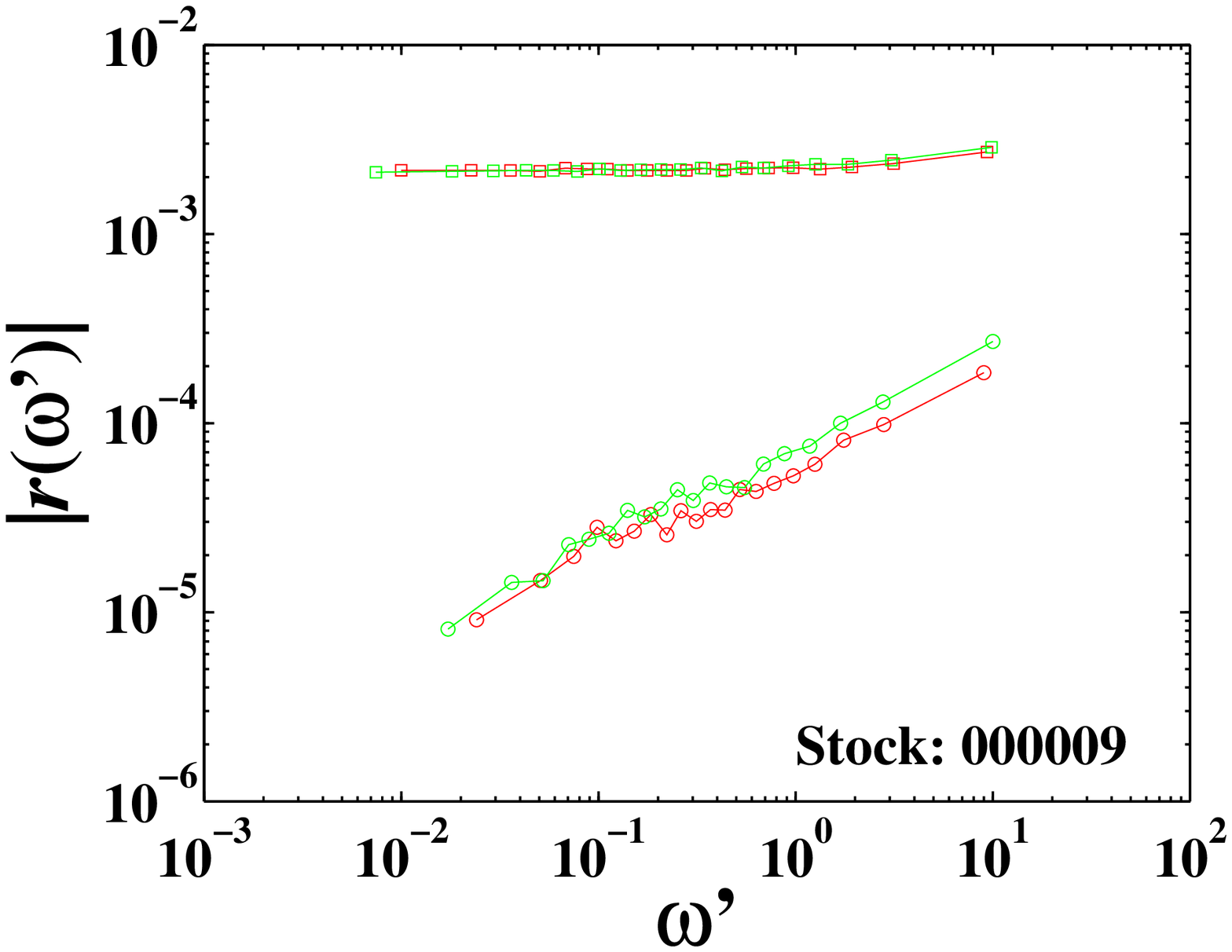}
\includegraphics[width=5cm]{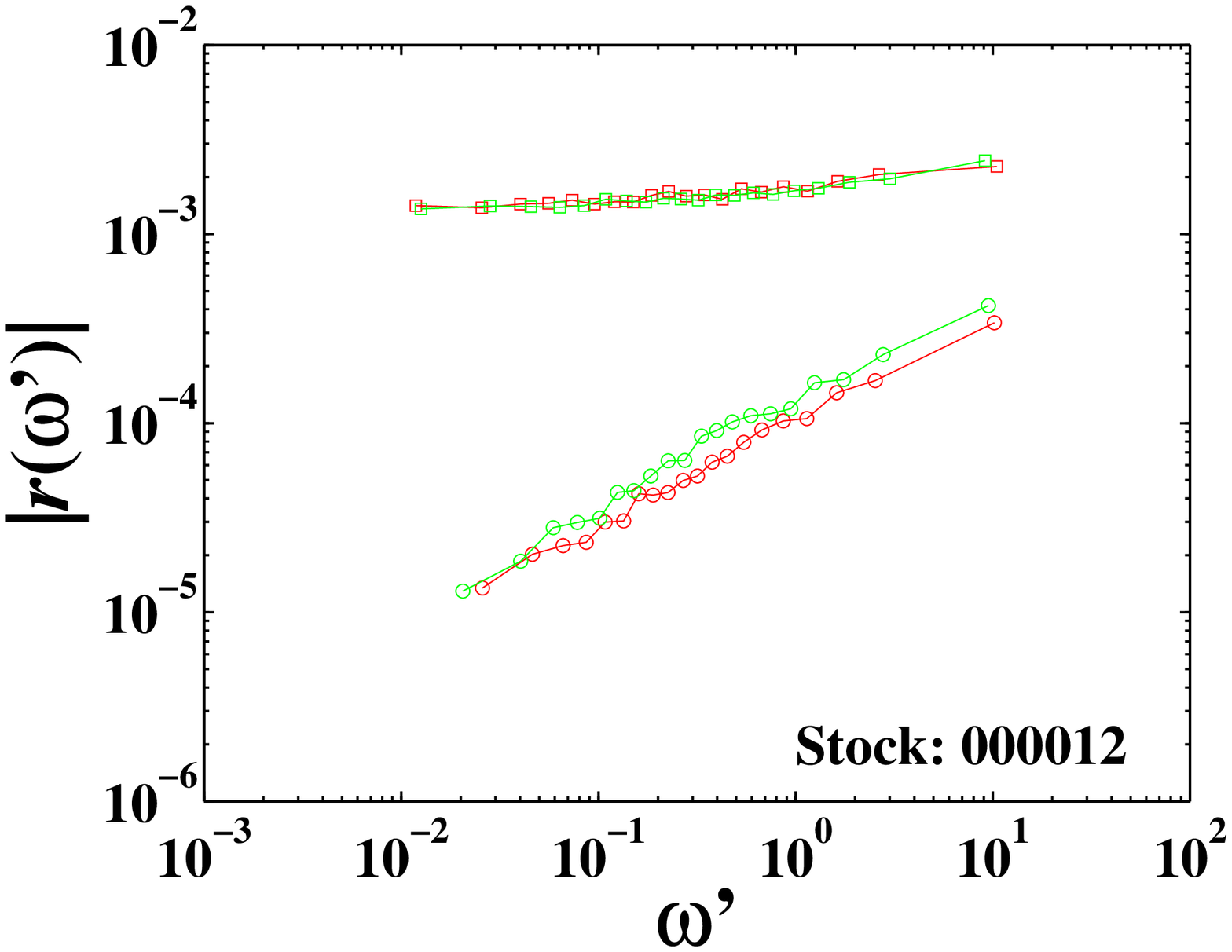}
\includegraphics[width=5cm]{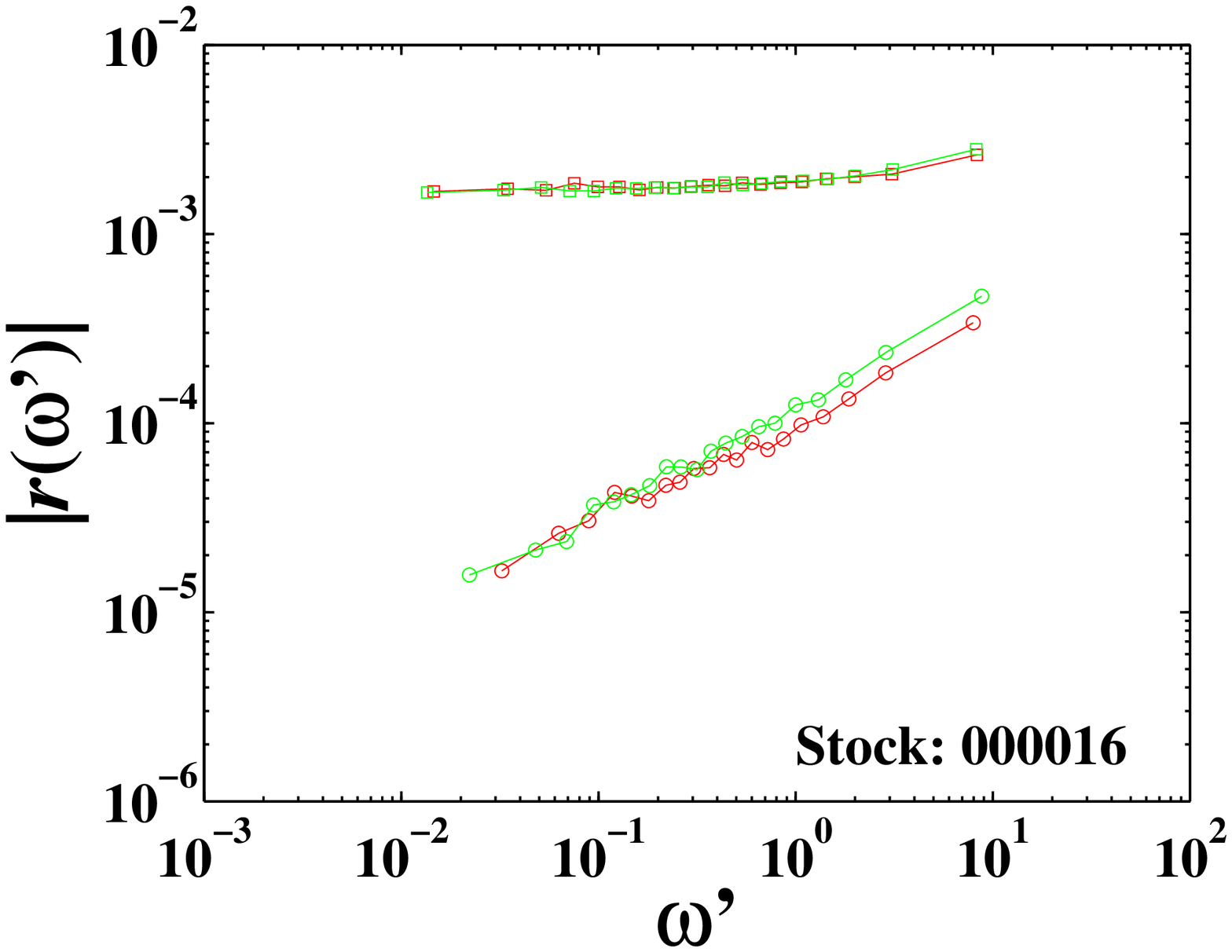}
\includegraphics[width=5cm]{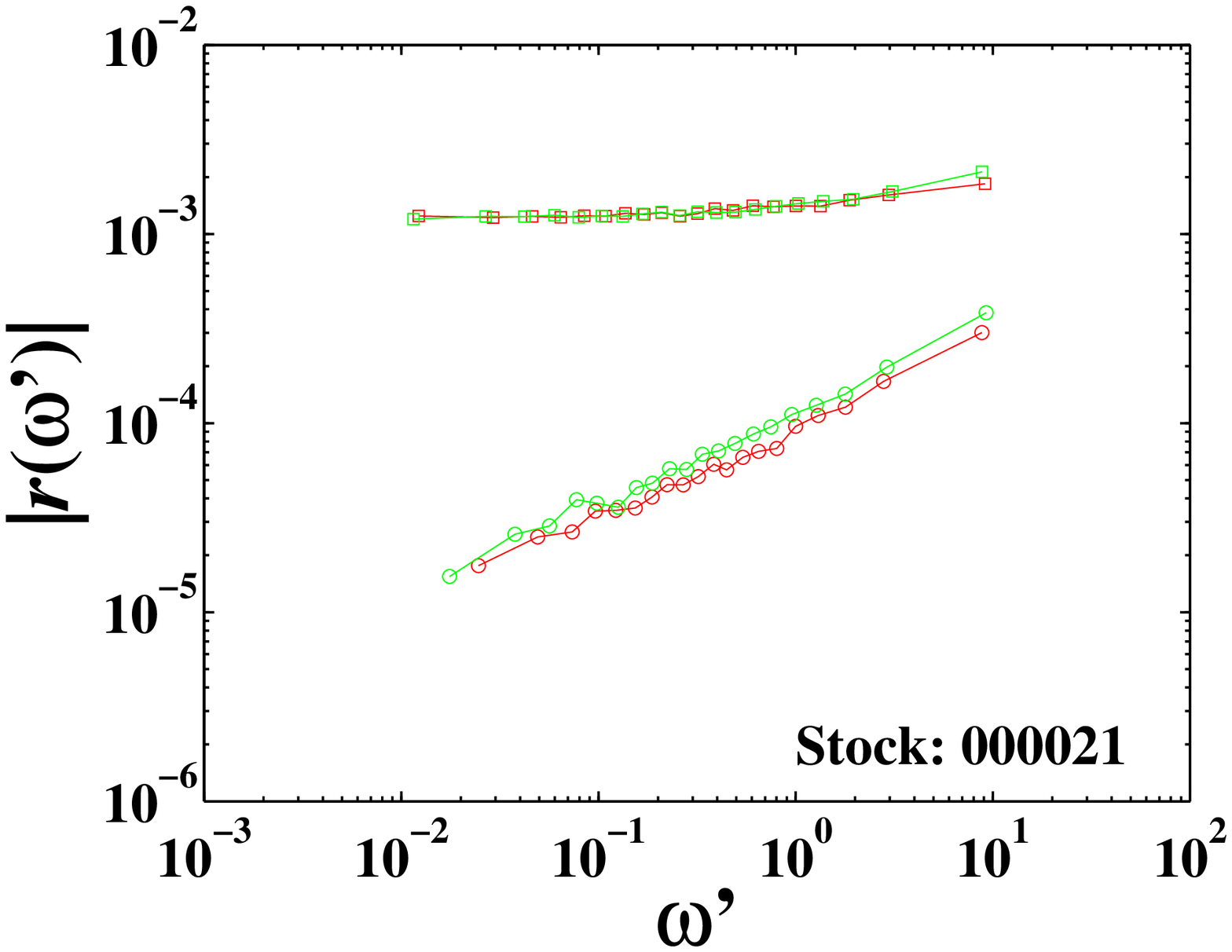}
\includegraphics[width=5cm]{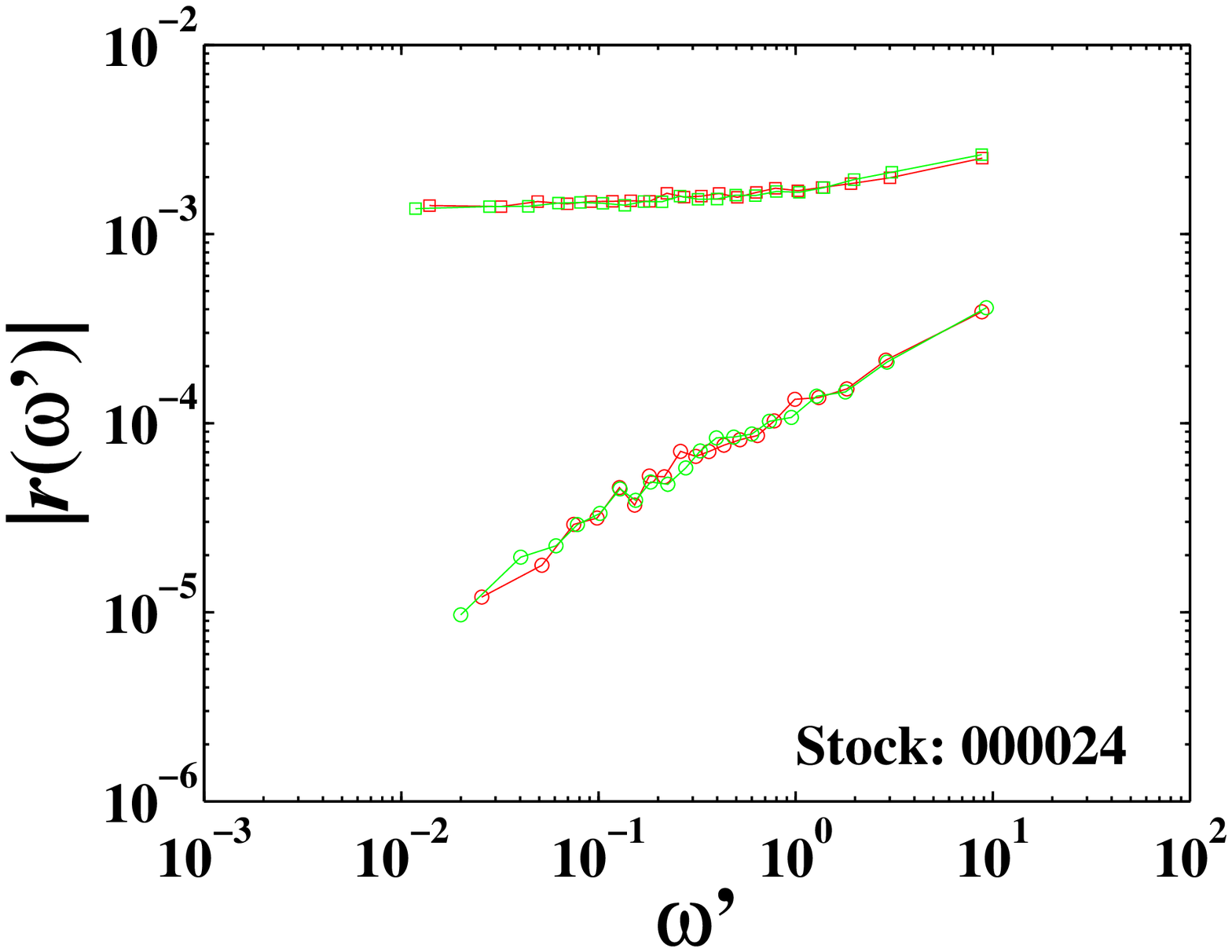}
\includegraphics[width=5cm]{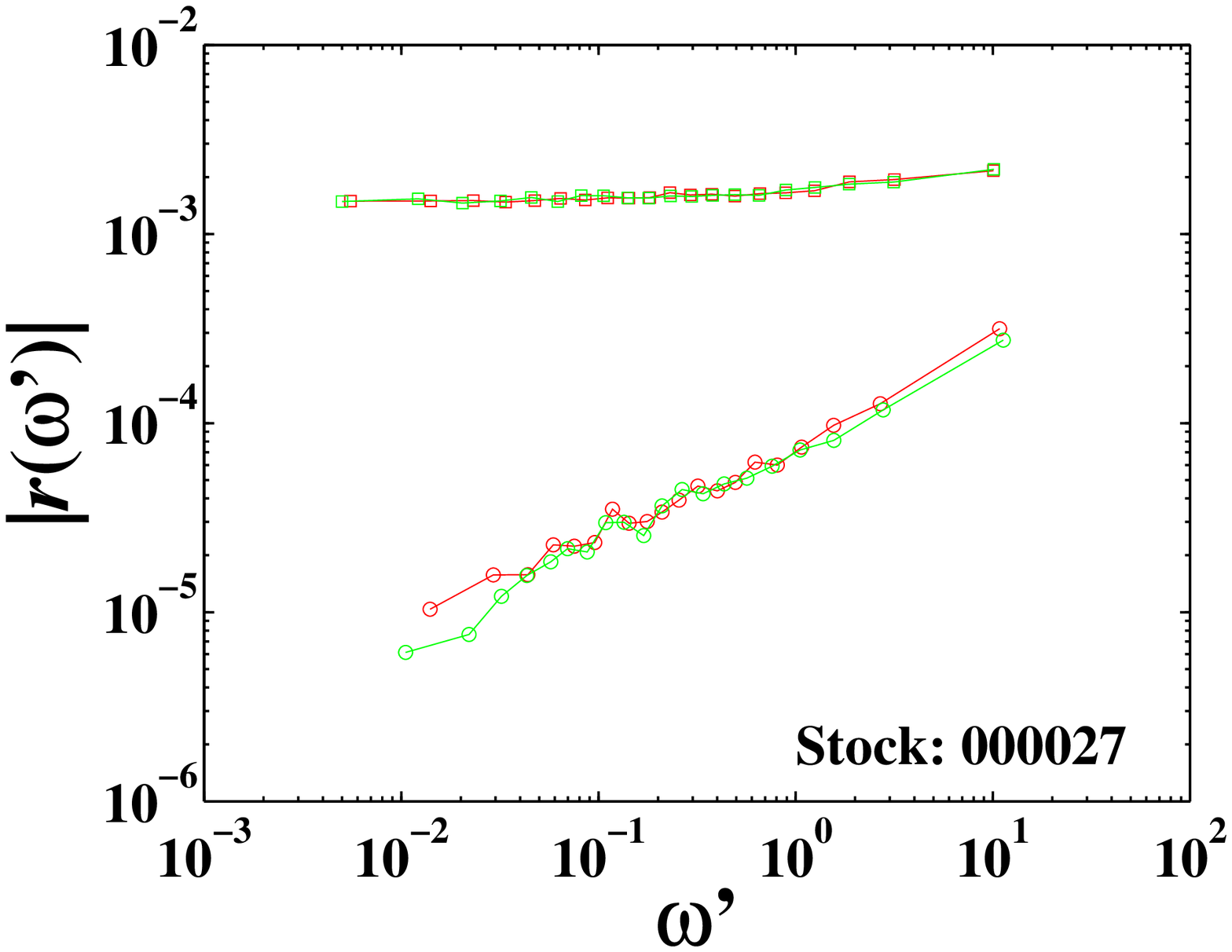}
\includegraphics[width=5cm]{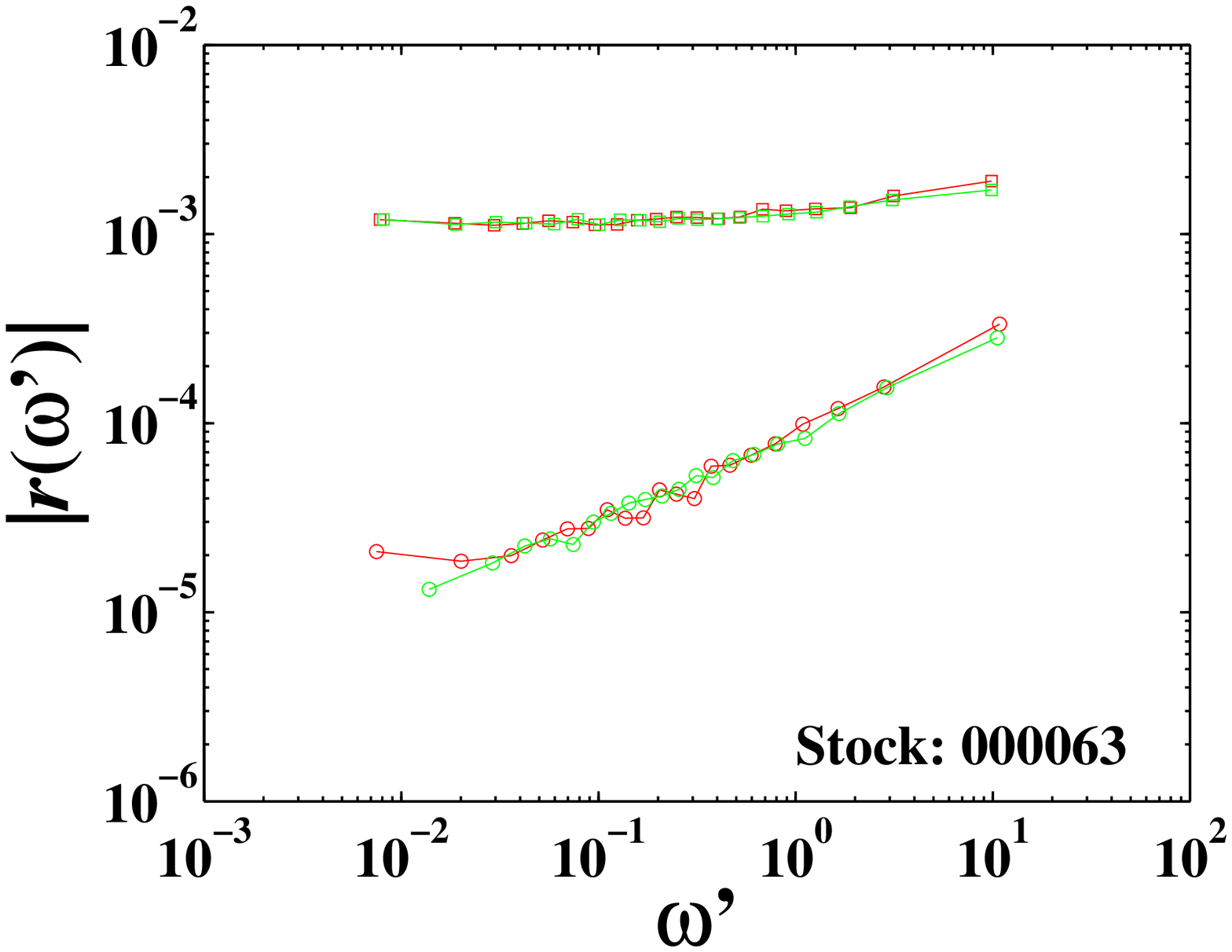}
\includegraphics[width=5cm]{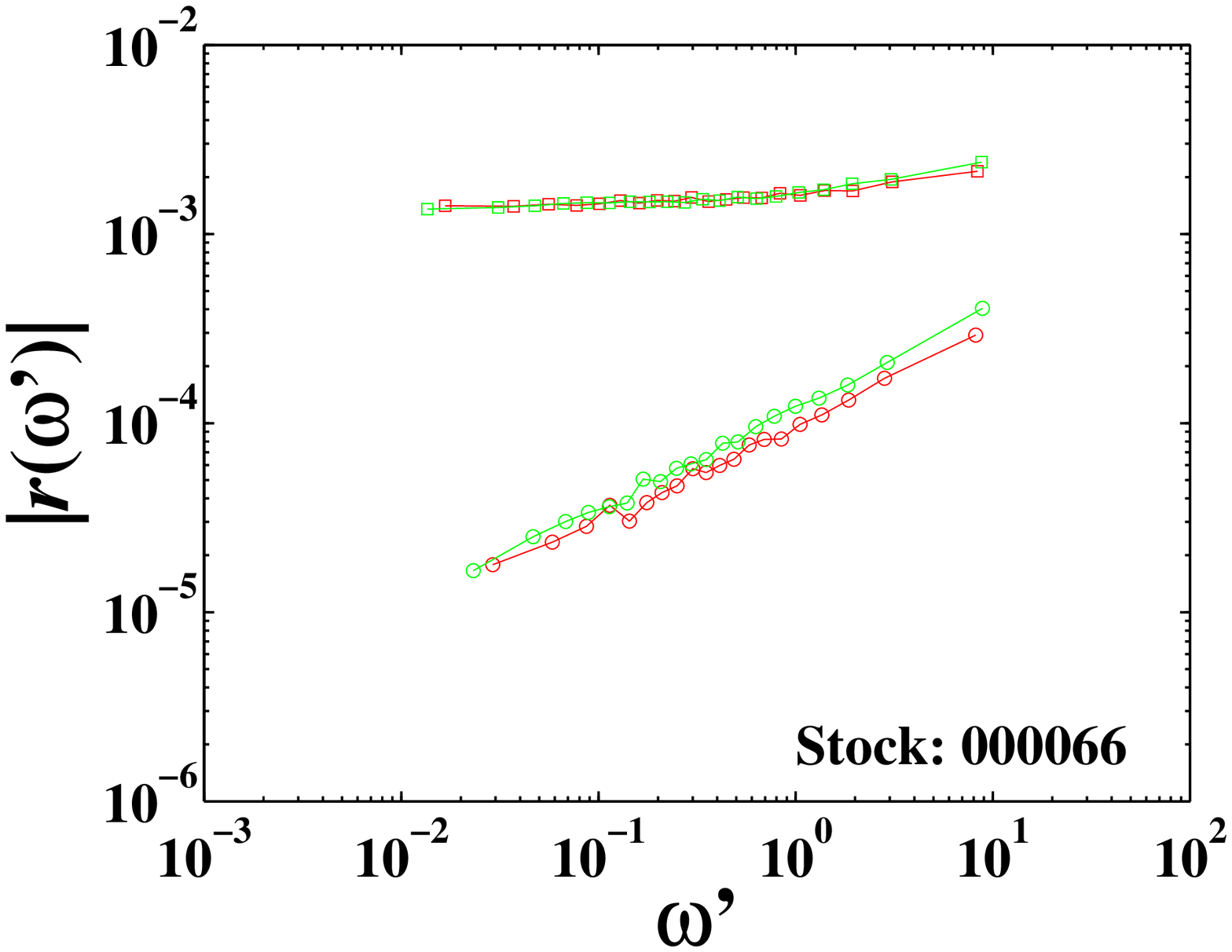}
\includegraphics[width=5cm]{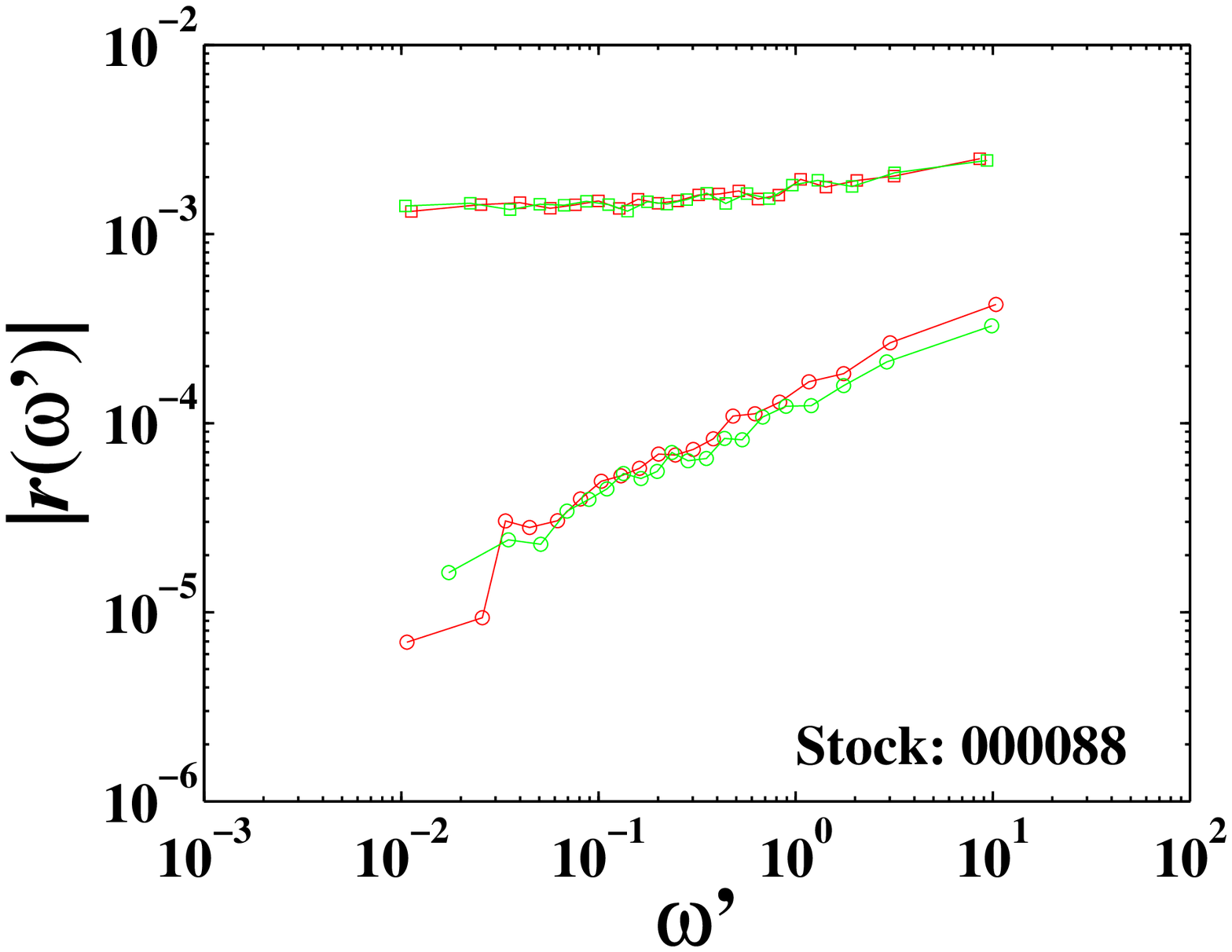}
\includegraphics[width=5cm]{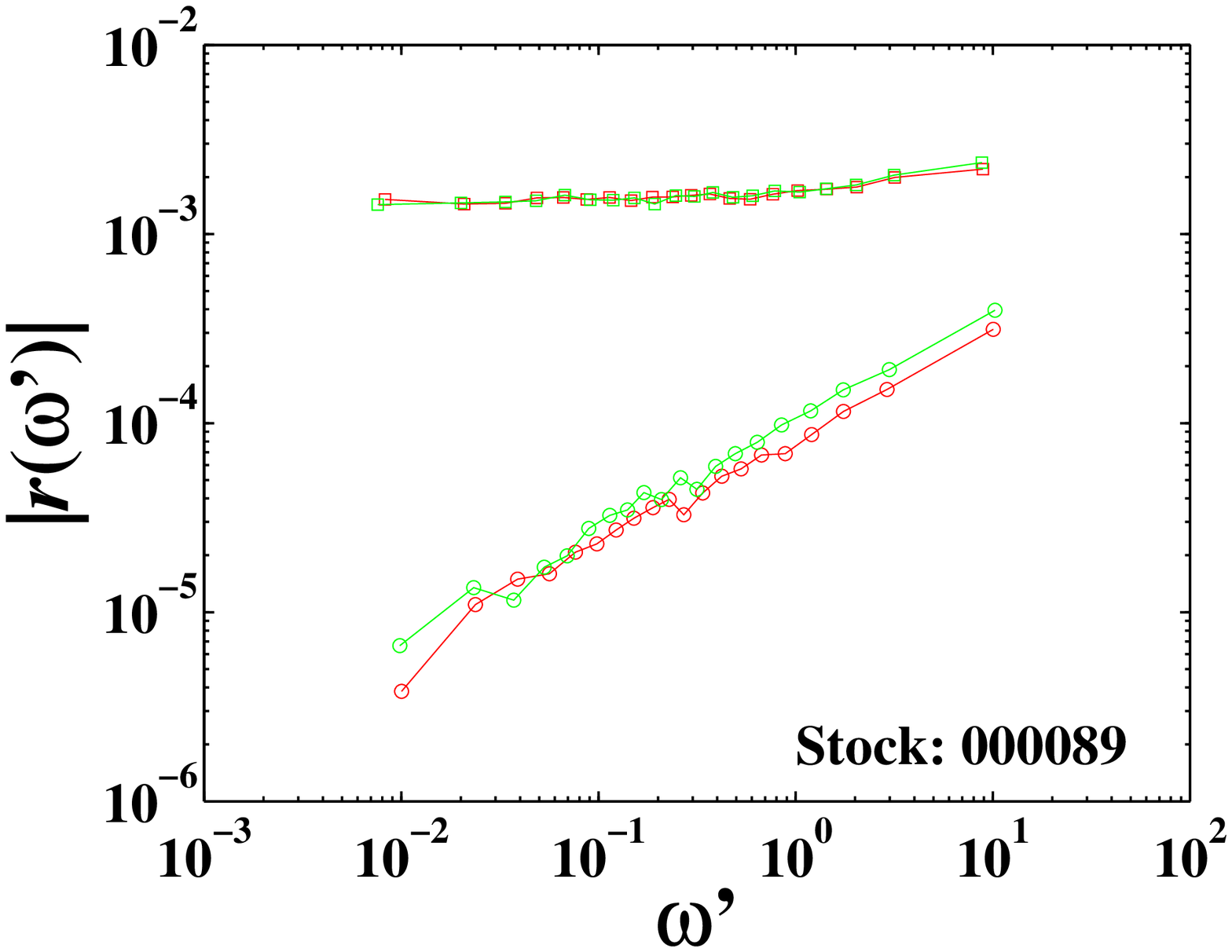}
\centering \caption{Dependence of $|r|$ with respect to $x$ for the
four types of trades for individual stocks. The normalized trade
size $x=\omega'$ is obtained following \citet{Lim-Coggins-2005-QF}.
Note that the price impact of buyer- and seller-initiated trades is
symmetric and unfilled trades have greater price impact than filled
trades.} \label{Fig:MPM:LC:Raw1}
\end{figure}

\clearpage
\begin{figure}
\centering
\includegraphics[width=5cm]{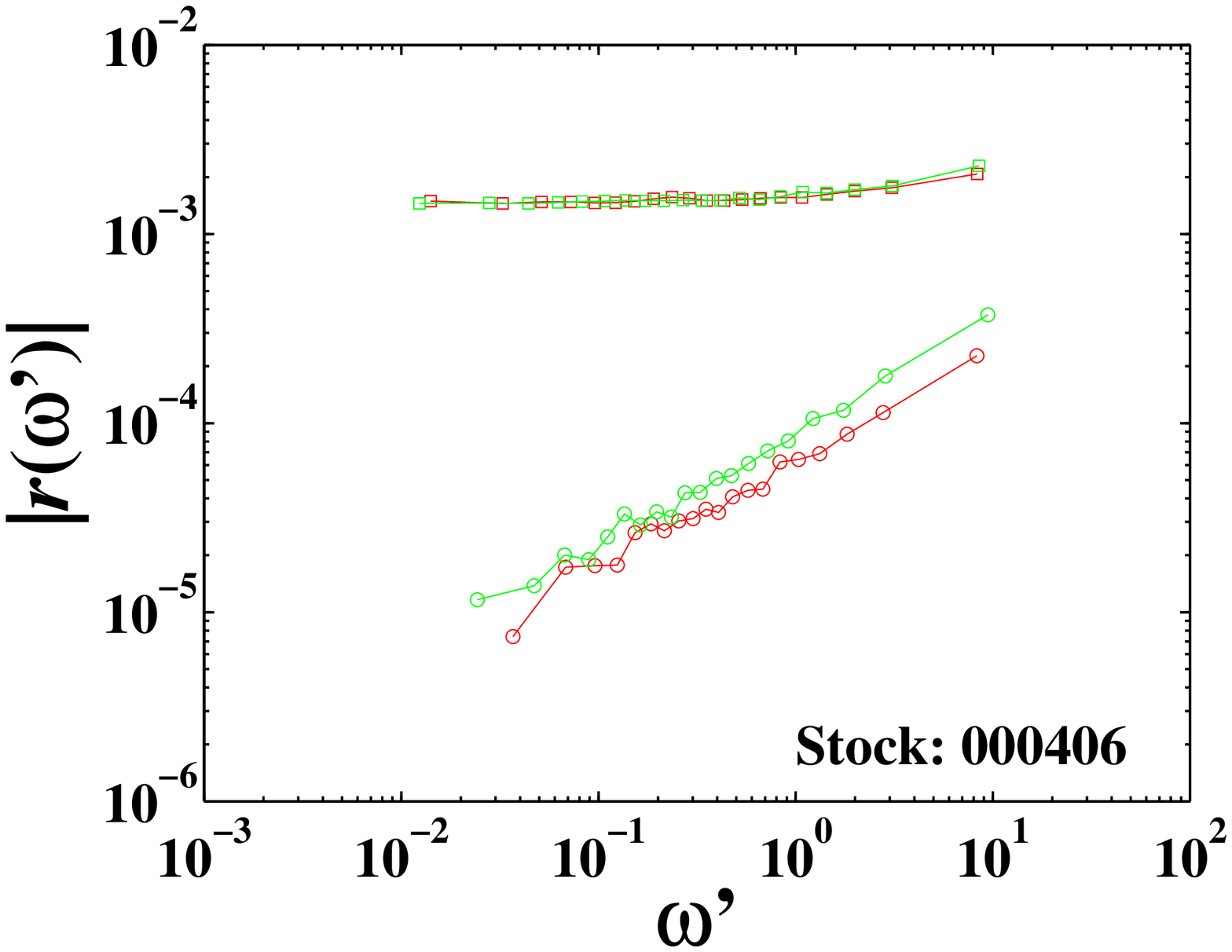}
\includegraphics[width=5cm]{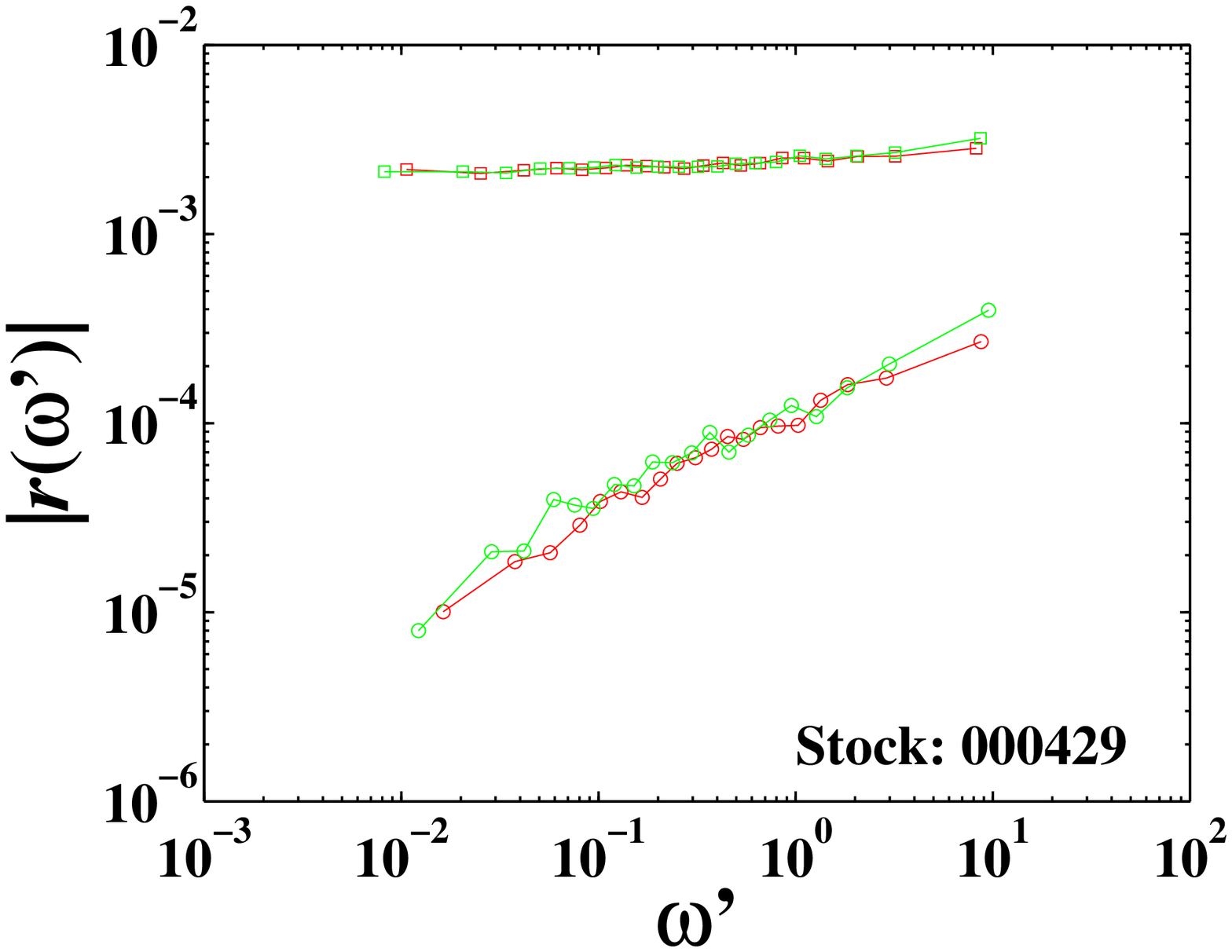}
\includegraphics[width=5cm]{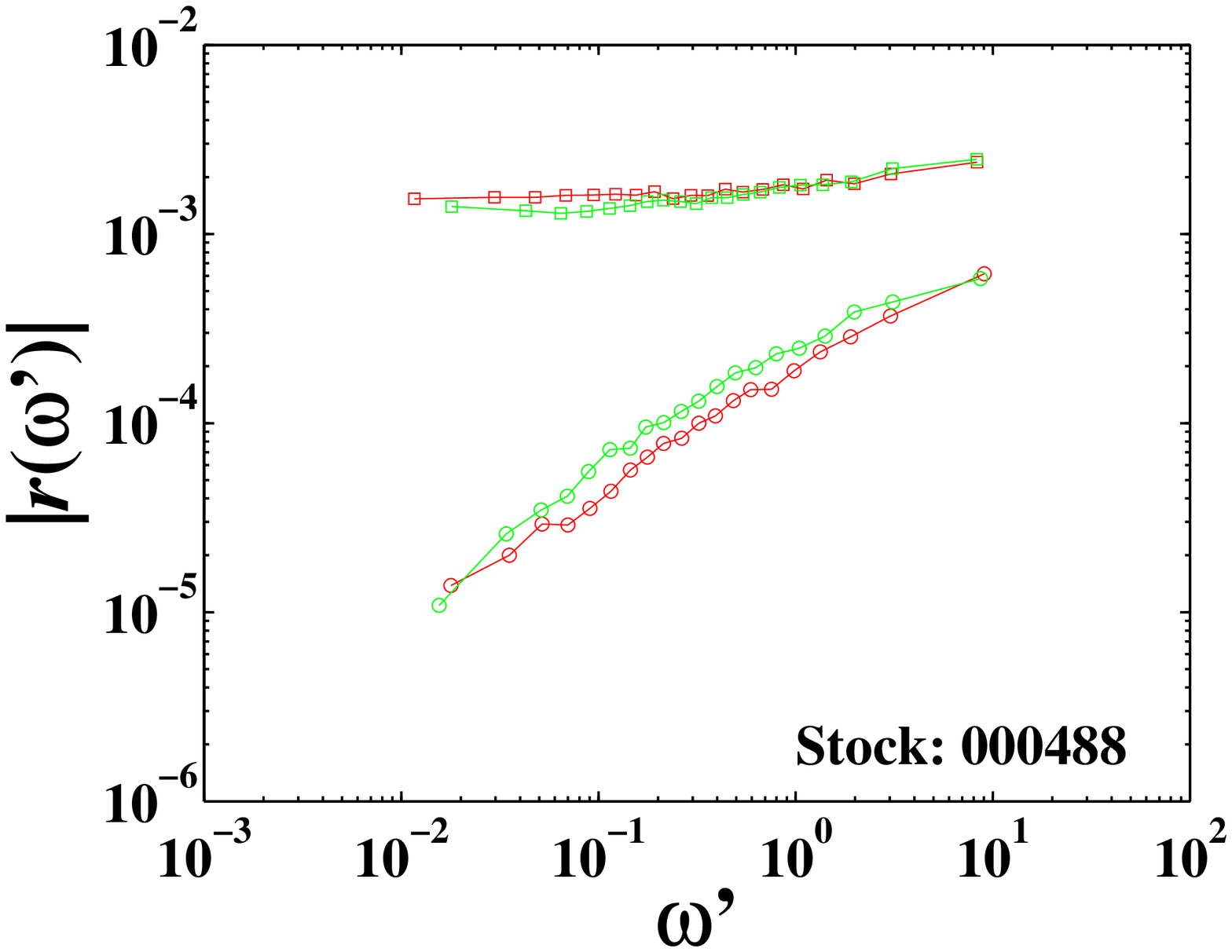}
\includegraphics[width=5cm]{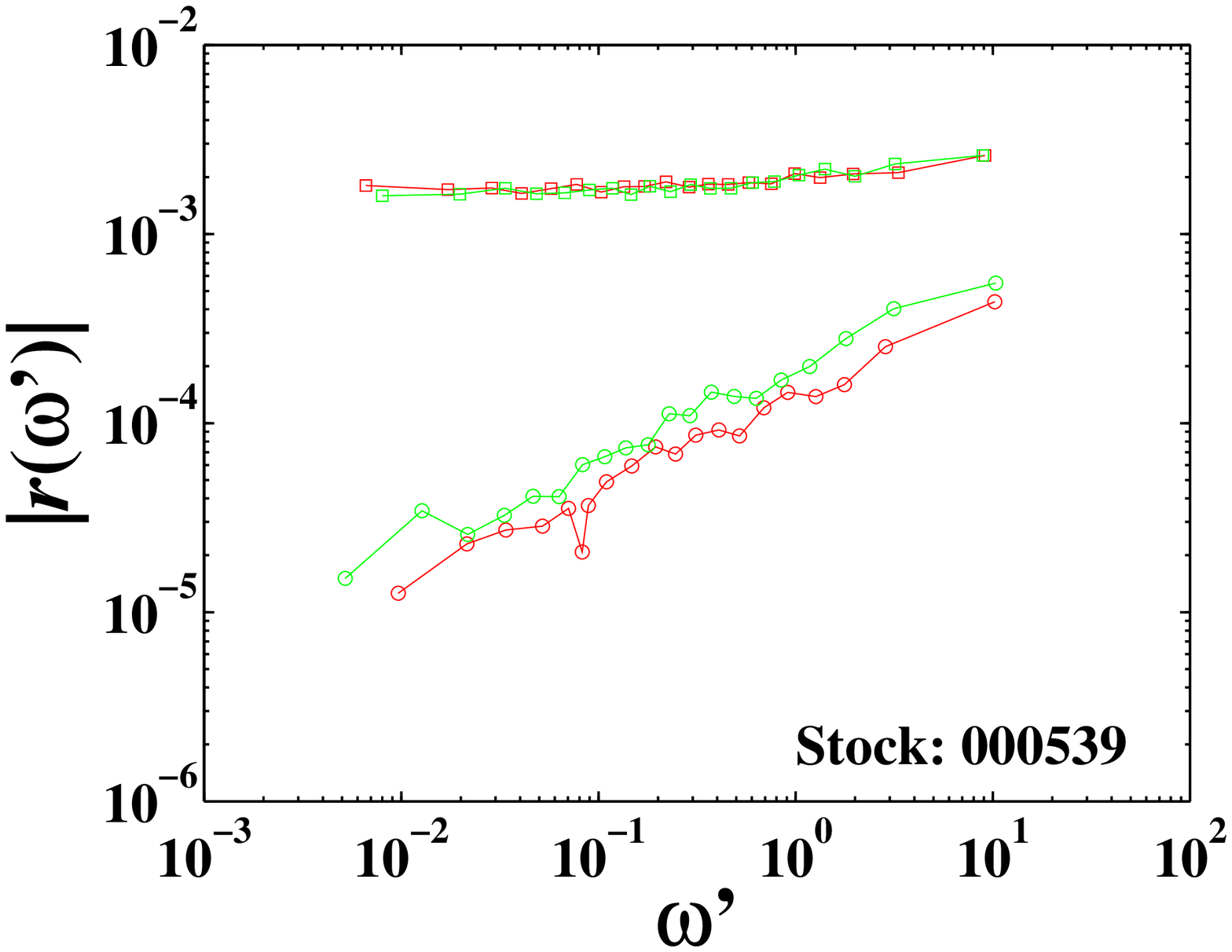}
\includegraphics[width=5cm]{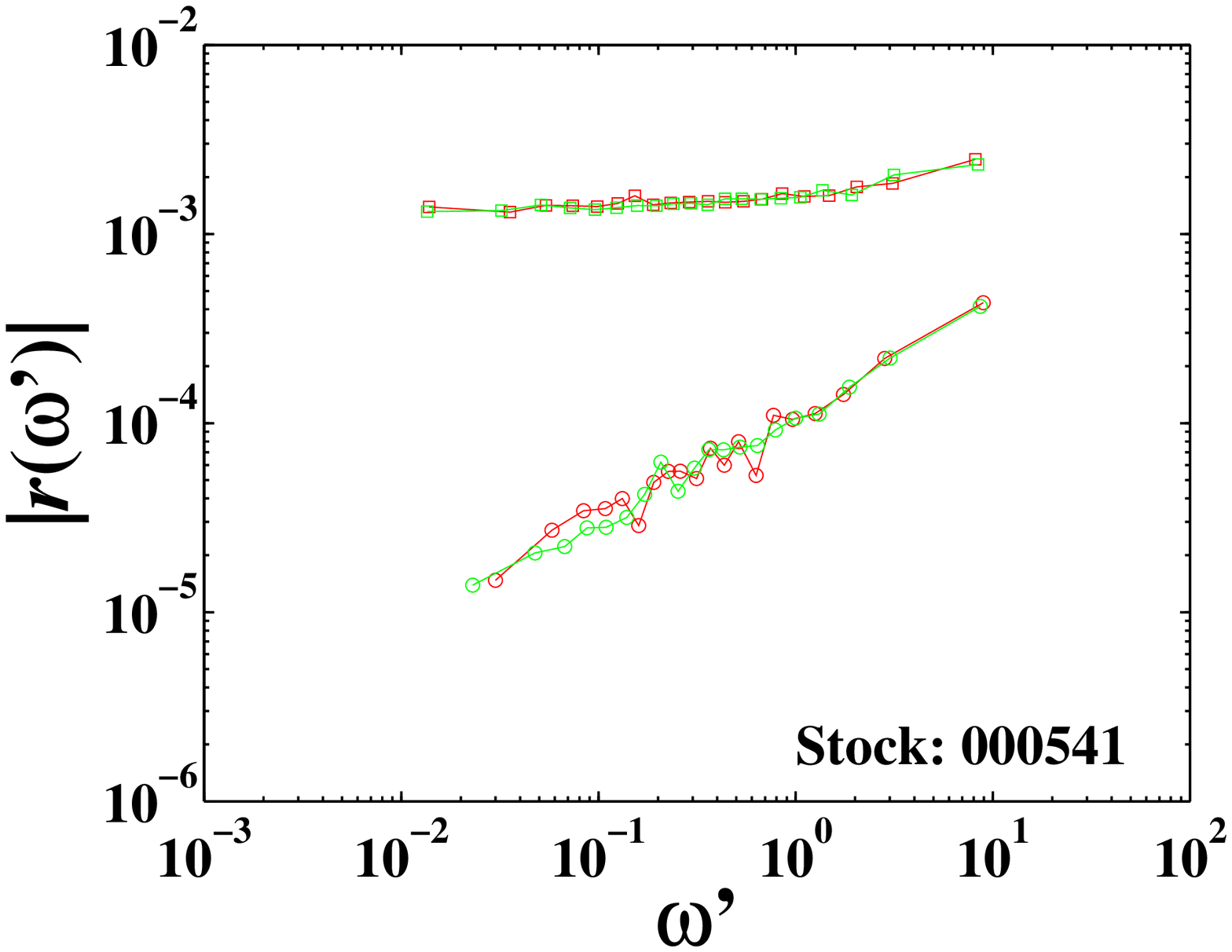}
\includegraphics[width=5cm]{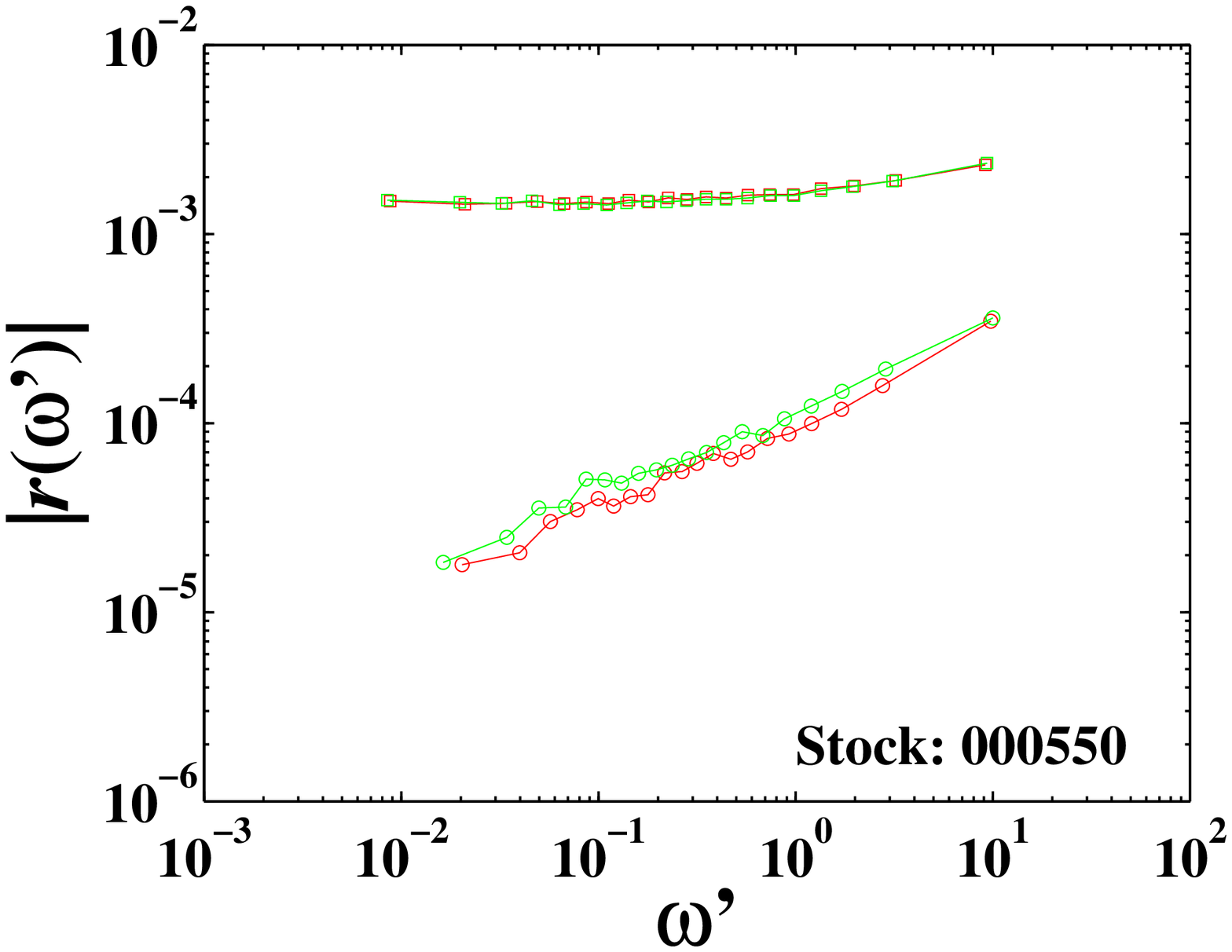}
\includegraphics[width=5cm]{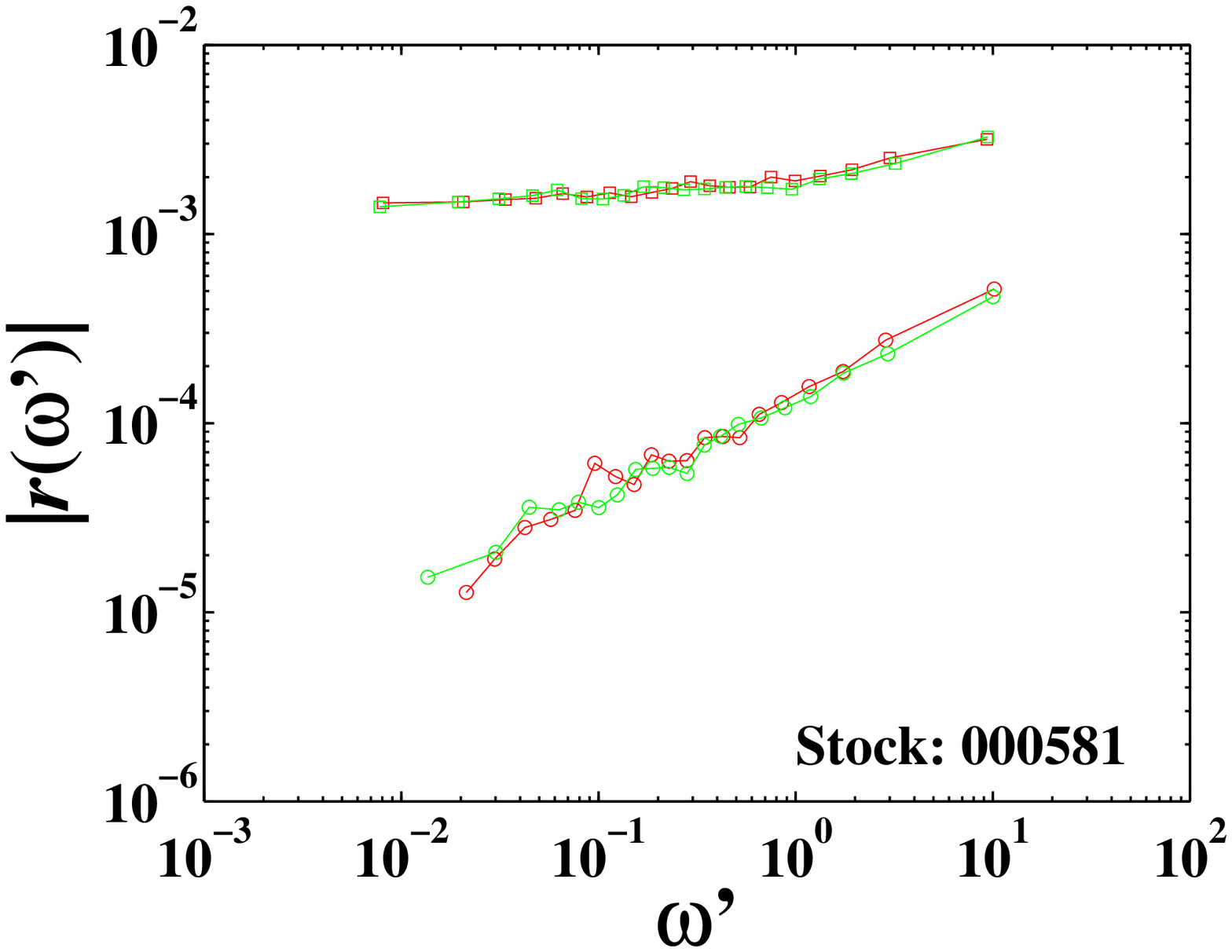}
\includegraphics[width=5cm]{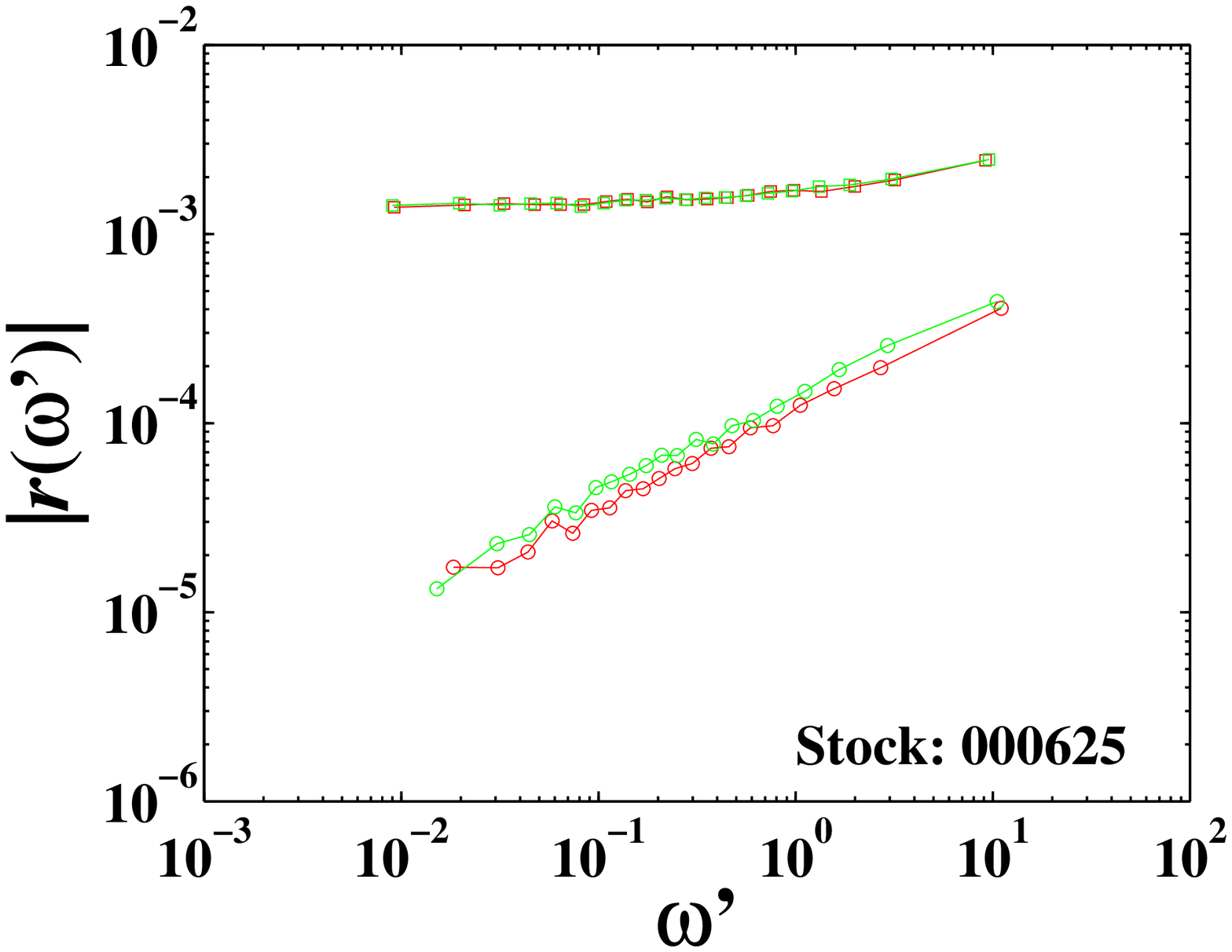}
\includegraphics[width=5cm]{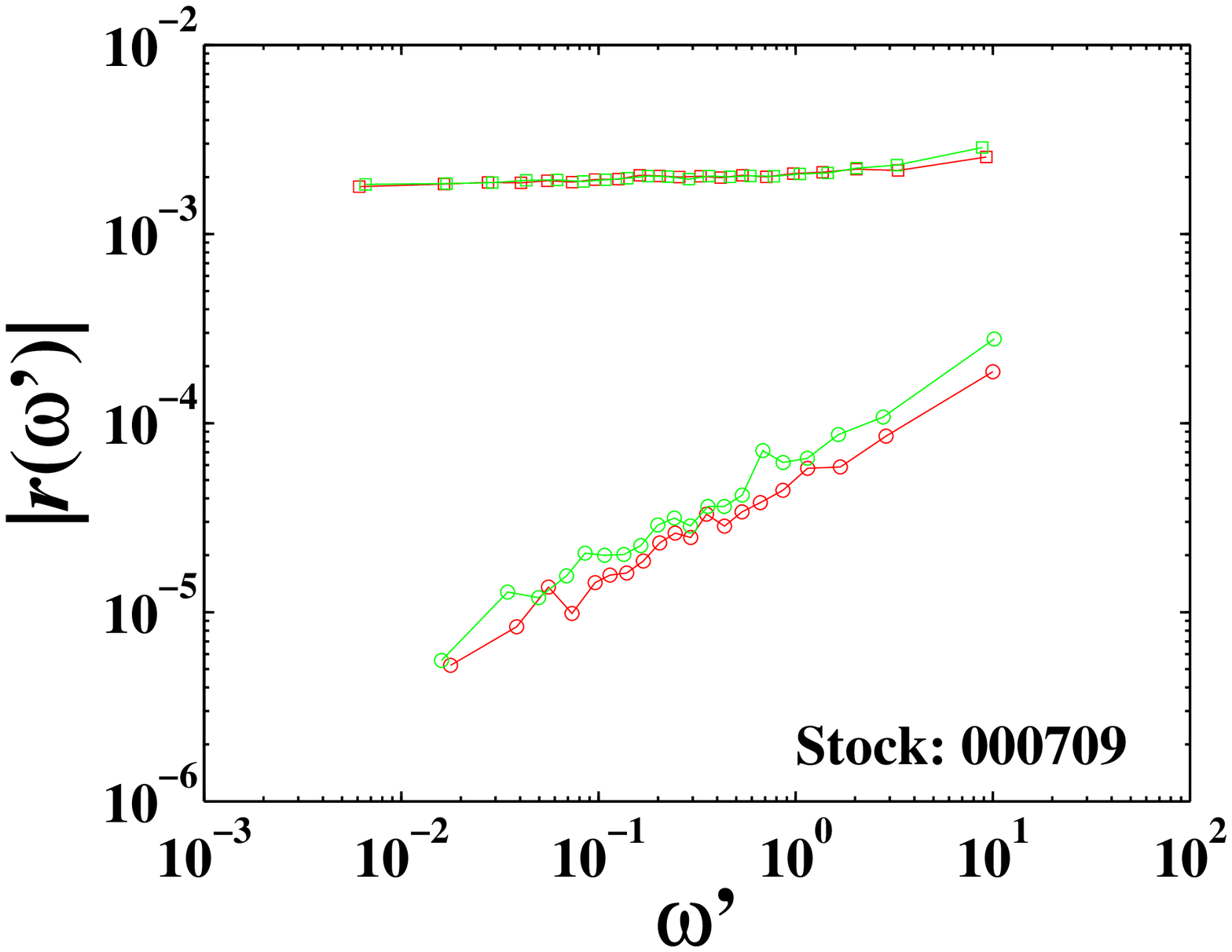}
\includegraphics[width=5cm]{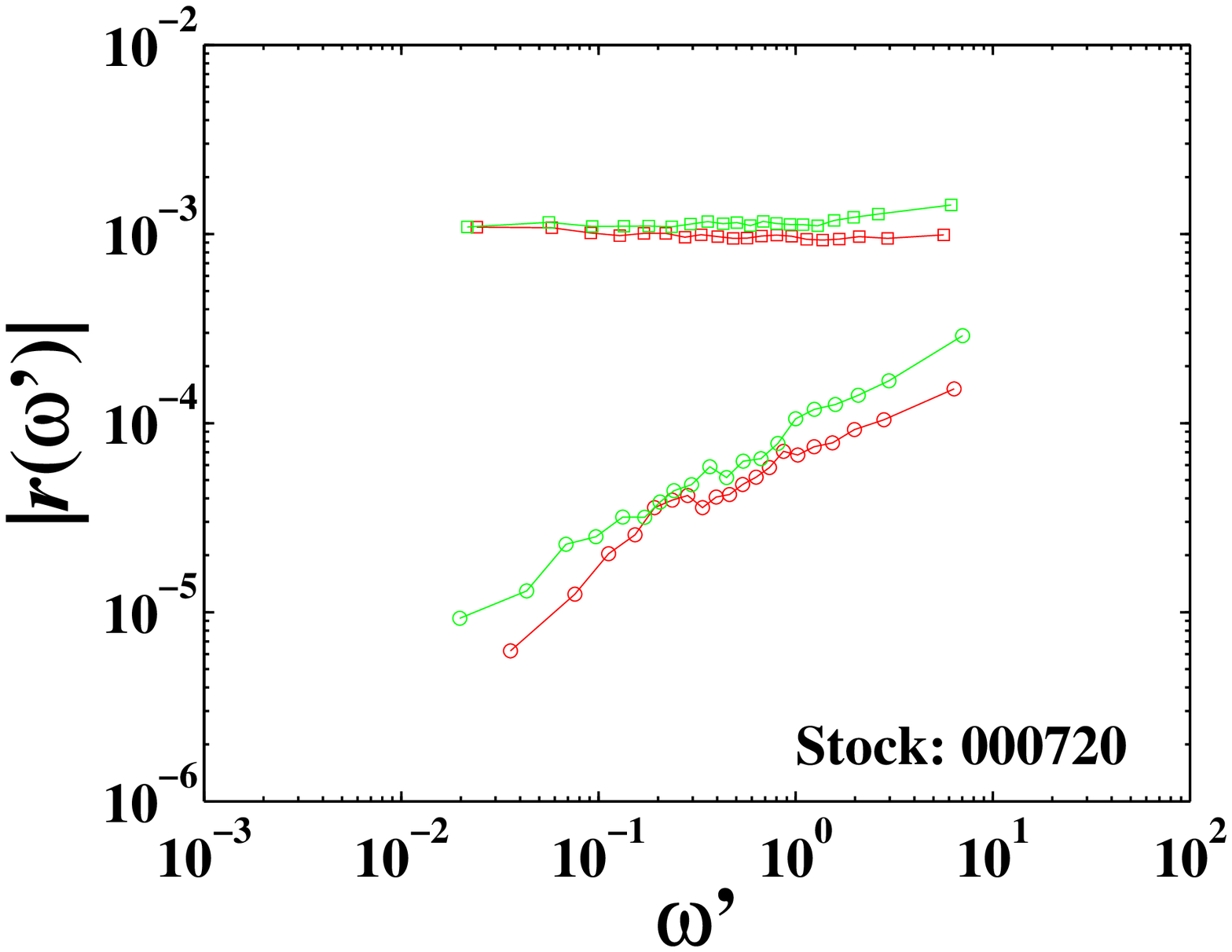}
\includegraphics[width=5cm]{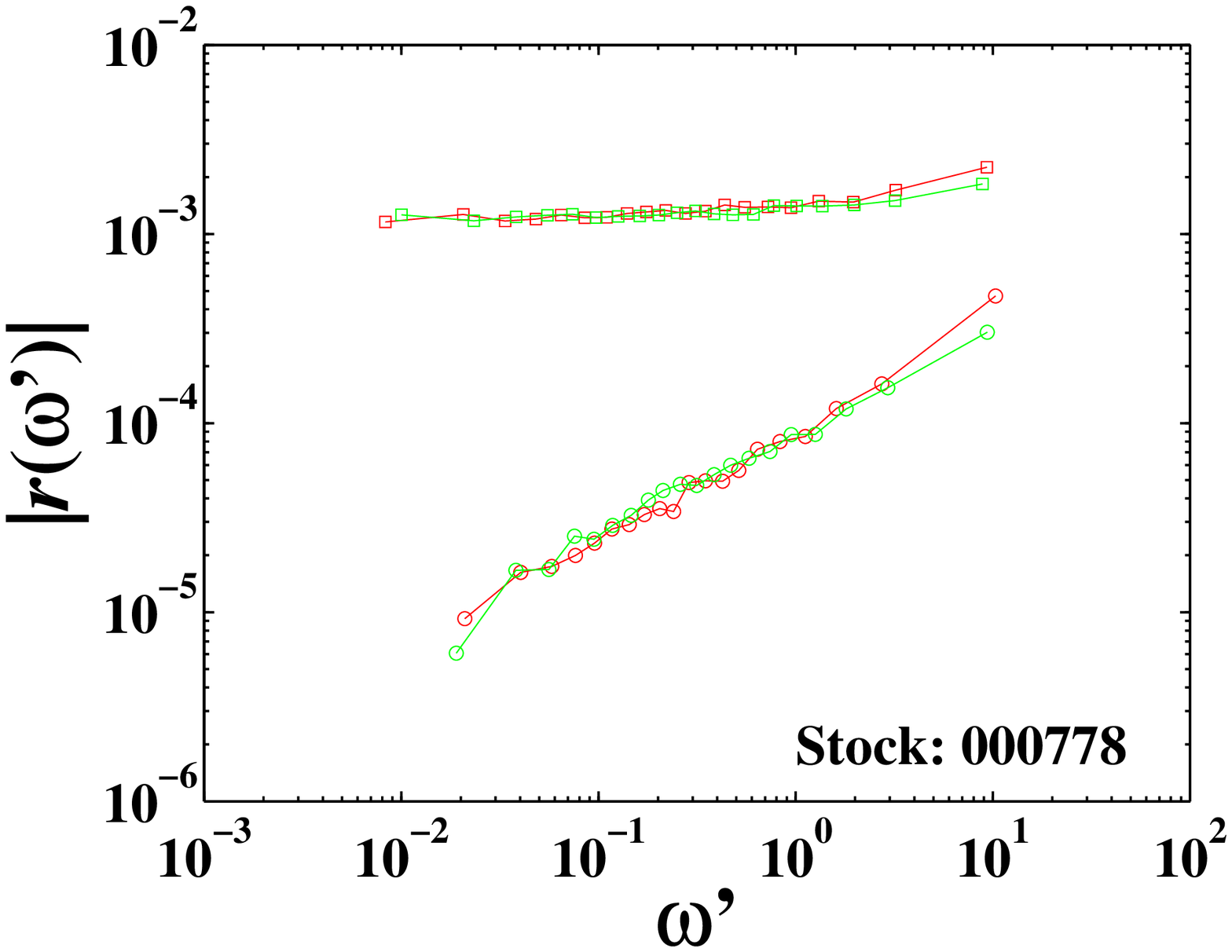}
\caption{Dependence of $|r|$ with respect to $x=\omega'$ for the
four types of trades for individual stocks. The normalized trade
size $x$ is obtained following \citet{Lim-Coggins-2005-QF}. Note
that the price impact of buyer- and seller-initiated trades is
symmetric and unfilled trades have greater price impact than filled
trades ({\em{continued}}).} \label{Fig:MPM:LC:Raw2}
\end{figure}

\newpage
\begin{figure}[htb]
\centering
\includegraphics[width=8cm]{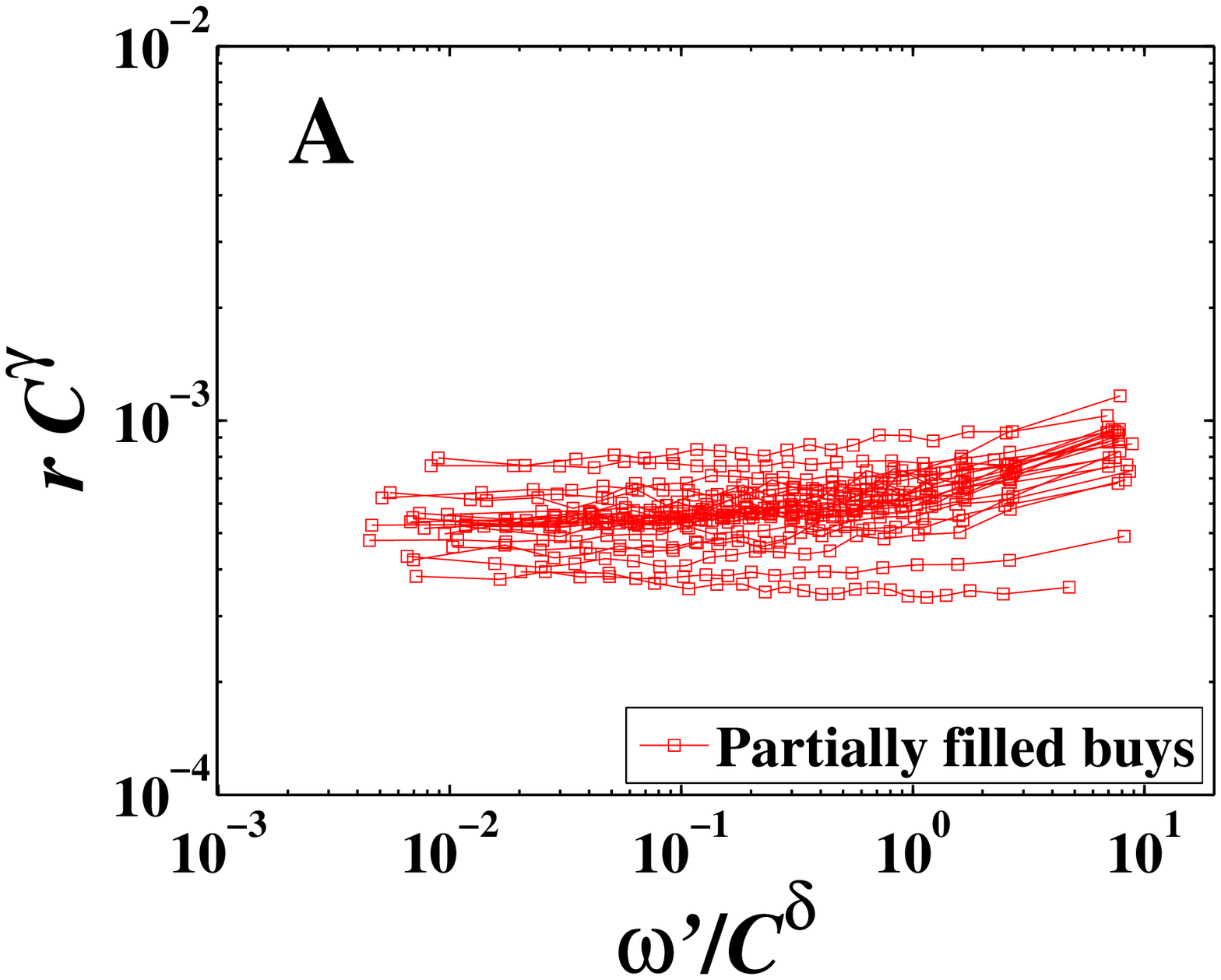}
\includegraphics[width=8cm]{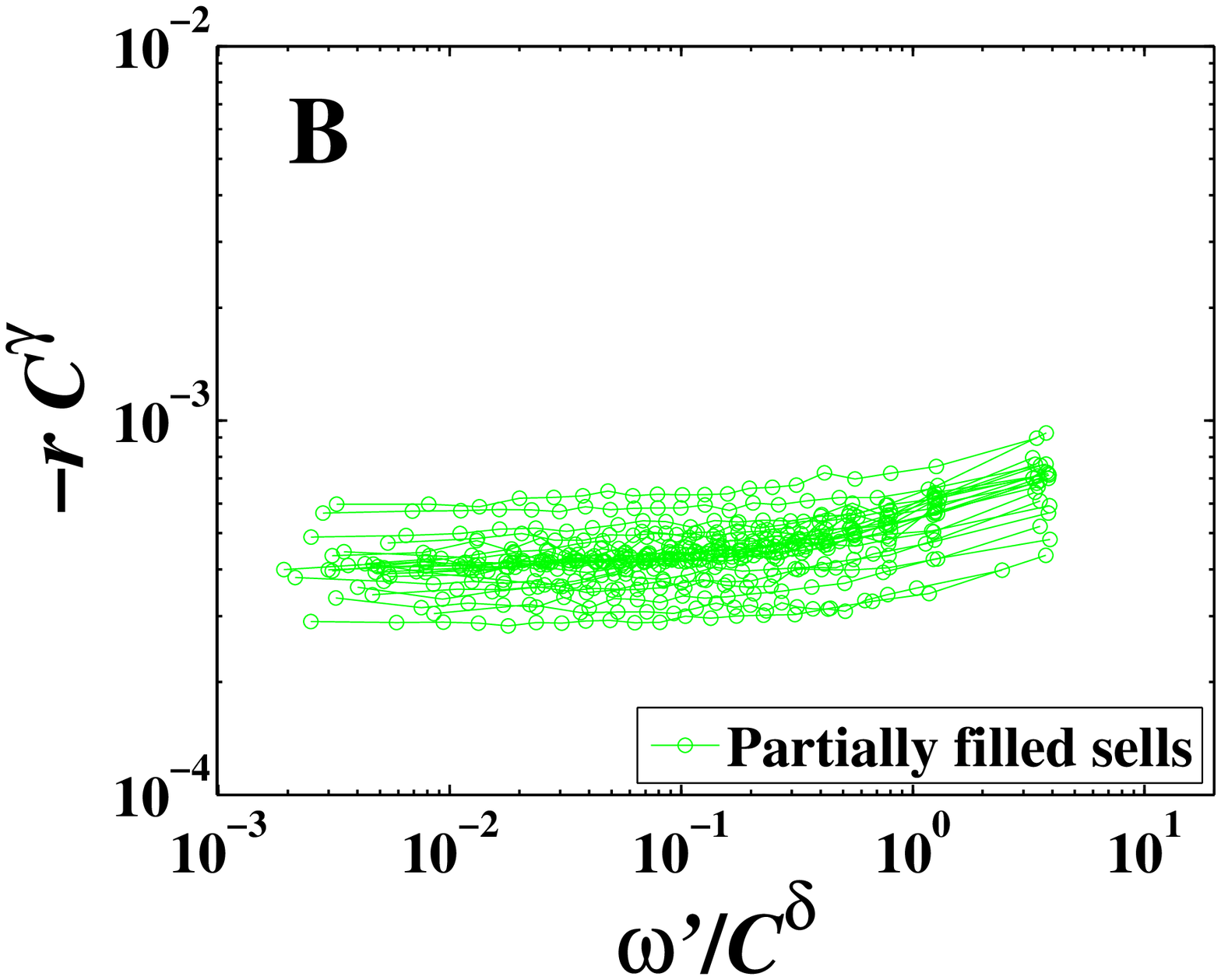}\\
\includegraphics[width=8cm]{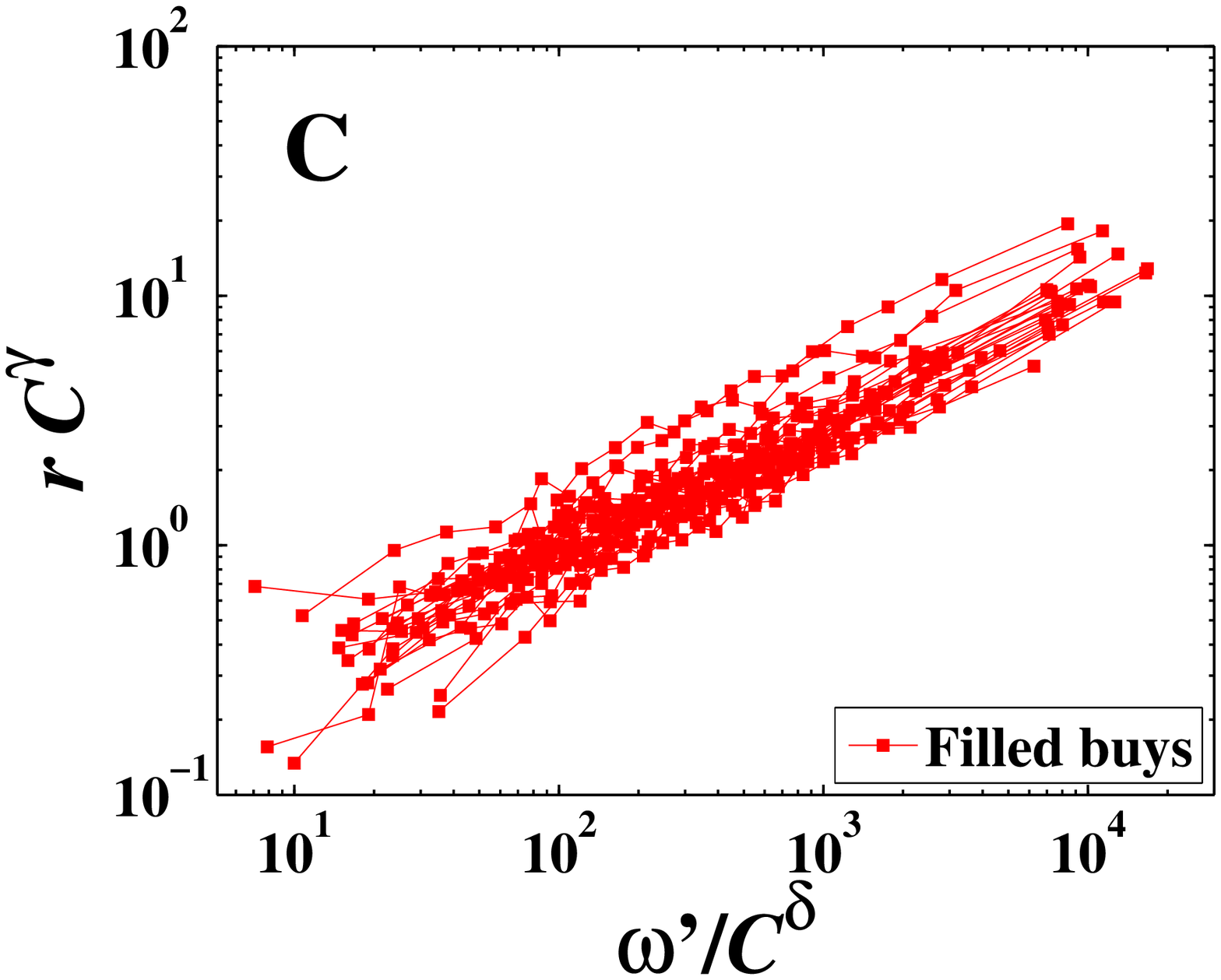}
\includegraphics[width=8cm]{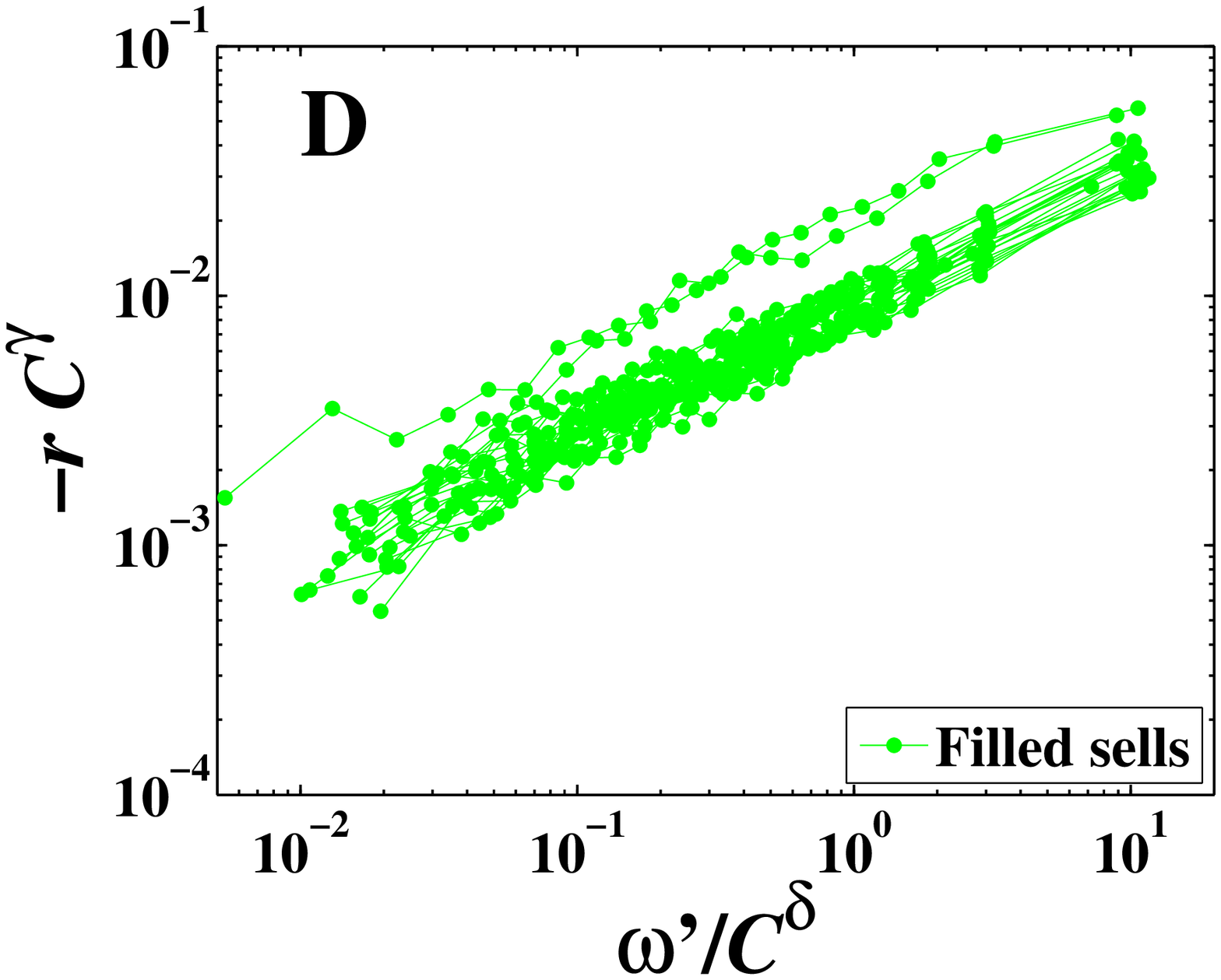}
\caption{Scaling analysis of the price impact functions of different
types of trades for the 23 individual stocks, following
\citet{Lim-Coggins-2005-QF}. The trades are classified into four
types due to their directions and aggressiveness. The values of
$\delta$ and $\gamma$ for each type of trades are listed in table
\ref{Tb:LC}.} \label{Fig:MPM:LC:Trades:20-23}
\end{figure}

\begin{figure}[tb]
\centering
\includegraphics[width=8cm]{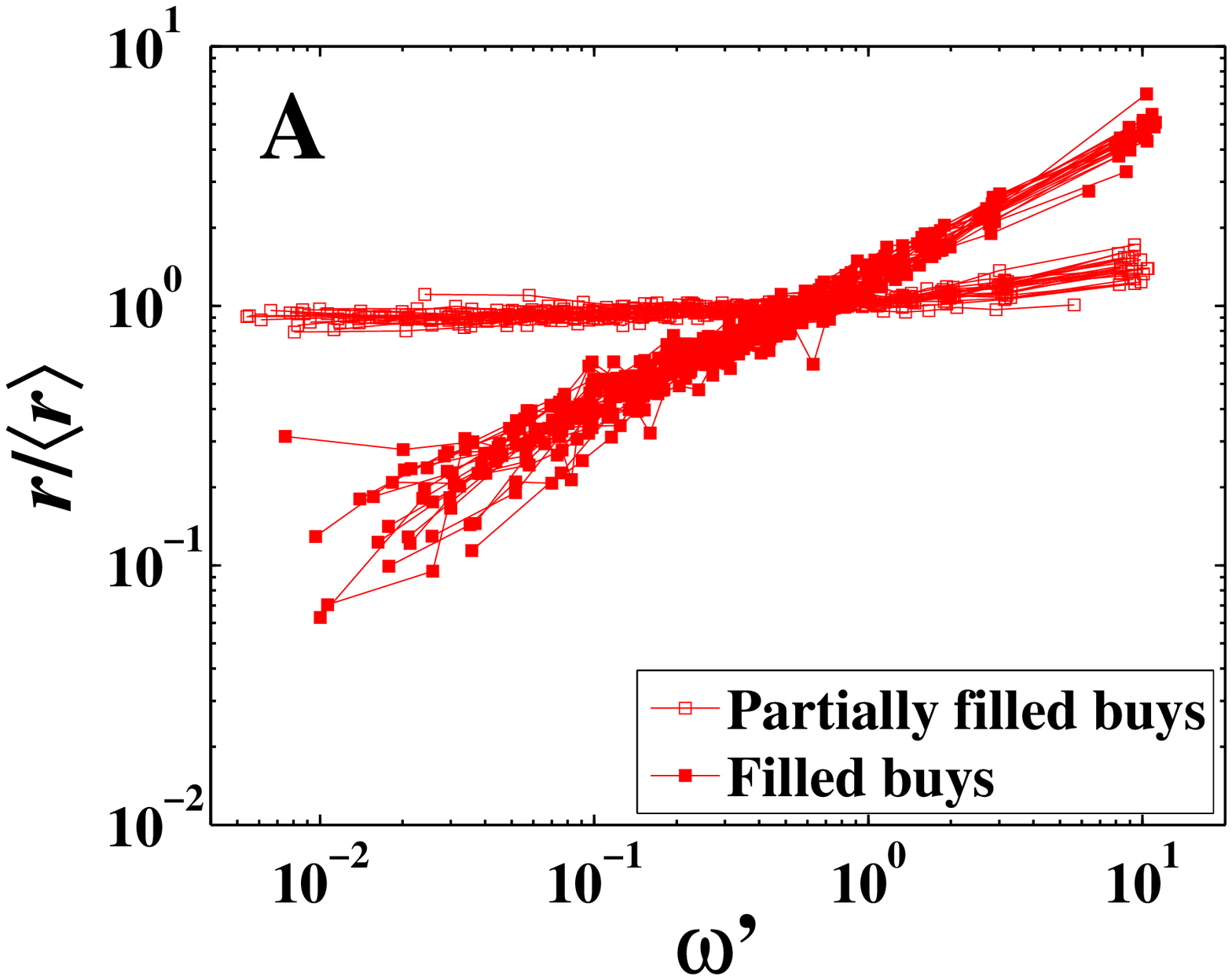}
\includegraphics[width=8cm]{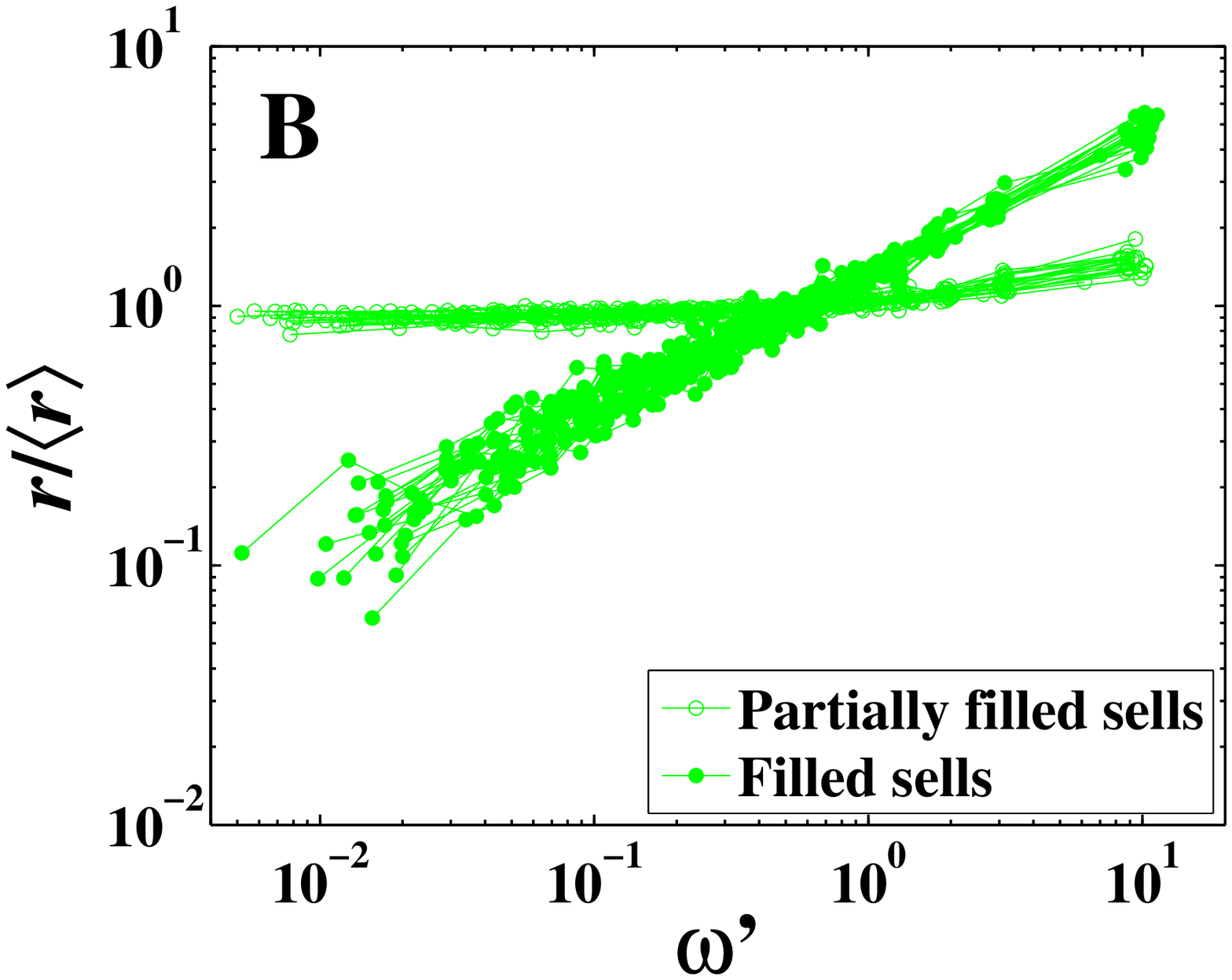}
\caption{Scaling of price impact functions for different types of
trades. The trade sizes are normalized following
\citet{Lim-Coggins-2005-QF}.} \label{Fig:MPM:LC:Trades:Normalized}
\end{figure}

\begin{figure}[htp]
\centering
\includegraphics[width=8cm]{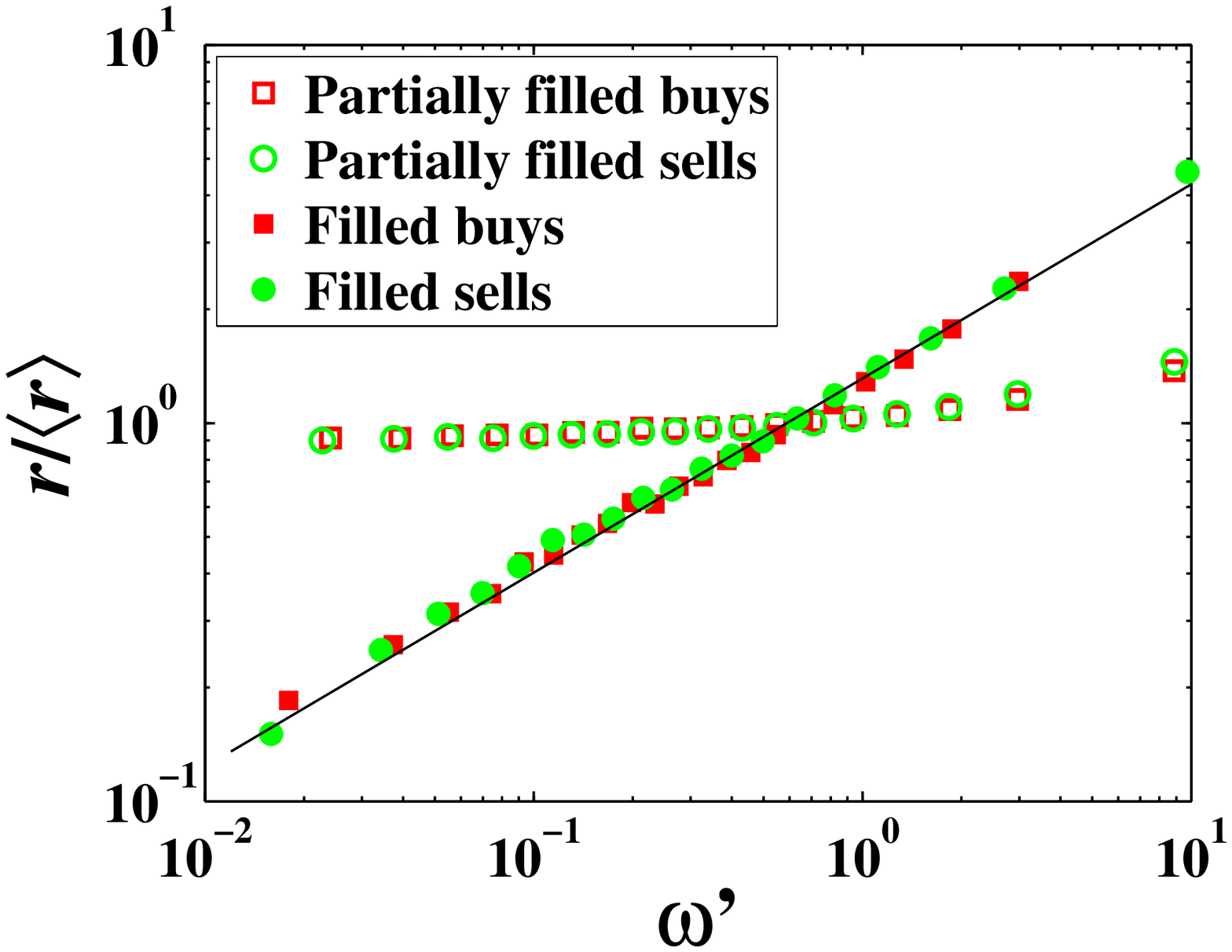}
\caption{Market averaged price impact curves for the four types of
trades. The trade sizes are normalized following
\citet{Lim-Coggins-2005-QF}. The solid line is a power-law function
with the exponent $\alpha=0.50$ to guide the eye.}
\label{Fig:MPM:LC:Trades}
\end{figure}

\end{document}